\RequirePackage{fix-cm}
\RequirePackage{rotating}

\documentclass[smallextended,natbib]{svjour3}       \smartqed  \usepackage[T1]{fontenc}
\usepackage[utf8]{inputenc}
\usepackage[english]{babel}

\usepackage{graphicx}
\usepackage{multicol}
\usepackage{caption}
\usepackage{subcaption}

\usepackage{color}
\usepackage{textcomp}
\usepackage{listings}

\definecolor{source}{gray}{1}\newcommand{\myCommentStyle}[1]{{\footnotesize\sffamily\color{gray!100!white} #1}}

\newcommand{\myStringStyle}[1]{{\footnotesize\sffamily\color{violet!100!black} #1}}

\newcommand{\mySymbolStyle}[1]{{\footnotesize\sffamily\color{violet!100!black} #1}}

\newcommand{\myKeywordStyle}[1]{{\footnotesize\sffamily\color{green!70!black} #1}}

\newcommand{\myGlobalStyle}[1]{{\footnotesize\sffamily\color{blue!100!black} #1}}

\makeatletter
\newcommand*\idstyle[1]{\expandafter\id@style\the\lst@token{#1}\relax }
\def\id@style#1#2\relax{\ifnum\pdfstrcmp{#1}{\#}=0\mySymbolStyle{\the\lst@token}\else \edef\tempa{\uccode`#1}\edef\tempb{`#1}\ifnum\tempa=\tempb \myGlobalStyle{\the\lst@token}\else \the\lst@token \fi \fi }
\makeatother

\lstnewenvironment{code}{\lstset{frame=single,
    framerule=0pt,
    mathescape=false
    }\noindent \minipage{\linewidth}}{\endminipage }

\lstnewenvironment{codeWithLineNumbers}{\lstset{frame=single,
    framerule=0pt,
    mathescape=false,
    numbers=left
    }\noindent \minipage{\linewidth}}{\endminipage }

\usepackage{float} \usepackage{hyperref} \usepackage{amssymb} \usepackage{amsmath} \usepackage{xspace}\usepackage{color}\usepackage{xcolor}\usepackage{fancybox}\usepackage{fancyhdr}\usepackage{ifthen} \usepackage{paralist} \usepackage{enumitem} \usepackage[normalem]{ulem} \usepackage{listings} \usepackage{multicol} \usepackage[linesnumbered]{algorithm2e} \usepackage[noend]{algpseudocode} \usepackage{fixltx2e} \usepackage[breakall]{truncate}
\usepackage{collcell}
\usepackage{xfrac} \usepackage{tikz} \usetikzlibrary{trees} \usepackage{pgfplots} \pgfplotsset{compat=1.8} \usepgfplotslibrary{statistics} \usepackage{booktabs} \usepackage{longtable} \usepackage{rotating} \usepackage{array} \usepackage{caption}
\usepackage{lineno}

\usepackage{multirow}

\DeclareCaptionFormat{cont}{#1 (cont.)#2#3\par}

\colorlet{BLUE}{blue}

\algnewcommand{\LineComment}[1]{\State \(\triangleright\) #1}

\newcommand{\hypobox}[1]{
        \begin{center}\noindent\thicklines\setlength{\fboxsep}{8pt}\cornersize{0.1}
        \ovalbox{\begin{minipage}{0.9\linewidth}#1
                \end{minipage}}
        \end{center}}

\newcommand{\addressedRvw} [1] {#1}

\newenvironment{Spoonfeed}{\begin{compactitem} \item [$\implies$]\begin{itshape}}
	{\end{itshape}\end{compactitem}}

\newcommand{\toolname}[1]{\textsc{\textsf{#1}}\xspace}        
\newcommand{\smallamp}[0]{\textsc{\textsf{Small-Amp}}\xspace}
\newcommand{\smallampurl}[0]{https://github.com/mabdi/small-amp}
\newcommand{\experimentsurl}[0]{https://github.com/mabdi/SmallAmp-evaluations}
\newcommand{\dspot}[0]{\textsc{\textsf{DSpot}}\xspace}

\newcommand{\inlinecode}[1]{\texttt{#1}}
\newcommand{\comment}[1]{``\textit{#1}''}

\newcommand{\secref}[1]{Section~\ref{#1}\xspace}
\newcommand{\figref}[1]{Figure~\ref{#1}\xspace} 
\newcommand{\lstref}[1]{Listing~\ref{#1}\xspace} 
\newcommand{\algoref}[1]{Algorithm~\ref{#1}\xspace} 
\newcommand{\tabref}[1]{Table~\ref{#1}\xspace}

\newcommand{\secrefpage}[1]{Section~\ref{#1}~--~p.~\pageref{#1}\xspace}

\newcommand{\lstrefpage}[1]{Listing~\ref{#1}~--~p.~\pageref{#1}\xspace}

\newcommand{\algorefpage}[1]{Algorithm~\ref{#1}~--~p.~\pageref{#1}\xspace}
\newcommand\testClassNameShorten[1]{\truncate{4cm}{#1}}

\newcolumntype{W}{>{\collectcell{\testClassNameShorten}}c<{\endcollectcell}}
\newcommand{\rowhighlightOne}[0]{\(^*\) }

\newcommand{\noProfilingText}[0]{$\circ\circ\circ$}

\newcolumntype{$}{>{\global\let\currentrowstyle\relax}}
\newcolumntype{^}{>{\currentrowstyle}}
\newcommand{\rowstyle}[1]{\gdef\currentrowstyle{#1}#1\ignorespaces
}
\newcommand{\noProfiling}[0]{\rowstyle{\itshape}\noProfilingText}

\newcommand{\paperTitle}{\smallamp: Test Amplification in a Dynamically Typed Language}
\newcommand{\myAbstract}{Rewrite abstract later}
\newcommand{\submitTo}{EMSE Journal}

\hypersetup{
    colorlinks,citecolor=black,filecolor=black,linkcolor=black,urlcolor=gray,
    linktocpage=true,
    bookmarks=true,
    bookmarksopen=true,
    pdfpagemode=UseOutlines,
pdftitle={\paperTitle},    pdfauthor={Mehrdad Abdi et. al.},     pdfsubject={Draft for \submitTo},   pdfcreator={PDF LaTex},   }
 
\begin{document}

\pagenumbering{arabic}
\setcounter{page}{1}

\newcommand{\nSentPR}{11~}
\newcommand{\nMergedPR}{8~}
\newcommand{\percPR}{$\approx$72\%}
\newcommand{\nAmplified}{28~}

\renewcommand{\myAbstract}{
Some test amplification tools extend a manually created test suite with additional test cases to increase the code coverage.
The technique is effective, in the sense that it suggests strong and understandable test cases, generally adopted by software engineers.
Unfortunately, the current state-of-the-art for test amplification heavily relies on program analysis techniques which benefit a lot from explicit type declarations present in statically typed languages.
In dynamically typed languages, such type declarations are not available and as a consequence test amplification has yet to find its way to programming languages like Smalltalk, Python, Ruby and Javascript.
We propose to exploit profiling information ---readily obtainable by executing the associated test suite--- to infer the necessary type information creating special test inputs with corresponding assertions.
We evaluated this approach on 52 selected test classes from 13 mature projects in the Pharo ecosystem containing approximately 400 test methods.
We show the improvement in killing new mutants and mutation coverage at least in \nAmplified out of 52 test classes ($\approx$53\%).
Moreover, these generated tests are understandable by humans: \nMergedPR out of \nSentPR pull-requests submitted were merged into the main code base (\percPR).
These results are comparable to the state-of-the-art, hence we conclude that test amplification is feasible for dynamically typed languages.
}

\title{\paperTitle}

\titlerunning{\smallamp: \ldots (submitted to \submitTo)}        

\author{Mehrdad Abdi \and Henrique Rocha \and Serge Demeyer \and Alexandre Bergel
}

\institute{Mehrdad Abdi, Serge Demeyer \at
	University of Antwerp -- Flanders Make vzw, Belgium\\
	\email{mehrdad.abdi@uantwerpen.be, serge.demeyer@uantwerpen.be}
	\and
	Henrique Rocha \at
	Loyola University Maryland\\
	\email{henrique.rocha@gmail.com}
	\and
	Alexandre Bergel \at
	RelationalAI \\
	ISCLab, Department of Computer Science (DCC), University of Chile \\
	\email{abergel@dcc.uchile.cl}
}

\date{Received: date / Accepted: date}

\maketitle
  
\begin{abstract}
\myAbstract

\end{abstract}

\section{Introduction}

Modern software projects contain a considerable amount of hand-written tests which assure that the code does not regress when the system under test evolves.
Indeed, several researchers reported that test code is sometimes larger than the production code under test~\citep{DanielASE2009,Tillman2006,ZaidmanEMSE2011}.
More recently, during a large scale attempt to assess the quality of test code, Athanasiou \etal\xspace reported six systems where test code takes more than 50\% of the complete codebase~\citep{Athanasiou2000TSE}.
Moreover, Stack Overflow posts mention that test to code ratios between 3:1 and 2:1 are quite common~\citep{StackOverflowTestToCodeRatio}.

\emph{Test amplification} is a field of research which exploits the presence of these manually written tests to strengthen existing test suites~\citep{Danglot2019JSS}.
The main motivation of test amplication is based on the observation that manually written test cases mainly exercise to the default scenarios and seldom cover corner cases.
Nevertheless, experience has shown that strong test suites must cover those corner cases in order to effectively reveal failures~\citep{Li2016testoracle}.
Test amplification therefore automatically transforms test-cases in order to exercise the boundary conditions of the system under test.

Danglot \etal\xspace conducted a literature survey on test amplification, identifying a range of papers that take an existing test suite as the seed value for generating additional tests~\citep{Fraser2012seeds, rojas2016seeding, yoo2012test}.
This culminated in a tool named \emph{\dspot} which represents the state-of-the-art in the field~\citep{Baudry2015corr, Danglot2019EMSE}.
In these papers, the authors demonstrate that \dspot is effective, in the sense that the tool is able to automatically improve 26  test classes (out of 40)  by triggering new behaviors and adding valuable assertions.
Moreover, test cases generated with \dspot are well perceived by practitioners --- 13 (out of 19) pull requests with amplified test have been incorporated in the main brach of existing open source projects~\citep{Danglot2019EMSE}.

Unfortunately, the current state-of-the-art for test amplification heavily relies on program analysis techniques which benefit a lot from explicit type declarations present in statically typed languages. 
Not surprisingly, previous research has been confined to statically typed programming languages including Java, C, C++, C\#, Eiffel~\citep{Danglot2019JSS}.
In dynamically typed languages, performing static analysis is difficult since source code does not embed type annotation when defining variable.
As a consequence test amplification has yet to find its way to dynamically-typed programming languages including Smalltalk, Python, Ruby, Javascript, etc.

In this paper, we demonstrate that test amplification is feasible for dynamically typed languages by exploiting profiling information readily available from executing the test suite.
As a proof of concept, we present \smallamp which amplifies test cases for the dynamically typed language Pharo \citep{black2010pharo,Berg13f}; a variant of Smalltalk \citep{Goldberg-Smalltalk-80}.
We argue that Pharo is a good vehicle for such a feasibility study, because it is purely object-oriented and it comes with a powerful program analysis infrastructure based on metalinks \citep{costiou2020sub}. Pharo uses a minimal computation model, based on object and message passing, thus reducing possibilities to experiences biases due to some particular and singular language constructions. 
Moreover, Pharo has a growing and active community with several open source projects welcoming pull requests from outsiders.
Consequently, we replicate the experimental set-up of \dspot~\citep{Danglot2019EMSE} by including a quantitative and qualitative analysis of the improved test suite.

\paragraph{}This paper is an extension of a previous paper presenting the proof-of-concept to the Pharo community~\citep{abdi2019test}.
As such, we make the following contributions:
\begin{compactitem}
\item \emph{Small-Amp}, a test amplification algorithm and tool, implemented in Pharo Smalltalk. To the best of our knowledge this is the first test amplification tool for a dynamically typed language.
\item Demonstrating the use of \emph{dynamic type profiling}
as a substitute for type declarations within a system under test.
\item \emph{Quantitative evaluation} of our test amplification for the Pharo dynamic programming language on 13 mature projects with good testing and maintenance practices.
We repeated the experiment three times.
For \nAmplified out of 52 test classes we see an improvement in killing new mutants and consequently the mutation score.
Our evaluation shows that generated test methods are focused (i.e. they do not overwhelm the developer) and all amplification steps are necessary to obtain strong and understandable tests.
\item \emph{Qualitative evaluation} of our approach by submitting pull requests containing amplified tests on \nSentPR active projects. \nMergedPR of them (\percPR) were accepted and successfully merged into the main branch. 
\item We contribute to \emph{open science} by releasing our tool as an open-source package under the MIT license  (\url{\smallampurl}). The experimental data is publicly available as a replication package (\url{\experimentsurl}). 
\end{compactitem}

\noindent
The remainder of this paper is organised as follows.
\secref{sec:background}, provides the necessary background information on test amplification and the Pharo ecosystem.
\secref{sec:smallamp} and \secref{sec:extrasDSpot} explain the inner workings of \smallamp, including the use of dynamic profiling as a substitute for static type information. 
\secref{sec:replication} discusses the quantitative and qualitative evaluation performed on 13 mature open source projects; a replication of what is reported by Danglot et. al.~\citep{Danglot2019JSS}.
\secref{sec:threats} enumerates the threats to validity while  \secref{sec:relatedworks} discusses related work and \secref{sec:futureworks} lists limitations and future work. 
\secref{sec:conclusion} summarizes our contributions and concludes our paper.

\section{Background}\label{sec:background}

\subsection{Test amplification}\label{sec:testamp}

In their survey paper, Danglot \etal\xspace define test amplification as follows: 
\begin{quote}
\textit{Test amplification consists of exploiting the knowledge of a large number of test cases, in which developers embed meaningful input data and expected properties in the form of oracles, in order to enhance these manually written tests with respect to an engineering goal (e.g., improve coverage of changes or increase the accuracy of fault localization).} \citep{Danglot2019JSS}
\end{quote}

\noindent
Test amplification is a not replacement for other test generation techniques and should be considered as a complementary solution. 
The main difference between test generation and test amplification is the use of an existing test suite.
Most work on test generation accept only the program under test or formal specifications and ignore the original test suite which is written by an expert.

\vspace{1em}
\noindent
A typical test amplification tool is based on two complementary steps.
\begin{compactenum}
\item[(i)] \emph{Input amplification.}
The existing test code is altered in order to force previously untested paths.
This involves changing the set-up of the object under test, providing parameters that represent boundary conditions.
Additional calls to state-changing methods of the public interface are injected as well.
\item[(ii)] \emph{Assertion amplification.}
Extra assert statements are added to verify the expected output of the previously untested path.
The system under test is then used as an oracle: while executing the test the algorithm inspects the state of the object under test and asserts the corresponding values.
\end{compactenum}

\noindent
The input amplification step is typically governed by a series of \emph{amplification operators}.
These operators represent syntactical changes to the test code that are likely to force new paths in the system under test.
To verify that this is indeed the case, the amplification tool compares the (mutation) coverage before and after the amplification operator.
It is beyond the scope of this paper to explain the details of mutation coverage; we refer the interested reader to the survey by \cite{Papadakis2019}.

We illustrate the input and assertion amplification steps via an example based on \inlinecode{SmallBank}\footnote{Available at: \url{https://github.com/mabdi/smalltalk-SmallBank}} and its test class \inlinecode{SmallBankTest} in \lstref{listing:amplificationcodesample}. 
In this example \inlinecode{testWithdraw} is the original test method while \inlinecode{testWithdrawAll} and \inlinecode{testWithdrawOnZero} are two new test methods derived from it.
In the \inlinecode{testWithdrawAll}, the input amplification has changed the literal value of \inlinecode{100} with \inlinecode{30} (line 19), and the assertion-amplification step regenerated the assertions on the balance (line 20) and added a missing assertion on the status of the operation (line 22).
The \inlinecode{testWithdrawAll} test method thus verifies the boundary condition of withdrawing by an amount equal to the balance.
In the \inlinecode{testWithdrawOnZero}, on the other hand, an input amplifier has removed the call to the \inlinecode{deposit:} method in line 11. This test method now verifies the boundary condition that calling a \inlinecode{withdraw:} with an amount more than zero when the balance is zero is not allowed. This is illustrated by the extra assertions in line 29 and 30.

\begin{multicols}{2}[\captionof{lstlisting}{\inlinecode{testWithdraw} amplified into \inlinecode{testWithdrawOnAll} and \inlinecode{testWithdrawOnZero}}]
\begin{lstlisting}[label=listing:amplificationcodesample]
SmallBank >> withdraw: amount
	balance >= amount
		ifTrue: [ 
			balance := balance - amount.
			^ true ].
	^ false

SmallBankTest >> testWithdraw
   | b |
   b := SmallBank new.
   b deposit: 100.
   self assert: (b balance = 100).
   b withdraw: 30.
   self assert: (b balance = 70).

SmallBankTest >> testWithdrawAll
   | b success |
   b := SmallBank new.
   b deposit: 30.
   self assert: (b balance = 30).
   success := b withdraw: 30.
   self assert: success.
   self assert: (b balance = 0).

SmallBankTest >> testWithdrawOnZero
   | b success |
   b := SmallBank new.
   success := b withdraw: 30.
   self deny: success.
   self assert: (b balance = 0).
\end{lstlisting}
\end{multicols}

\subsection{Pharo}\label{sec:pharo}

Pharo [\url{http://www.pharo.org/}] is a pure object oriented language based on Smalltalk~\citep{black2010pharo,Berg13f}.
It is dynamically typed; i.e. there are no type declarations for variables, parameters, nor return values statically, but dynamically, the environment enforces that all objects to have a type and only respond to messages part of the interface.
It includes a run-time engine and an integrated development environment with code browsers and live debugging.
Pharo users work in a live environment called \textit{Pharo image} where writing code and executing it is tied seamlessly together.

Invoking a method in Pharo is called \textit{message sending}.
As a pure language, every action in Pharo is achieved by sending messages to objects. 
There are no predefined operators, like \inlinecode{+} or \inlinecode{-}, nor control structures like \inlinecode{if} or \inlinecode{while}.
Instead, a Pharo program sends the message \inlinecode{\#+} or \inlinecode{\#-} to a number object, a  \inlinecode{\#ifTrue:ifFalse:} message to a boolean object, or the message \inlinecode{\#whileTrue:}  to a boolean returning block object.
Any message can be sent to any object.
In case the message is not part of the object interface, instead of a compile-time syntax error, the system raises a \inlinecode{MessageNotUnderstood} exception in runtime.
Thus, when transforming test code, a test amplification tool should be attentive to not create faulty test codes.

Like Java, all ordinary classes inherit from the class \inlinecode{Object} and every class can add instance variables and methods.
Unlike Java, all instance variables are private and all methods are public.
Pharo encourages programmers to write short methods with intention revealing names so that the code becomes self explanatory.

\paragraph{Protocols.}\label{sec:protocols}
Pharo, and Smalltalk in general, features \textit{protocols} to organize the methods defined in classes.
The notion of protocol is a tag of a method and it acts like a metadata provided by the integrated development environment.
As such, classifying a method under a particular protocol has no impact on the behavior.

Since all instance variables are private in Pharo, in order to make them accessible by the external world, accessor methods should be provided which are typically grouped into the protocol \inlinecode{accessing}.
In a similar vein, all methods used to set the content of an object upon initialization are grouped into the protocol \inlinecode{instance creation}.
Long lived classes who evolve over time, use the \inlinecode{deprecated} protocol, signalling that these methods will be removed from the public interface in the near future.
And while all methods are public, Pharo uses the protocol \inlinecode{private} to mark methods which are not expected to be used from the outside.
However, as we mentioned earlier, protocols are a tag and Pharo does not block an access to a private method.

The most similar concepts to protocols in other languages are naming conventions, annotations and also access modifiers.
For instance, a Java equivalent for methods in \textit{accessing} protocol is following a naming convention like \inlinecode{setVar()} and \inlinecode{getVar()}.
In a similar vein, Java uses \inlinecode{@Deprecated} annotation to identify the deprecated methods.
An equivalent for methods in \textit{private} protocol in Python is the naming convention of using underscore before the name of private methods, but Java uses \textit{access modifiers} for this purpose.

\subsection{Coding conventions in dynamically typed languages }\label{sec:pharo_coding}

In this section, we describe typical coding conventions that are used by programmers to compensate for the lack of type declarations. When we transform code (like we do when amplifying tests), special care must be taken to adhere to such coding conventions otherwise the code will look artificial and will decrease chances to be adopted by test engineers. Our perspective comes from Pharo / Smalltalk (as documented in~\cite{ducasse2019pharostyle}), but similar coding conventions must be adhered to when amplifying tests in Python, Ruby or Javascript.

\paragraph{Untyped parameters.}
In dynamically typed languages, when defining a method which accepts a parameter, the type of the parameter is not specified.
However, it is a convention to name the parameter after the class one expects or the role it takes.
This is illustrated by the code snippet in \lstref{listing:untypedParameters}.
Line 1 specifies that this is a method \inlinecode{drawOn:} defined on the class \inlinecode{Morph} which expects one parameter.
The parameter itself is represented by an untyped variable \inlinecode{aCanvas} however the name of the variable suggests that the method expects an instance of the class \inlinecode{Canvas}, or one of its subclasses.
Line 7 on the other hand specifies that the method \inlinecode{withdraw:} expects one parameter and its role is to be an amount.
There is no clue on the type of the parameter (integer, longinteger, float, \ldots); all we can infer from looking at the code is that we should be allowed to pass it as an argument when invoking the messages \inlinecode{>=} (line 8) and \inlinecode{-} (line 10) on balance.

\begin{minipage}{\linewidth}
\begin{multicols}{2}[\captionof{lstlisting}{Examples of naming conventions for untyped parameters.}]
\begin{lstlisting}[
	label=listing:untypedParameters,
	xleftmargin=.1\textwidth]
Morph >> drawOn: aCanvas
	aCanvas fillRectangle: self bounds
		fillStyle: self fillStyle
		borderstyle: self BorderStyle.
	
	
SmallBank >> withdraw: amount
	balance >= amount
		ifTrue: [ 
			balance := balance - amount.
			^ true ].
	^ false
\end{lstlisting}
\end{multicols}
\end{minipage}

\begin{Spoonfeed}
When passing a parameter to a method, a test amplification tool has no guaranteed way of knowing the expected type.
The name of the parameter only hints at the expected type, hence during input assertion special care must be taken.
\end{Spoonfeed}

\paragraph{No return types.}
In dynamically typed languages, there is no explicit declaration of the return type of a method.
In Pharo, all computation is expressed with objects sending messages and a message sends always returns an object.
By default a method returns the receiving object, which is the equivalent of the \inlinecode{void} return type in Java.
However a program can explicitly return another value using \textasciicircum~followed by an expression.

\begin{multicols}{2}[\captionof{lstlisting}{Example of a void method (left) or function method (right)}]
\begin{lstlisting}[
	label=listing:returntypes, xleftmargin=.1\textwidth]
Object >> printOn: aStream
	| title |
	title := self class name.
	aStream nextPutAll:
		(title first isVowel
				ifTrue: [ 'an ' ]
				ifFalse: [ 'a ' ]);
		nextPutAll: title

Object>>printString
	| aStream |
	aStream := WriteStream on:
	       (String new: 16).
	self printOn: aStream.
	^aStream contents	
\end{lstlisting}
\end{multicols}

This is illustrated in \lstref{listing:returntypes}, showing the methods \inlinecode{printOn:} (displays the receiver on a given stream) and \inlinecode{printString} (which returns a string representation of the receiver).
\inlinecode{printOn:} is the equivalent of a void call thus returns the receiving object; however the method is declared on \inlinecode{Object} so the receiver object can be anything.
\inlinecode{printString} on the other hand returns the result of sending the message \inlinecode{contents} to \inlinecode{aStream}.
The exact type of what is returned is difficult to infer via static code analysis.
Smalltalk programmers would assume that the return type is a \inlinecode{String} because of the intention revealing name of the method.
However, there is no guarantee that this is indeed the case.
Thus, when a test amplification tool manipulates the result of a method, it cannot easily infer the type of what is returned.

\begin{Spoonfeed}
The lack of explicit return types makes it hard to manipulate the result of a method call while ensuring that no \inlinecode{MessageNotUnderstood} exceptions will be thrown.
\end{Spoonfeed}

\paragraph{Different return types.}\label{subsec:return-types}
In addition to the lack of return type declarations, it is also possible to write a method that can returns different types of values.
For example, in \lstref{listing:different-returns} the method \inlinecode{someMethod:} can return an instance of the classes Integer, Boolean or Object (the default return value is \inlinecode{self}).

As a result, removing the return operator (a common mutation operator) will not cause a syntax error yet may cause a change in the return type of a methods.
For example, in \lstref{listing:different-returns} the method \inlinecode{width} (lines 5 and 6),  if the return operator is removed in the mutation testing, the type of the return value will be converted from a number to a \inlinecode{Shape} object.

\begin{multicols}{2}[\captionof{lstlisting}{Examples of a changing the return type.}]
\begin{lstlisting}[label=listing:different-returns, xleftmargin=.1\textwidth]
Example >> someMethod: anInt
   anInt = 1 ifTrue: [ ^ 1 ]
   anInt = 0 ifTrue: [ ^ false ]
   
Shape >> width
   ^ width   
\end{lstlisting}
\end{multicols}

\begin{Spoonfeed}
Methods in dynamically typed languages can return various types.
Test amplification tools must be aware that a small change in the code may lead to changes in the returned type.
Consequently, assertions verifying the result of a method call must be adapted.
\end{Spoonfeed}

\paragraph{Accessor methods.}
In Pharo, all instance variables are private and only accessible by the object itself.
If one wants to manipulate the internal state of an object one should implement a method for it, as illustrated in \lstref{listing:gettersetter}which shows the setter method \inlinecode{x:} and the getter method \inlinecode{x}.
In Pharo, such accessor methods are typically collected in the protocol \inlinecode{accessing} and are a convenient way for programmers to look for ways to read or write the internal state of an object.

\begin{multicols}{2}[\captionof{lstlisting}{Example of a getter (left) and a setter method (right).}]
\begin{lstlisting}[label=listing:gettersetter, xleftmargin=.1\textwidth]
Point >> x
	^x
Point >> x: anInteger
	x := anInteger
\end{lstlisting}
\end{multicols}

Such accessor methods are especially relevant for all test generation algorithms~\citep{Fischer2020VST}.
For test amplification in particular, the setter methods are necessary in the input amplification step to force the object into a state corresponding to a boundary condition.
The getter methods are necessary in the assertion amplification step to verify whether the object is in the appropriate state.
However, there is no explicit declaration for the type of the parameter passed to the setter method \inlinecode{x:} nor for the type to be returned by the getter method \inlinecode{x}.
\begin{Spoonfeed}
When manipulating the state of an object one cannot rely on type declarations to infer which parameter to pass to a setter method and which result to expect from a getter method.
\end{Spoonfeed}

\paragraph{Pass-by-reference.}
In dynamic languages including Pharo, when sending messages, all arguments are passed by reference.
This may imply that sometimes the state is changed and sometimes it is not.
This is illustrated by the method \inlinecode{r} in \lstref{listing:pureaccessor}, which returns the radius in polar coordinates.
This involves some calculation (the invocation of \inlinecode{dotProduct:}) which passes the receiver object as a reference.
There is no ``pass-by-value'' type declaration for \inlinecode{dotProduct:}, so one cannot know whether the internal state is changed or not.
If  \inlinecode{dotProduct:} does not alter the internal state it may be used as a pure accessor method during assertion amplification anywhere in the test.
However, if the accessor method does change the internal state the order in which the accessor methods are called has an effect on the outcome of the test.

\begin{lstlisting}[
         captionpos=t,   caption=Is \inlinecode{r} a pure accessor method that does not alter the internal state?,
	label=listing:pureaccessor,
	xleftmargin=.4\textwidth]
Point >> r
	^(self dotProduct: self) sqrt
\end{lstlisting}

\begin{Spoonfeed}
The pass-by-reference parameter passing makes it difficult to distinguish pure accessor methods.
Pure accessor methods can be inserted anywhere during assertion amplification, for accessor methods changing the internal state one must take into account the calling order.
\end{Spoonfeed}

\paragraph{Cascading.}
\lstref{listing:cascade} shows the archetypical \inlinecode{Hello World} example.
Line 2 specifies that this is a method \inlinecode{helloWorld} defined on a class \inlinecode{HelloWorld}.
Line 4 and 6 each sends the message \inlinecode{cr} (a message without any parameters) to the global variable \inlinecode{Transcript} which emits a carriage return on the console.
Line 5 sends the message \inlinecode{show:} with as parameter the string \inlinecode{`hello world'} to the global variable \inlinecode{Transcript} which writes out the expected message.

However, a Pharo programmer would never write this piece of code like that.
When a series of messages is being sent to the same receiver, this can be expressed more succinctly as a cascade.
The receiver is specified just once, and the sequence of messages is separated by semi-colons as illustrated on lines 7---10.

\begin{multicols}{2}[\captionof{lstlisting}{A sequence of messages sent to the same receiver object (left) is written as a cascade (right)}]
\begin{lstlisting}[
	label=listing:cascade,
	xleftmargin=.05\textwidth]
HelloWorld >> helloWorld
	Transcript cr.
	Transcript show: 'hello world'.
	Transcript cr.
	
HelloWorld >> helloWorldCascading
	Transcript
		cr;
		show: 'hello world';
		cr.
\end{lstlisting}
\end{multicols}

\paragraph{Instance creation.}
Cascading is frequently used when creating instances of a class as illustrated by the  \inlinecode{createBorder} example in the left of \lstref{listing:instancecreation}.
In line 2 it creates a new \inlinecode{SimpleBorder} object and then initialises the object with color blue (line 3) and width 2 (line 4).
During input amplification we need to change the internal state of the object under test, hence it is tempting to inject extra calls in such a cascade.
However, because we cannot distinguish between state-changing and state-accessing methods, we risk injecting errors.
The code snippet to the right illustrates that injecting an extra \inlinecode{isComplex} call (a call to a state-accessing method) at the end of the cascade erroneously returns a boolean instead of an instance of SimpleBorder.
This will eventually result in a run-time type error via a \inlinecode{messageNotUndersood} exception when the program tries to use the result of \inlinecode{createBorderErroneous}.

\begin{multicols}{2}[\captionof{lstlisting}{Injecting extra statements may result in type errors.}]
\begin{lstlisting}[
	label=listing:instancecreation,
	xleftmargin=.05\textwidth]
TestBorder >> createBorder
	^ SimpleBorder new
		color: Color blue;
		width: 2. "returns self"

TestBorder >> createBorderErroneous
	^ SimpleBorder new
		color: Color blue;
		width: 2;
		isComplex. "returns a boolean"
\end{lstlisting}
\end{multicols}

\begin{Spoonfeed}
When injecting additional calls during instance creation, one runs the risk of returning an inappropriate value.
\end{Spoonfeed}

\hypobox{
Like most dynamically typed languages, Pharo has a lot of coding conventions.
When transforming code (for instance, when amplifying tests) we must adhere to these conventions.
However, the lack of explicit type information hinders static analysis, needed to identify relevant code constructs.
}

\section{\smallamp Design}\label{sec:smallamp}

In this section, we explain  the design of the \smallamp which is an adaption and extension of \dspot \citep{Baudry2015corr,Danglot2019EMSE} for the Pharo ecosystem.
\dspot is an opensource\footnote{\url{https://github.com/STAMP-project/DSpot}} test amplification tool to amplify tests for Java programs. 
Our \smallamp implementation is also publicly available\footnote{\url{\smallampurl}} on GitHub.

\subsection{Main algorithm}\label{sec:alg}

The main amplification algorithm is presented in Algorithm \ref{alg:main} and represents a search-based test amplification algorithm.
The algorithm accepts a class under test (CUT) and its related test class (TC) and returns the set of amplified test methods (ATM).
In addition, the algorithm needs a set of input amplification operators (AMPS) and is governed by a series of hyperparameters: 
\begin{compactitem}
\item
N\textsubscript{iteration} -- 
This parameter specifies the number of iterations and shows the maximum number of transformations on a test input.
The default value for this parameter is 3.
\item
N\textsubscript{maxInputs} -- 
This parameter specifies the maximum number of generated test inputs that algorithm keeps. 
It discards other test inputs.
The default value for this parameter is 10.
\end{compactitem}

\begin{algorithm}
\SetKwFunction{amplifyInputs}{amplifyInputs}
\SetKwFunction{amplifyAssertions}{amplifyAssertions}
\SetKwFunction{inputReducing}{reduce}
\SetKwFunction{testRunner}{testRunner}
\SetKwFunction{strip}{removeAssertions}
\SetKwFunction{profileCollect}{profileCollect}
\SetKwFunction{postProcess}{improveReadability}
\SetKwInOut{Input}{input}\SetKwInOut{Output}{output}
 \Input{CUT: class-under-test}
 \Input{TC: original test class}
 \Input{AMPS: a set of input amplification operators}
 \Input{hyperparameters \{N\textsubscript{iteration}, N\textsubscript{maxInputs}\}}
 \Output{ATM: set of amplified test methods}
 $ATM\leftarrow \{\}$\;
 $extraInfo \leftarrow \profileCollect{CUT, TC}$\;\label{lst:line:profile}
\For{\textbf{each} $t \in TC$}{\label{lst:line:for1}
$V\leftarrow\{\strip{t}\}$\;\label{lst:line:strip}
$U\leftarrow \amplifyAssertions{V}$\;\label{lst:line:assertamp1}
$ATM\leftarrow ATM \cup \{x \in U |$ x improves mutation score$\} $\;\label{lst:line:selection1}
\For{$i\leftarrow 0$ \KwTo $N\textsubscript{iteration}$}{\label{lst:line:for2}
	$TMP \leftarrow \{\}$ \;
	\For{\textbf{each} amp $\in$ AMPS}{
		$TMP\leftarrow TMP \cup \amplifyInputs{amp, V, extraInfo}$\;\label{lst:line:inputamp}
	}
	
	$V\leftarrow \inputReducing{TMP, N\textsubscript{maxInputs}}$\;\label{lst:line:reduce}
	$U\leftarrow \amplifyAssertions{V}$\;\label{lst:line:assertamp2}
	$ATM\leftarrow ATM \cup \{x \in U |$ x improves mutation score$\} $\;\label{lst:line:selection2}
	}\label{lst:line:for2end}
}
$ATM \leftarrow \postProcess{ATM}$\;\label{lst:line:postProcess}
\KwRet{ATM}
 \caption{\smallamp amplification algorithm
 }\label{alg:main}
\end{algorithm}

\noindent
Initially, the code of CUT and TC is instrumented to allow for dynamic profiling.
The test class is executed, all required information is collected and then the instrumentation is removed again.
This extra information including the type information allows us to perform  input amplification more efficiently and circumvent the lack of type information in the source code (line \ref{lst:line:profile}). 
We discuss about the profiling in \secref{sec:profiling}.

The main loop of the algorithm amplifies all test methods one by one (line \ref{lst:line:for1}).
V is the set of test inputs, thus test methods without assertion statements.
In the beginning, V has only one element which is obtained from removing assertion statements in the original test method (line \ref{lst:line:strip}).
U is the set of generated test methods which are generated by adding new assertion statements to the elements in V (lines \ref{lst:line:assertamp1}).
Then the coverage is calculated using the generated test methods accumulated in U and the tests increasing the coverage are added to the final result.
\smallamp uses mutation score as a coverage criteria (line \ref{lst:line:selection1})

In the inner loop of the algorithm  (lines \ref{lst:line:for2} to \ref{lst:line:for2end}), \smallamp generates additional tests by repeating the following steps N\textsubscript{iteration} times:
\begin{compactenum}
\item
\smallamp applies different input amplification operators on V (the current test inputs) to create new variants of test methods accumulated in the variable TMP (line \ref{lst:line:inputamp}).
We discuss input amplification in \secref{sec:input-amp}.
\item
\smallamp reduces TMP by keeping only N\textsubscript{maxInputs} of current inputs and discarding the rest (line \ref{lst:line:reduce}).
We discuss input reduction in \secref{sec:input-reducing}.
\item
\smallamp injects assertions on the remaining test inputs in V and stores the result in U (line \ref{lst:line:assertamp2}).
We discuss assertion amplification in \secref{sec:assert-amp}.
\item
\smallamp selects all test methods in U that increase mutation score and adds them to the final result ATM (line \ref{lst:line:selection2}).
We discuss about test selection in \secref{sec:selection}.
\end{compactenum}

\noindent
After both loops have terminated, \smallamp applies a set of post-processing steps to increase the readability of the generated tests (line \ref{lst:line:postProcess}).
We discuss these steps in \secref{sec:post-processing}.

\algoref{alg:main} is heavily inspired by \dspot, but not entirely the same.
In other words, we have added a pre-process step (line \ref{lst:line:profile}) to collect the necessary information about CUT and TC before entering the main loop.
We also have added a post-processing step (line \ref{lst:line:postProcess}) to make the output more readable.
We discuss about extras to \dspot algorithm in \secref{sec:extrasDSpot}.

\subsection{Input amplification} \label{sec:input-amp}

During input amplification, existing test code is altered to force previously untested paths.
Input amplification involves changing the set-up of the object under test, passing arguments which represent boundary conditions.
Additional calls to state-changing methods of the public interface are injected as well.
Such changes are bound to fail the original assertions of TC, therefore \smallamp removes all assertions from a test $t$ in TC. 

The test code itself is transformed via a series of \emph{Input Amplification Operators}.
These change the code in such a way that they are likely to force untested paths and cover boundary conditions.
Input amplification operators are based on the genetic operators introduced in Evolutionary Test Classes \citep{Tonella_2004}.
Below we explain the \emph{Input Amplification Operators} adopted from \dspot.

\paragraph{Amplifying literals.}
This input amplifier scans the test input source code to find literal tokens (numbers, booleans, strings). 
Then it transforms the literal to a new literal based on its type according to \tabref{table:literal-input-amp}.
For example, test input shown in \lstref{listing:testVectorGreater} is transformed into \inlinecode{testVectorGreater\_L} by manipulating the second element from the literal array.

\begin{multicols}{2}[\captionof{lstlisting}{Example Literals Amplification Operator (line 3 vs. line 7)}]
\begin{lstlisting}[label=listing:testVectorGreater,
	xleftmargin=.1\textwidth]
testVectorGreater
	| u w |
	u := #(-1 0 1) asPMVector.
	w := u > 0.
testVectorGreater_L
	| u w |
	u := #(-1 1 1) asPMVector.
	w := u > 0.
\end{lstlisting}
\end{multicols}

\begin{table}[t!]
\begin{center}

\begin{tabular}{ |c|c| } 
 \hline
 \textbf{Type} & \textbf{Transformation} \\ 
 \hline
\textbf{Numbers} & 0,\\
& increased and decreased values (\(+1\) and \(-1\)), \\
 & doubled and halved values (\(\times 2\) and \(\div 2\)), \\
 & negated value ($\times -1$)\\
 & replacing with an existing number from the test body\\ 
\hline
  \textbf{Booleans} & negate via $not$  \\ 
 \hline
\textbf{Strings} & add a new random character to a random position \\
  \textbf{ } & remove a character randomly \\
  \textbf{ } & change a character randomly \\ 
  \textbf{ } & replace by a random string in the same size \\ 
\hline
\end{tabular}
\caption{Transformations in literal amplification}
\label{table:literal-input-amp}
\end{center}
\end{table}

\paragraph{Amplifying method calls.}
This input amplifier scans the test input source code to find the method invocations on an object.
Then it transforms the source code by \textit{duplicating} or \textit{removing} the method invocations.
It also \textit{adds new method invocations} on the objects.
If the method requires new values as arguments, the amplifier creates new objects.
For primitive parameters, a random value is chosen from the profiled values.
For object parameters, the default constructor is used i.e it creates a new instance by sending \inlinecode{\#new} message to the class.
\smallamp ignores \textit{private} and \textit{deprecated} methods (regarding to their protocol) when it adds a new method call.
The type information required do safely apply these transformations is obtained in the profiling step explained in \secrefpage{sec:profiling}.

\subsection{Assertion amplification}\label{sec:assert-amp}

During the assertion amplification step, we inject assertion statements which verify the state of the object under test.
The object under test is then used as an oracle: while executing the test the algorithm inspects the state of the object under test and asserts the corresponding values.
The assertion amplification step is based on Regression Oracle Checking~\citep{Xie_2006}.

Note that assertion amplification is applied twice during the amplification algorithm (\algoref{alg:main}, in lines \ref{lst:line:assertamp1} and \ref{lst:line:assertamp2}).
There are two reasons for this seemingly redundant design.
(1) We assure that the original test method is assertion amplified as well.
Since the test inputs are reduced in line \ref{lst:line:reduce}, there is a possibility that the original test method is discarded and never reaches the assertion amplification in line \ref{lst:line:assertamp2}.
(2) We can run only assertion amplification by setting the value of $N_{iteration} = 0$.
This way no new tests will be generated, but existing tests may become stronger because they check more conditions.

\paragraph{Observing state changes via object serialisation.}
\smallamp manipulates the test code and surrounds each statement with a series of what we call ``observer meta-statements'' (see \lstref{listing:assert-amp-manipulate}).
Such meta-statements include a surrounding block to capture possible exceptions (lines 19--20 and 24--25) and calls to observer methods to capture the state of the receiver (line 17 and line 18) and the return value (line 18 and line 23).
When necessary, temporary variables are added to capture intermediate return values (tmp1 on line 21 and line 23).

\begin{minipage}{\textwidth}
\begin{multicols}{2}[\captionof{lstlisting}{Injection of observer meta-statements}]
\begin{lstlisting}[label=listing:assert-amp-manipulate,
	xleftmargin=.02\textwidth]
testDeposit
   | b |
   b := SmallBank for: 'JDoe'.
   b deposit: 100.
   








testDeposit_instrumented
   | b tmp1 |
   [ b := SmallBank for: 'JDoe'.
     self observe: SmallBank.
     self observeRetVal: b.
    ] on: Error do: [ :ex | 
        self observeException: ex].
   [ tmp1:= b deposit: 100.
     self observe: b.
     self observeRetVal: tmp1.
    ] on: Error do: [ :ex | 
        self observeException: ex]
   
\end{lstlisting}
\end{multicols}
\end{minipage} 

\noindent
After manipulating the test method, \smallamp runs the test to capture the values by the observer methods.
\smallamp serializes objects by capturing the values from its accessor methods.
If the return value of an accessor method is another object, it recursively repeats the object serialization up to N\textsubscript{serialization} times.
N\textsubscript{serialization} is a configurable value (default value is 3).
The output of this step of assertion amplification is a set of trace logs which reflect the object states.

\paragraph{Identifying accessor methods.}
\smallamp relies on the Pharo/Smalltalk coding conventions and therefore selects methods if they belong to protocols \inlinecode{\#accessing} or \inlinecode{\#testing} or when their name is identical to one of the instance variables.
From the selected methods, all methods lacking an explicit return statement and all methods in the protocols \inlinecode{\#private} or \inlinecode{\#deprecated} are rejected and the remaining are considered as accessors.

\paragraph{Preventing flaky tests via trace logs.}
A flaky test is a test that may occasionally succeed (green) or fail (red).
This may happen if the test is asserting a non-deterministic value.
\smallamp tries to detect non-deterministic values before making assertions on them.
The assertion amplification module, repeats collecting the trace logs for N\textsubscript{flakiness} (default value is 10) times.
Then it compares the observed values.
If a value is not identical between all collected logs, \smallamp marks it as non-deterministic.

\paragraph{Recursive assertion generation.}
Based on the type of the observed value, zero, one or more assert statements are generated.
If the type is a variant of collection or an object, which include other internal values, the assertion generator uses a recursive method to build valid assertion statements.
For non-deterministic values, the value is not asserted and only its type is asserted.
The output of the assertion amplification step is a passing (green) test with extra assertions.

\paragraph{Intended values versus actual values.} \label{sec:assert-amp-intended}
During assertion amplification, the assertion statements should the expected value which is deduced from an oracle.
We assume that the current implementation of the program is correct, and therefore we deduce the oracle from the current state of the object under test. 
However, when there is a defect in the method under test, the generated assertions would verify against an incorrect oracle.
This is an inherent limitation for both \dspot and \smallamp, inherited from Regression Oracle Checking~\citep{Xie_2006}. 

\paragraph{Example.}
\lstref{listing:assert-amp-recursive} shows an example of a trace log collected by line 22 from \lstref{listing:assert-amp-manipulate} (left) and its recursive assertion statements (right).
In this example, we point out that the method \inlinecode{timestamp} is an accessor method in \inlinecode{SmallBank} class which returns a timestamp value.
Since this value differs in different executions, it has been marked as a flaky value (line 8) hence only its type is asserted (line 24).

\begin{multicols}{2}[\captionof{lstlisting}{An example of a trace log and its assertion statements}]

\begin{lstlisting}[mathescape=true, label=listing:assert-amp-recursive,
	xleftmargin=.04\textwidth]
b:
   type $\rightarrow$ SmallBank,
   accessors:
      balance:
         flaky $\rightarrow$ false,
         value $\rightarrow$ 100
      timestamp:
         flaky $\rightarrow$ true,
         value $\rightarrow$ 1624
       user:
          type $\rightarrow$ SmallBankUser,
          accessors:
             name:
                flaky $\rightarrow$ false,
                value $\rightarrow$ 'JDoe'

testDeposit_withAssertions
   | b tmp1 |
   b := SmallBank for: 'JDoe'.
   "..."
   tmp1 := b deposit: 100.
   self assert: b class equals: SmallBank.
   self assert: b balance equals: 100.
   self assert: b timestamp class equals: Integer. "flaky"
   self assert: b user class equals: SmallBankUser.
   self assert: b user name equals: 'JDoe'.
   "..."         
\end{lstlisting}
\end{multicols}

\subsection{Test selection -- prefer focussed tests}\label{sec:selection}

During each iteration of the inner loop (lines \ref{lst:line:for2} to \ref{lst:line:for2end} in \algorefpage{alg:main}) \smallamp generates N\textsubscript{maxInputs} new tests with their corresponding assertions.
In the test selection step (lines \ref{lst:line:selection1} and \ref{lst:line:selection2}) the algorithm selects those tests which kill mutants not killed by other tests.

First of all, \smallamp performs a mutation testing analysis on CUT and TC and creates a list of live and uncovered mutants.
Then \smallamp selects those test methods from U (the set of amplified test methods) which increase the mutation score, thus killing a previously live or uncovered mutant.
If multiple tests are killing the same mutant, the shortest test is chosen.
If there are multiple short tests, the test with the least changes is chosen.
In the \dspot paper, a similar heuristic is chosen, which the authors refer to as \emph{Focused Test Cases Selecting}.

\section{\smallamp Extras compared to \dspot}\label{sec:extrasDSpot}

While the design of \smallamp was inspired by \dspot, the lack of explicit type information forced us to make major changes but also permitted us to make improvements.
This section describes additional and diverging aspects of \smallamp compared to \dspot.

\subsection{Dynamic profiling to collect type information}\label{sec:profiling}

\addressedRvw{At the very beginning of the main algorithm~(\algoref{alg:main} line \ref{lst:line:profile}), dynamic type profiling is done only once by executing the original test methods and observing the actual type information of variables.}

In dynamically typed languages like Pharo, type annotations are not provided in the source code.
So, performing static analysis which depend on types are challenging.
In the context of \smallamp, the most important step that relies on static code analysis is input amplification.
The other steps are either based on dynamic analysis like assertion amplification, or depend on a third-party library such as selection based on mutation-testing.

\vspace{0,5em}
\noindent
In input-amplification, we can group operators into two classes as: 
\vspace{-0,5em}
\begin{enumerate}
\item
\textit{Type sensitive operators.}
These operators heavily depend on the type information and without type information they are ineffective or impossible.
An important type sensitive input amplifier in \smallamp is method call addition. 
The types of variables defined in a test method must be inferred when adding a valid method call. 
In addition, it needs the type information of parameters in the newly called method.

\item 
\textit{Type insensitive operators.}
These are all operators that are still applicable without the type information.
An example is the operators amplifying literals. 
These operators are easy to adapt to a dynamic language because literals are distinguishable from a token representation of the source code.
\end{enumerate}

To obtain accurate type information we rely on the presence of manually written tests, which should be representative for the normal behaviour of the program under test.
We exploit profiling tools (commonly available in modern program environments) to extract accurate type information from the variables present in the program.
The profiler is configured to attach hooks to the relevant elements in the code.
When these important code elements are executed, the hooks are triggered, the profiler reads the information from the program state and logs it.

\vspace{0,5em}
\noindent
In \smallamp, we rely on two distinct profilers:
\vspace{-0,5em}
\begin{itemize}
\item
\textit A \textit{Method-proxy} profiler, which collects the type of parameters in Class-Under-Test methods.
\item
\textit A \textit{Metalink} based profiler which collects the type of variables in the test methods
\end{itemize}

To apply test amplification to other dynamically typed languages one needs comparable profiling technology.
Some languages provide reflexive facilities that can be exploited. Python metaclasses for example allow one to transparently hook into the code proxy objects similar to the method-proxies adopted in \smallamp.
If such reflexive facilities are not available, one can resort to the debugger APIs to inspect values of variables at run-time.

\paragraph{Profiling by Method-proxies.}

For gathering the type of parameters in methods, \smallamp uses method proxies~\citep{ducasse1999evaluating, peck2015ghost}.
Proxies are methods wrapping the methods in the class under test and trigger instead of the original methods.
They first log the arguments and then pass the control to the original method (\lstref{listing:proxy-profiler}).

\begin{lstlisting}[mathescape=true, caption={Wrapper method to log the types of the parameters},captionpos=t, label=listing:proxy-profiler,
	xleftmargin=.04\textwidth]
ProfilingProxy >> run: aSelector with: anArray in: aReceiver 
	self logCalled: aSelector withArguments: anArray inType: aReceiver.
	^ aReceiver withArgs: anArray executeMethod: method.

\end{lstlisting}

The main drawback of the Method-proxy profiler is that when a method is not covered by the test class, it will not be profiled.
\smallamp reports the list of such uncovered methods as one of its outputs.
Using this report, a developer can decide to add new tests for uncovered methods, make them private (using an adequate protocol / method tag), or remove them.

\paragraph{Profiling by Metalinks.}

Pharo provides Metalinks as a fine-grained behavioral reflection solution~\citep{Denker07Sub, costiou2020sub}.
For collecting the type of variables in the test method, \smallamp uses Metalinks.

A metalink contains an action to perform which is defined by providing a meta-object, a selector, and also a control.
Metalinks can be installed on one or more nodes in the abstract syntax tree.
\lstref{fig:metalink} shows how metalink is defined and installed on all variable nodes in the test method.

\begin{minipage}{\textwidth}
\begin{multicols}{2}[\captionof{lstlisting}{Defining a metalink to log the variable node type after execution}]
\begin{lstlisting}[label=fig:metalink]
link := MetaLink new
	metaObject: self;
	control: #after;
	selector: #'logNode:context:object:';
	arguments: #(node context object).
		
nodes := testMethod ast allChildren 
	select: #isVariable.
nodes do: [ :node |  
	node link: link ]

\end{lstlisting}
\end{multicols}
\end{minipage}

Line 1 to 5 shows how Metalink is initialized.
It says that after execution the AST node containing this link, the method \inlinecode{logNode:context:object:} will be called with the following arguments: 

\begin{itemize}
\item
\inlinecode{node}: 
The static representation of the AST node. 
It is used to get information such as name and the position in the code.
\item
\inlinecode{context}:
The context of execution including dynamic values of the variables and stack. 
It is used to access to the values of temporary variables.
\item
\inlinecode{object}:
The state of the object on which the metalink is installed (in this case the test class).
It is used to access to the values of instance variables.
\end{itemize}

In lines 7 to 10, all variable nodes in the test method are selected and then the link is installed on them.
After installing the metalinks, the test method is executed.
When the execution passes each variable node, the metalink is triggered and the logger method is called.
The logger method extracts the type information from the context, logs them and returns.
Then, the execution on the test method continues until the end or another metalink is triggered.

\paragraph{How the collected data is used.}
The collected data from each profiler is stored as a dictionary object mapping the identifier of the profiled data to its type and a list of sample values (only for primitive types).
In \smallamp, there are two dictionaries, for the type of method parameters and the type of variable nodes.
During the input amplification, when type information is needed, the corresponding dictionary is consulted.

\subsection{Test input reduction}\label{sec:input-reducing}

The input amplification step quickly produces a large number of new test inputs with the inner loop of \algorefpage{alg:main} -- lines \ref{lst:line:for2} to \ref{lst:line:for2end}.
For instance, if the number of inputs in the first iteration is \(|v|\), this number in the second iteration grows to \(|v| \times |v|\), and in iteration \(i\) reaches \(|v|^i\).
We refer to this problem as \textit{test-input explosion}.
Since the number of test inputs grows exponentially, either the number of transformations ($N_{iteration}$) needs to be chosen as small values for being feasible to try all generated test input, or we need to reduce the number of inputs by using a heuristic to select a limited number of them.

\smallamp uses a random selection heuristic which maximises diversity in order to select a maximum number (N\textsubscript{maxInputs}) of test inputs.
This selection is different from selection by mutation score (\secref{sec:selection}); we name it \textit{reduction}.

\noindent
\smallamp reduction considers two techniques:
\begin{itemize}
\item
\textit{a competitive selection.}
a portion of test inputs (by default half of N\textsubscript{maxInputs}) are selected completely randomly from the output of all input amplifiers.
\item
\textit{a balanced selection}
in the remaining portion, \smallamp assures that all input amplifiers are contributed by selecting from their outputs regarding an assigned weight.
In \smallamp, all input amplifiers are assigned a weight (it is 1 by default for all amplifiers).
This maintains a diversity in the selected test inputs.
\end{itemize}

\paragraph{Why diversity is important?}
Each input-amplifier algorithm performs transformations based on different considerations.
As a result, the number of generated tests is different for input amplifiers.
If the test inputs are selected purely random, the result will be dominated by generated tests from amplifiers generating more outputs.
Therefore, we need to have a balance between the outputs from each amplifier. 

As an example, we compare the number of new test inputs from a \textit{statement-removal amplifier} and a \textit{statement-addition amplifier}.
The former has a $O(S)$ complexity where $S$ is the number of statements in the test method.
It means that if the number of lines in the test is increased, the number of new test inputs generated by this input-amplifier shows a linear increase.
However, the latter has a $O(S*M)$ complexity where $M$ is the number of methods in the class under test.
It means that the increase in the outputs depends on not only the numbers of statements in the test, but also the number of methods in the class-under-test.
Now, if we select a number of generated test methods randomly, the outputs from the latter operator is more likely to be selected; so the result will be dominated by the result from the second input-amplifier.

\subsection{Improving readability via post-processing}\label{sec:post-processing}
In order to make the generated tests more readable, \smallamp adds a few steps after finishing the main loop of the algorithm (line \ref{lst:line:postProcess} in \algorefpage{alg:main}).
These steps do not have any effect on the mutation score of the amplified test suite; they only make the test cases more readable for \smallamp users.

\paragraph{Assertion reduction.} 
As described in \secref{sec:assert-amp}, \smallamp generates all possible assertions for all observation points.
Consequently, the generated test methods easily include hundreds of new assertions most of which appear redundant.
The assertion reducer is a post-processing step that discards all assertions that do not affect the mutation score.

Each amplified test method encompasses the identifier of all newly killed mutants.
\smallamp surrounds all assertion statements by exception handling blocks to catch exceptions, especially  \inlinecode{AssertionFailure} raised from the assertion statements.
Then, the mutation testing framework is run using the newly killed mutants only.
When an \inlinecode{AssertionFailure} is caught, the identifier of the assertion is logged as important.
Finally, \smallamp keeps only important assertions and remove all other assertion statements.

In some cases, an assertion may call an \emph{impure} accessor methods, i.e. an accessor method that alters the internal state of the object.
When such assertions are removed, some of the next assertions may fail.
\smallamp runs each test method after removing unnecessary assertions, to confirm that they remain green and the mutants are still killed by the test.
If the confirmation failed, the assertion reduction is not successful and all removed assertions are reinserted.

\paragraph{Comply with coding conventions.} 
Before processing a test method, \smallamp  breaks complex statements (chains of method invocations and cascades) into an explicit sequence of message sends to permit observing state changes (see \lstrefpage{listing:assert-amp-manipulate}).
This is necessary to observe state changes during assertion amplification.
In this post-process step, \smallamp cleans up all unused temporary variables, and chooses a better name for the remaining variables based on the type of the variable.
If possible, it reconstructs message chains and cascades to make the source code more readable and conform to Pharo coding conventions.

\section{Evaluation}\label{sec:replication}
To evaluate \smallamp, we replicated the experimental protocol introduced for \dspot \citep{Danglot2019EMSE}.
We adopted a qualitative experiment by sending pull-requests in GitHub for evaluating whether the generated tests are relevant to the developers or not (RQ1).
Next, we use a quantitative experiment to evaluate the effectiveness of \smallamp (RQ2, RQ3 and RQ4).
The order of RQ1 to RQ4 is exactly the same order in \citep{Danglot2019EMSE} to facilitate the comparing and it does not reflect the importance of the research questions.
In RQ5, we make a detailed comparison of our results versus the ones in the original experiment.
Finally, in RQ6, we report the time cost of the running \smallamp, with special attention to the performance penalty induced by the additional steps (profiling and oracle reduction).

\newcommand\rqOneKey{Pull Requests}
\newcommand\rqTwoKey{Focus}
\newcommand\rqThreeKey{Mutation Coverage}
\newcommand\rqFourKey{Amplification Steps}
\newcommand\rqFiveKey{Comparison}
\newcommand\rqSixKey{Time Costs}

\newcommand\rqOne{Would developers be ready to permanently accept amplified test cases into the test repository?}
\newcommand\rqTwo{To what extent are improved test methods considered as focused?}
\newcommand\rqThree{To what extent do improved test classes kill more mutants than developer-written test classes?}
\newcommand\rqFour{What is the contribution of input amplification and assertion amplification (the main steps in the test amplification algorithm) to the effectiveness of the generated tests?}
\newcommand\rqFive{How does \smallamp compare against \dspot?}
\newcommand\rqSix{What is the time cost of running \smallamp, including its steps?}

\begin{enumerate}[label=\textbf{RQ\arabic*: },leftmargin=*]
\item \textbf{\rqOneKey}. \textit{\rqOne}
We create pull-request on mature and active open-source projects in the Pharo ecosystem.
We propose the improvement as a pull request on GitHub, comprising improvements on an existing test (typically due to assertion amplification) or new tests (typically the result of input plus assertion amplification).
We interpret the statements where extra mutants are killed to provide a manual motivation on why this pull request is an improvement.
The main contributors then review, discuss and decide to merge, reject or ignore the pull request.
The ratio of accepted pull requests gives an indication of whether developers would permanently accept amplified test cases into the test repository.
More importantly, the discussions raised during the review of the pull request provides qualitative evidence on the perceived value of the amplified tests.
\item \textbf{\rqTwoKey}. \textit{\rqTwo}
We assess whether the amplified tests don't overwhelm developers, by assessing how many extra mutants the amplified tests kill.
Ideally, the amplified test method kills only a few extra mutants as then we consider the test \emph{focussed} (cfr. \secrefpage{sec:selection}).
We present and discuss the proportion of focused tests out of all proposed amplified tests.
An amplified test case is considered focus if, compared to the original, at least 50\% of the newly killed mutants are located in a single method.
\item \textbf{\rqThreeKey}. \textit{\rqThree}
We assess whether the amplified tests cover corner cases by using a proxy --- the improvement in mutation score via the mutation testing tool \toolname{mutalk} \citep{esug:mutalk}.
We first run \toolname{mutalk} on the original class under test (CUT) as tested by the test class (TC) to compute the original mutation score.
We distinguish between strong tests and weaker tests, by splitting the set of test classes in half after sorting according to the mutation score.
Next, we amplify the test class and compute the new mutation score.
We report the relative improvement (in percentage).
\item \textbf{\rqFourKey}. \textit{\rqFour}
Here as well we use mutation score as a proxy for the added value of both the input and assertion amplification step and here as well we distinguish between strong and weak test classes.
Therefore, we compare the relative improvement (in percentage) of assertion amplification against the relative improvement of input and assertion amplification combined.
We report separately which amplification operators have the most impact, paying special attention to the ones which are sensitive to type information.
\item \textbf{\rqFiveKey}. \textit{\rqFive}
To analyse the differences in result between \smallamp and \dspot, we compare the qualitative and quantitative results reported in the \dspot paper against the results we obtained for RQ1 to RQ4.
\item \textbf{\rqSixKey}. \textit{\rqSix}
To study the applicability of \smallamp, we analyse the time cost of all runs in the quantitative analysis.
We compare the relative time cost of each step, paying special attention to the extra overhead of profiling and oracle reduction.
 \end{enumerate}

\subsection{Dataset and metrics}\label{sec:dataset}

\begin{sidewaystable}
\begin{center}
\caption{Descriptive Statistics for the Dataset Composed of 13 Pharo Projects and the selected test classes}
\label{table:projects}

\resizebox{\textwidth}{!}{
\begin{tabular}{ ccccp{5cm}p{7.5cm} } 
 \textbf{Project} & \textbf{commit} & \textbf{\#test} & \textbf{\#test} & \textbf{Description} & \textbf{Test classes} \\
 \textbf{} & \textbf{Id} & \textbf{classes} & \textbf{methods} & \textbf{(Based on Gihub page)} & \textbf{in the experiment} \\ 
 \hline
Bloc & f69c94b & 6 &  44 & UI infrastructure \& framework for Pharo & BlLayoutExactResizerTest$^l$, BlShortcutTest$^l$, BlInsetsTest$^l$, BlCompulsoryCombinationTest$^l$\\ 
DataFrame & 7b222e2 & 6 &  497 & Tabular data structures for data analysis  & DataFrameCsvReaderTest$^h$, DataFrameJsonWriterTest$^h$, DataFrameTypeDetectorTest$^h$, DataFrameJsonReaderTest$^h$\\ 
DiscordSt & 44be9b7 & 43 & 532 & An API wrapper for Discord & DSUserTest$^h$, DSDetectChannelCommandTest$^h$, DSEmbedTest$^l$, DSSendUserTextMessageItemTest$^l$ \\ 
GraphQL & 3e57ca8 & 9 &  409 & A Smalltalk GraphQL Implementation & GQLSchemaGrammarTest$^h$, 
GQLSingleAnonimousQueryEvaluatorTest$^h$, 
GQLRequestGrammarTest$^h$, GQLArgumentsTest$^l$ \\ 
MaterialDesignLite & 45f2f4d & 77 &  484 & Binds the google's Material Design Lite project to Seaside and build widgets on top of Material Design & MDLPanelSwitcherButtonTest$^h$, 
MDLPaginationComponentTest$^h$, 
MDLNestedListTest$^l$, 
MDLDialogTest$^l$\\
openponk & bfcb84b & 19 &  128 & Modeling platform & OPRTElementsConstraintTest$^h$, 
OPNullSerializerTest$^h$, 
OPProjectTest$^l$, 
OPNavigatorAdaptersTest$^l$\\ 
petitparser2 & 86243ea & 47 &  535 & A high-performance top-down parser & WebGrammarTest$^h$, 
PP2BufferStreamTest$^h$, 
PP2ParsingGuardTest$^l$, 
PP2BenchmarkTest$^l$\\ 
pharo-launcher & f1ce748 & 38 & 216 & Manager for pharo images & PhLAboutCommandTest$^h$, 
PhLCopyImageCommandTest$^h$, 
PhLDirectoryBasedImageRepositoryTest$^l$, 
PhLLocalTemplateTest$^l$\\ 
PolyMath & 205dcfc & 57 &  605 & Scientific Computing with Pharo & PMBinomialGeneratorTest$^h$, 
PMBernoulliGeneratorTest$^h$, 
PMFixpointTest$^l$, 
PMExponentialDistributionTest$^l$ \\ 
Roassal3 & 69b5645 & 18 &  152 & Visualization Engine & RSLabelGeneratorTest$^h$, 
RSUMLClassBuilderTest$^h$, 
RSDraggableCanvasTest$^l$, 
RSAthensRendererTest$^l$\\ 
Seaside & 3038b49 & 55 &  919 & Framework for developing sophisticated web applications & WAKeyGeneratorTest$^h$, 
WACookieTest$^h$, 
WAXmlCanvasTest$^l$, 
WAErrorHandlerTest$^l$\\ 
Telescope & a4e128a & 11 &  75 & Engine for efficiently creating meaningful visualizations & TLExpandCollapseNodesActionTest$^h$, 
TLHideActionTest$^h$, 
TLDistributionMapTest$^l$, 
TLLegendTest$^l$\\ 
zinc & ee0d071 & 27 &  292 & HTTP Components  to deal with the HTTP networking protocol & ZnEasyTest$^h$, 
ZnMessageBenchmarkTest$^h$, 
ZnStatusLineTest$^l$, 
ZnBivalentWriteStreamTest$^l$ \\ 

  \hline
\end{tabular}
}
\end{center}
\end{sidewaystable} 
\label{sec:dataset-projects}
\paragraph{Selecting a dataset.}
Firstly, we collected some candidate projects under test from different sources:
(1) We looked at the projects used in a recent paper focusing on testing in Pharo \citep{delplanque:hal-02002346}.
(2) We looked at the projects introduced in "Innovation Technology Awards" section of ESUG conference from year 2014.
(3) We used GitHub API to find the Pharo projects hosted in GitHub with more than 10 forks and 20 stars.

Then we applied a set of inclusion and exclusion criteria.
Our projects needs to be hosted in GitHub and written in Pharo.
They should include a test suite written in sUnit, and can run in Pharo 8 (stable version).
For not being overwhelmed with resolving dependencies, they need to support installation with \inlinecode{Metacello} and not depend on system level packages like databases or a special installation service.
We discarded all libraries that are part of the Pharo system such as collections, or compiler.

Based on the mentioned criteria, we selected randomly 20 projects.
Then, we rejected projects having less than 4 green test classes with known class under test and mutation coverage less than 100.

Similar to the experimental protocol in \dspot \citep{Danglot2019EMSE}, we select randomly 4 test classes, 2 high mutation coverage and 2 low, for each project.
If a project lacks at least 2 test classes having high (or low) mutation score, we select from lower (higher) covered classes instead.
As result, we have 52 test classes, 27 of them considered strong (high mutation score) and 25 considered weaker (lower mutation score).

\tabref{table:projects} shows the descriptive statistics of the selected projects with a short description, area of usage, number of test classes and test methods and their version based on git commit id, and selected test classes (a superscript $^h$ is used to indicate a test class with high mutation coverage, and $^l$ is used to indicate low mutation coverage).

\paragraph{Detecting the class under test.} \label{sec:tests-mapping}
\smallamp needs a test class and its class-under-test as inputs.
Finding a mapping between a test and a class can be challenging.
As the default mapping heuristic, we rely on the pattern used by Pharo IDE to detect a test method for a class.
The Pharo code browser finds a unit test for a class as follows: it adds the postfix \inlinecode{"Test"} to the name of the class.
If there is such class loaded in the system that is a subclass of \inlinecode{TestCase} it is considered as the unit test class.
If this heuristic is not followed in a project, one can explicitly define the class-under-test by overriding a hook method in test classes.

\paragraph{Metrics.}
We adopt the same metrics used in the experimental protocol in \citep{Danglot2019EMSE}:

\begin{compactitem}

\item \textbf{All killed mutants ($\#Mutants.killed$)}: The absolute number of mutants killed by a test class in a given class under test.

\item \textbf{Mutation score ($\%M.Score$)}: The ratio (in percentage) of killed mutants over the number of all mutants injected in the class under test.

\[ \%M.Score = 100 \times \frac{\#Mutants.killed}{\#Mutants.All} \]

\item \textbf{Newly killed mutants ($\#Mutants.killed_{new}$)}: The number of all new mutants that are killed in an amplified version of the test class.
\[\#Mutants.killed_{new} = \#Mutants.killed_{amplified} - \#Mutants.killed_{original}\]

\item \textbf{Increase killed ($\%Inc.killed$)}: The ratio (in percentage) of all newly killed mutants over the number of all killed mutants.

\[ \%Inc.killed = 100 \times \frac{\#Mutants.killed_{new}}{\#Mutants.killed_{original}} \] 

\end{compactitem}

\subsection{RQ1 --- \rqOneKey}\label{sec:PR}

In this experiment, we choose an amplified test method for each project and send a pull-request in GitHub.
Before the experiment, we sent a pilot pull-request to learn how developers deal with external contributions. 
Firstly, we explain the pilot pull-request and then all pull-requests are described one by one.
\tabref{table:pr} demonstrates the status as well as the url of each pull-request per project.

\begin{table*}[t!]
\begin{tabular}{lclc|c|}
\hline
\textbf{Project} & \textbf{Status} & \textbf{Pull request url}\\
\hline

PolyMath & Merged & \url{https://github.com/PolyMathOrg/PolyMath/pull/178}\\
Pharo-Launcher & Merged & \url{https://github.com/pharo-project/pharo-launcher/pull/500} \\
DataFrame & Merged & \url{https://github.com/PolyMathOrg/DataFrame/pull/132} \\
Bloc & Merged & \url{https://github.com/feenkcom/Bloc/pull/7} \\
GraphQL & Merged & \url{https://github.com/OBJECTSEMANTICS/GraphQL/pull/12} \\
Zinc & Merged & \url{https://github.com/svenvc/zinc/pull/58}\\
DiscordSt & Merged & \url{https://github.com/JurajKubelka/DiscordSt/pull/75}\\
MaterialDesignLite & Merged & \url{https://github.com/DuneSt/MaterialDesignLite/pull/308}\\
PetitParser2 & Open & \url{https://github.com/kursjan/petitparser2/pull/64} \\
OpenPonk & Open & \url{https://github.com/OpenPonk/openponk/pull/35} \\
Telescope & Open & \url{https://github.com/TelescopeSt/Telescope/pull/162}\\
\hline
\end{tabular}
\caption{Pull requests submitted on GitHub}
\label{table:pr}
\end{table*}
 
\subsubsection{Pull-requests preparation} \label{sec:PR-preparation}

\addressedRvw{Each pull-request contains a single amplified test method\footnote{Exception for the project \inlinecode{MaterialDesignLite} where we 2 very similar test methods in a single pull request}.
In order to attract the developers' interest, we try to select a test method testing an important class/method.
We select the class under test by scanning their name and relating the names to the context of the project.
For example, we know the project \inlinecode{Zinc} is a HTTP component, so the class \inlinecode{ZnRequest} should be a core class.
We run the tool on the selected test class, and then scan the generated test methods to select one of them.
In selecting an amplified test method, we still consider the vocabularies in the name of the original test method.
We also prioritize the tests with more mutants killed.}

In addition, we need to explain why a test is valuable in each pull-request.
Since some developers may not be familiar with the concept of mutation testing, we need to understand the test in advance and explain it in simpler words.
\addressedRvw{We inspect at the selected test method and try to understand the effect of the killed mutant and come up with an easy to understand explanation.
Examples of explanations are: increasing branch coverage (\inlinecode{PolyMath}), raising an exception (\inlinecode{pharo-project, DataFrame}), covering new state revealing methods (\inlinecode{Bloc, Zinc}), reducing technical debt (\inlinecode{GraphQL}).}

After selecting an amplified test method, we perform small corrections on the generated code, as a normal Pharo developer would do when see an auto-generated code.
These corrections include choosing more meaningful names for the test method, variables and string constants, or deleting the superfluous lines, and adding comments for small hints.

All preparation steps are performed by the first author and are reviewed by the second and third authors.
In the time of experiment, the familiarity of the first author about the projects was only studying parts of provided readme description in GitHub.
So, he was totally unfamiliar with the projects, he had not contributed to any of the projects, and had never reviewed their code.
In fact this shows although there might be more interesting tests for experts, a normal Pharo developer with limited knowledge about the projects is able to review the output and detect some useful test methods that are merged to the projects.
The preparation of the tests was quite straightforward and normally did not take more than one hour for each project.

\subsubsection{Pilot pull-request} \label{sec:PR-pilot}
Initially, we sent a pull-request\footnote{\url{https://github.com/SeasideSt/Seaside/pull/1215}} to \texttt{Seaside} project containing the suggestion for adding a set of new lines into an existing test method.
The main goal of this pull-request was to learn more about how developers deal with pull-requests from strangers.

We consider the fact that \texttt{Seaside} project is a framework for web application development and we scanned the name of classes and selected \inlinecode{WARequestTest} because we expected this test class is related to a core class-under-test which interacts with Http requests.
Then, we amplified the test class and selected the test method with the most mutants killed.

\begin{figure}[!h]
  \includegraphics[width=0.95\linewidth]{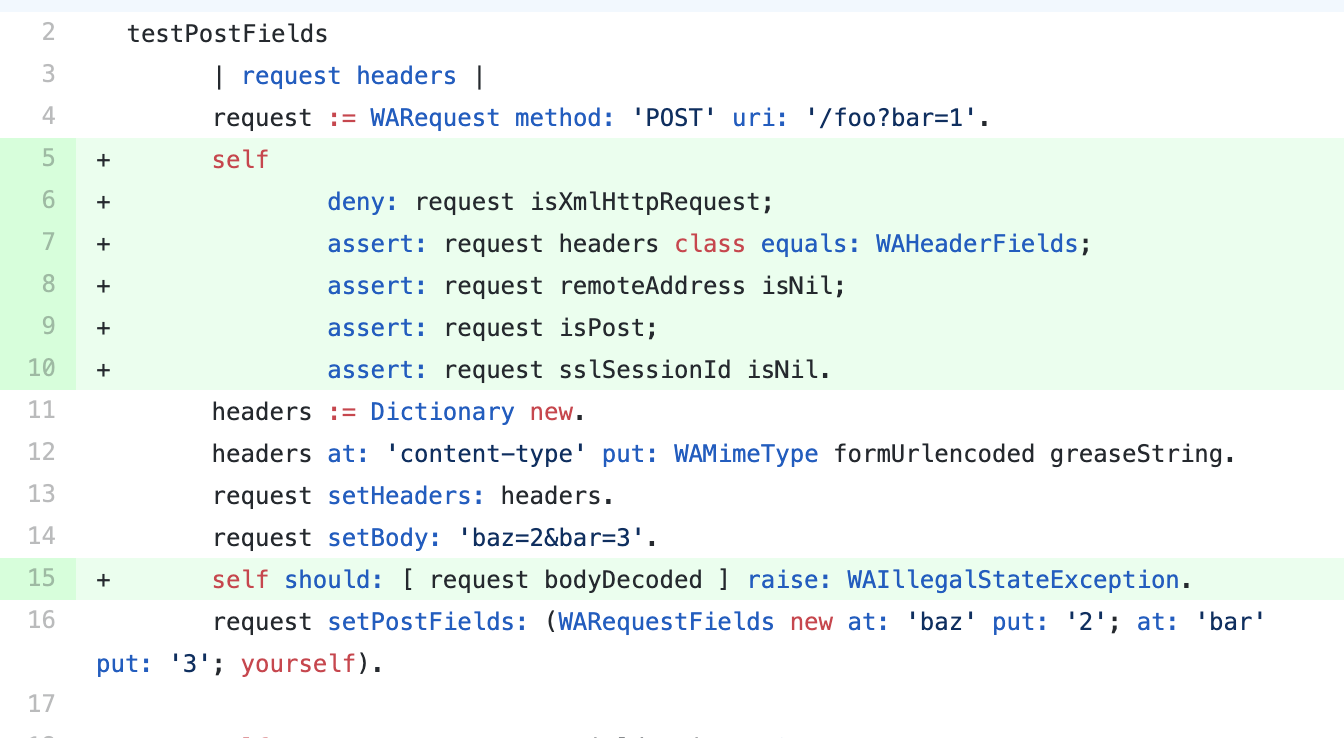}
  \caption{Improvements on an existing test method submitted to \toolname{Seaside}}
  \label{fig:pr-seaside}
\end{figure}

The selected new test method was able to kill 6 new mutants and was the result of a cooperation between assertion amplification (\secref{sec:input-amp}) and input amplification (section  \ref{sec:assert-amp}).
We merged the parts of amplified test into the original test method (\inlinecode{\#testPostFields}).
The test is shown in \figref{fig:pr-seaside}. 
Lines 5 to 10 are produced by assertion amplification on the original test method (\inlinecode{\#testPostFields}).
Line 15 is added by the \textit{method-call-adder} input-amplifier.

We wrote a description for the pull-request trying to explain why this test is useful.
We also expressed that the test method is the output of a tool, because it is important to inform developers in advance that they are participating in an experiment.

After a few days the test was merged by one of the project's developers.
Moreover, the developer left a valuable comment containing the following points:

\begin{compactitem}

\item 
\textbf{The suggestions do not fit this test method}: 
The developer said \comment{I expected the testPostfields unit test method to focus on testing the postFields}.
We agree with his remark.
If the suggested changes do not have a semantic relation to the original test method, it should be moved to another test or a new one.
We considered this advice in the subsequent pull-requests.

\item
\textbf{Usefulness of the result to refactoring the tests}: 
The developer also stated \comment{the result of the test amplification makes me evaluate the existing unit tests and refactor them to improve the test coverage and test factorization}.
This shows that even if the immediate results of test amplification are not tidy enough, they still help refactor existing tests.
\end{compactitem}

\subsubsection{Pull-request details}\label{sec:PR-all}

In the following parts we describe the details on the pull-requests on each project.

\begin{figure}[!h]
  \includegraphics[width=0.95\linewidth]{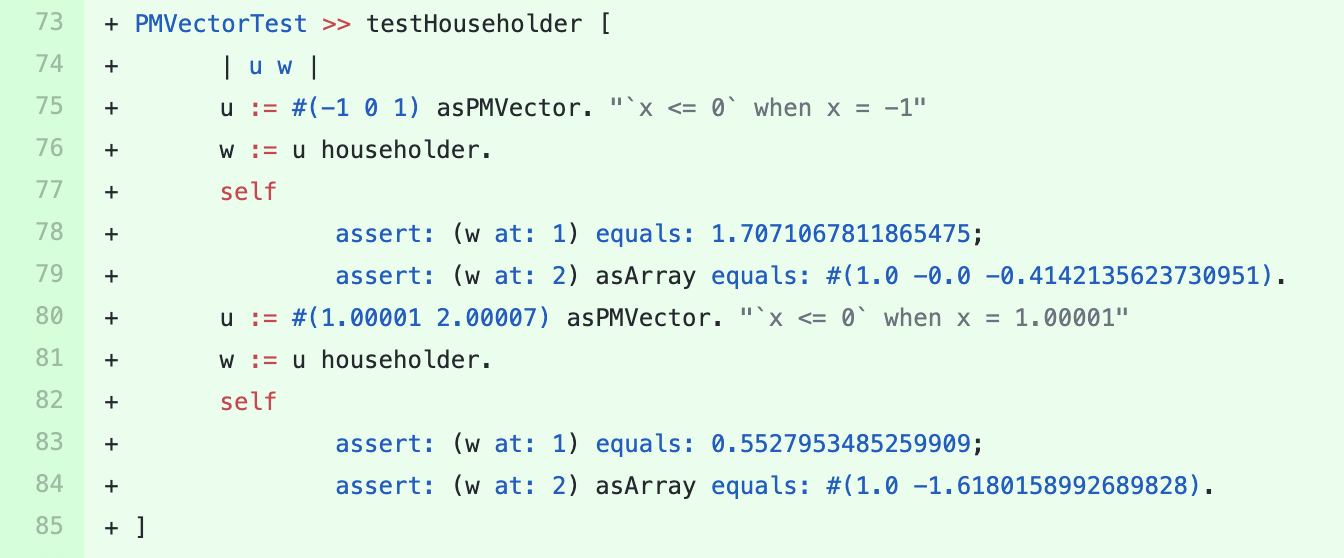}
  \caption{A new test method submitted to \toolname{PolyMath}}
  \label{fig:pr-polymath}
\end{figure}

\paragraph{PolyMath.}
We sent a pull-request\footnote{\url{https://github.com/PolyMathOrg/PolyMath/pull/178}} to this project containing the suggestion  for adding a new test method in the test class \inlinecode{PMVectorTest}.
The suggested test method is shown in \figref{fig:pr-polymath}. 

This test method is testing the call of the method \inlinecode{\#householder} on two different vectors.
Before this test, the method \inlinecode{\#householder} was not covered in this test class.

The \textit{method-call-adder} input amplifier adds calls to an existing method in the public interface of a class to the test to force the object under test in a new state.
We merged two of them in a new test method that execute two different branches in the test method (based on the condition \inlinecode{x $\leq$ 0}).
The former vector (line 75 in \figref{fig:pr-polymath}) forces the ifTrue branch and the latter vector (line 80 in \figref{fig:pr-polymath}) forces ifFalse branch.
Note that the comment text (line 75) is added manually to increase the readability of the test.

The original test method included two assertions to confirm the type of the returned value of the method (\inlinecode{self assert: w class equals: Array}).
The developers asked us to omit these assertion statements.
We changed the pull request accordingly and it was merged immediately.

\begin{figure}[!h]
  \includegraphics[width=0.95\linewidth]{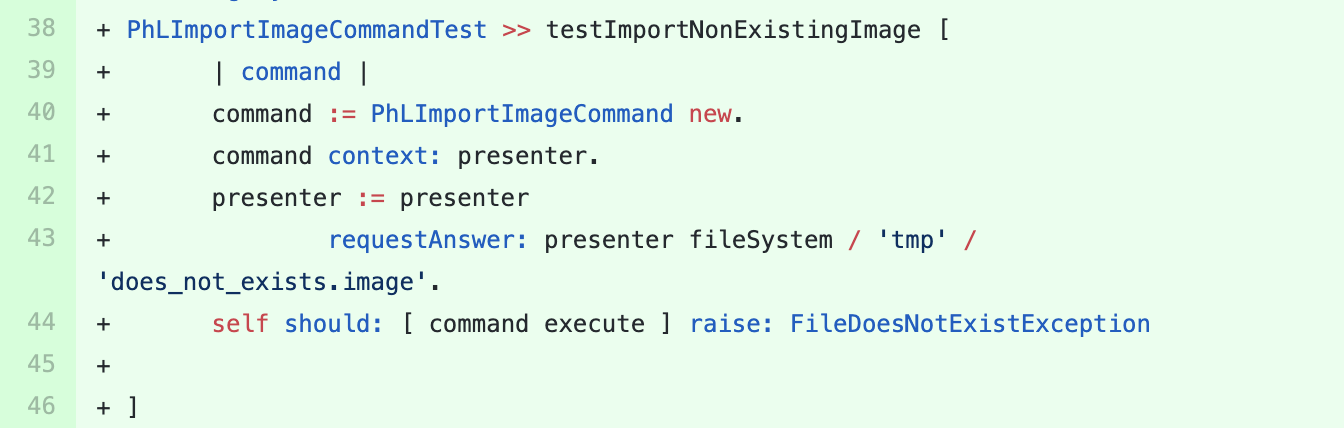}
  \caption{A new test method sent in a pull-request to the project Pharo-Launcher}
  \label{fig:pr-pharolauncher}
\end{figure}

\paragraph{Pharo-Launcher.}
We sent a pull-request\footnote{\url{https://github.com/pharo-project/pharo-launcher/pull/500}} to this project containing the suggestion  for adding a new test method in the test class \inlinecode{PhLImportImageCommandTest}.
The suggested test method is shown in \figref{fig:pr-pharolauncher}. 

This test is produced from the original test method of \inlinecode{testCanImportAnImage} which verifies an image can be imported using a valid filename.
\smallamp applies a literal mutation on the file name (\inlinecode{`foo.image'} changed to \inlinecode{`fo.image'}) that results an invalid filename and consequently raising a \inlinecode{FileDoesNotExistException} error.
While preparing the test for the pull-request, we modified the name of the test method and the file to be more meaningfull.

The pull-request was merged in the same day with this comment: \comment{Indeed, the test you are adding has a value. Good job SmallAmp}.

\paragraph{DataFrame.}

\begin{figure}[!h]
  \includegraphics[width=0.95\linewidth]{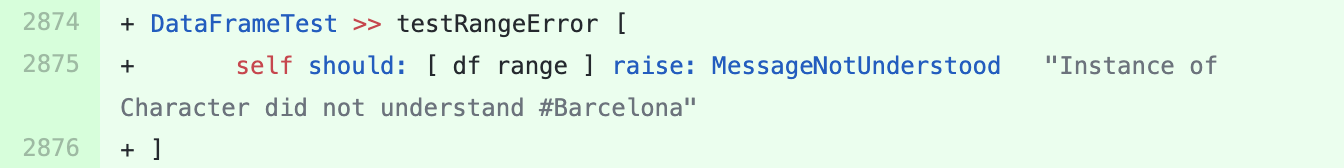}
  \caption{A new test method sent in a pull-request to the project DataFrame}
  \label{fig:pr-dataframe}
\end{figure}

We sent a pull-request\footnote{\url{https://github.com/PolyMathOrg/DataFrame/pull/132}} to this project containing the suggestion  for adding a new test method in the test class \inlinecode{DataFrameTest}.
The suggested test method is shown in \figref{fig:pr-dataframe}. 

The variable \inlinecode{df} is an instance variable that has been initialized in the \inlinecode{\#setUp} method.
It includes a tabular data mixed from numbers and texts.
The initial amplified test method was generated by adding the method \inlinecode{\#range} as the first statement in one of the original test methods.
We recognized the remaining statements as superfluous lines and removed all of them.
We also added a comment including the exception description.

This test method makes it explicit that calling the method \inlinecode{\#range} on a DataFrame object containing non-numerical columns throws an exception.
With this new test it becomes an explicit part of the contract for DataFrame.

The pull-request was merged after a few weeks.
A developer of the project commented: \comment{Small-amp seems to be a very valuable tool!}
   
\begin{figure}[!h]
  \includegraphics[width=0.95\linewidth]{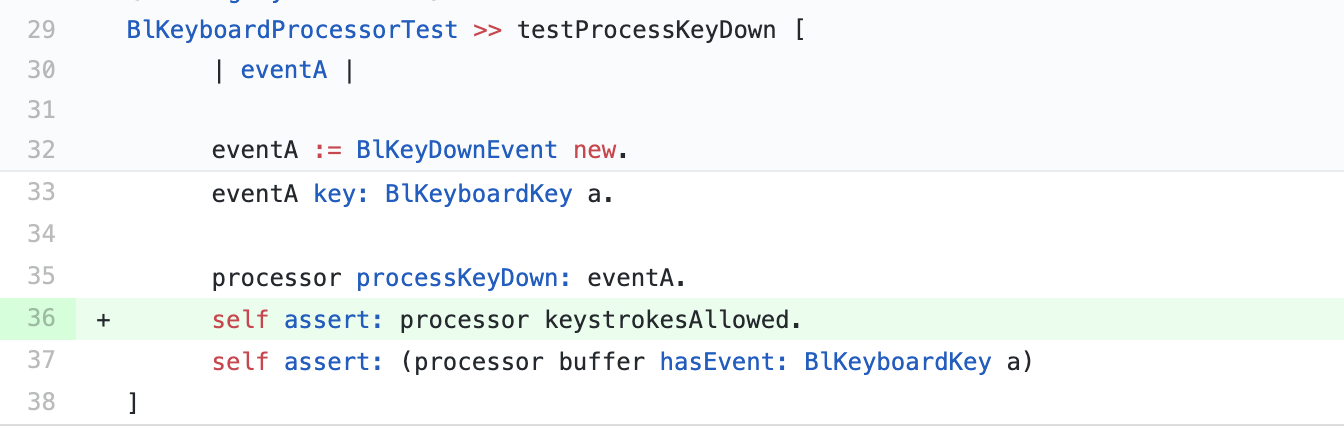}
  \caption{A new assertion suggested in a pull-request to the project Bloc}
  \label{fig:pr-bloc}
\end{figure}

\paragraph{Bloc.}
We sent a pull-request\footnote{\url{https://github.com/feenkcom/Bloc/pull/7}} to this project containing the suggestion  for adding a new assertion in an existing test method  in the test class \inlinecode{BlKeyboardProcessorTest}.
The suggested test method is shown in \figref{fig:pr-bloc}. 

By calling the state revealing method \inlinecode{\#keystrokesAllowed}, the assertion verifies the correctness of the object state after an \inlinecode{\#processKeyDown:} event.
This test is the result of combining assertion-amplification with oracle-reduction.
Normally, the assertions-amplification step generates lots of assertions, and the oracle-reduction module removes all assertion statements that do not kill any mutant.
So, the test code did not need any special preparation and we only need to provide a comment to explain the test method.

The pull-request was also merged after a few weeks with a positive comment.

\paragraph{GraphQL.}

\begin{figure}[!h]
  \includegraphics[width=0.95\linewidth]{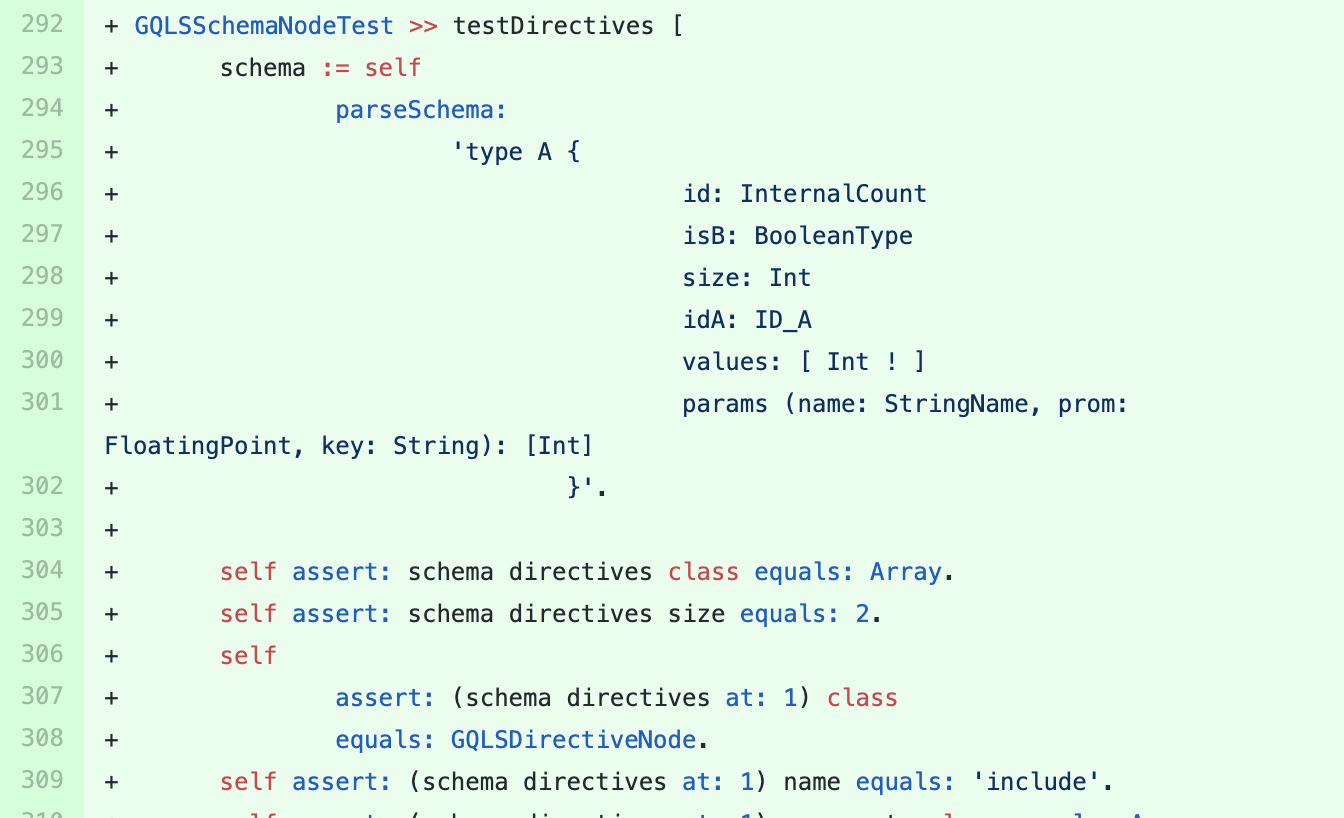}
  \caption{A new test method suggested in a pull-request to the project GraphQL}
  \label{fig:pr-graphql}
\end{figure}

We sent a pull-request\footnote{\url{https://github.com/OBJECTSEMANTICS/GraphQL/pull/12}} to this project containing the suggestion  for adding a new test method in the test class \inlinecode{GQLSSchemaNodeTest}.
The suggested test method is shown in \figref{fig:pr-graphql}. 

This test method verifies the return value of \inlinecode{directives} in an \inlinecode{schema} object.
The returned value is generated in the method \inlinecode{GQLSSchemaNode $>$$>$ initializeDefaultDirectives} and contains technical debt. 
This test method guards against future evolutions which may break assumptions made by clients.
We selected a meaningful name for the test and wrote a comment text.
We also added back some of the assertions removed by oracle-reduction step.
The pull-request was merged after a few days.

\paragraph{Zinc.}

\begin{figure}[!h]
  \includegraphics[width=0.95\linewidth]{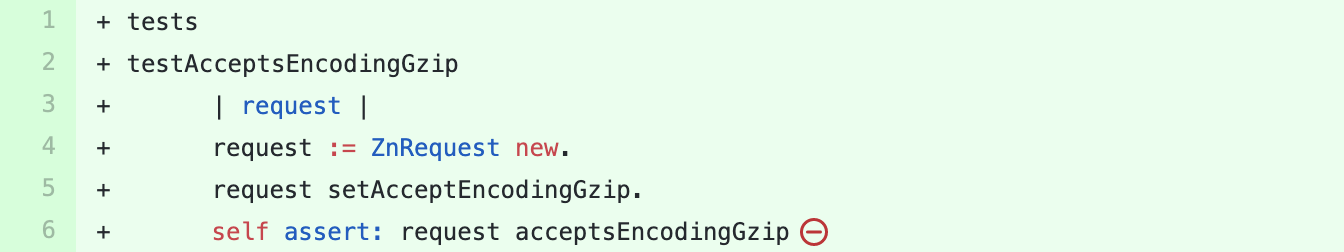}
  \caption{A new test method suggested in a pull-request to the project Zinc}
  \label{fig:pr-zinc}
\end{figure}

We sent a pull-request\footnote{\url{https://github.com/svenvc/zinc/pull/58}} to this project containing the suggestion  for adding a new test method in the test class \inlinecode{ZnRequestTest}.
The suggested test method is shown in \figref{fig:pr-zinc}. 

This test method calls the method \inlinecode{\#setAcceptEncodingGzip} on an \inlinecode{request} object.
Then calls another method \inlinecode{\#acceptsEncodingGzip} to verify the change.
Both of these methods were not covered in this test class before this test method.

This method is built by cooperating three module of \smallamp.
First, \textit{method-call-adder} input amplifier adds a new method call.
Then \textit{assertion amplification} inserts a set of new assertions after the added method call.
And finally, after the main amplification loop is finished, the oracle-reduction step rejects all superfluous assertion statements.
This test method did not need much preparation and we only selected a meaningful name for it.
The pull-request was merged in the same day.

\begin{figure}[!h]
  \includegraphics[width=0.95\linewidth]{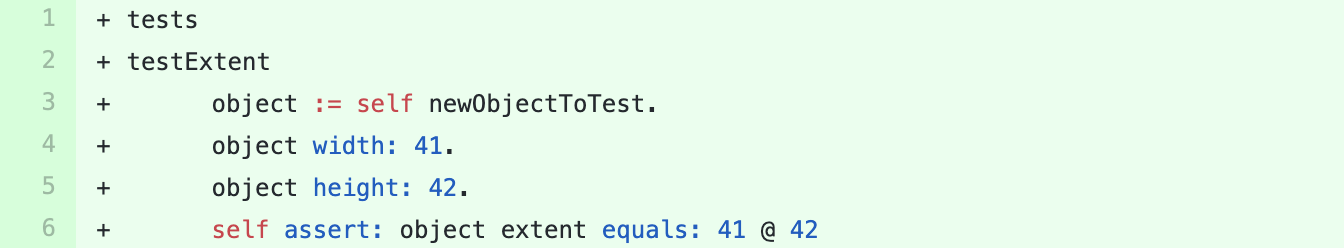}
  \caption{A new test method suggested in a pull-request to the project DiscordSt}
  \label{fig:pr-discordst}
\end{figure}

\paragraph{DiscordSt.}
We sent a pull-request\footnote{\url{https://github.com/JurajKubelka/DiscordSt/pull/75}} to this project containing the suggestion  for adding a new test method in the test class \inlinecode{DSEmbedImageTest}.
The suggested test method is shown in \figref{fig:pr-discordst}. 

The method covers the method \inlinecode{\#extent} which was not covered in the test class before.
This test method did not need much preparation and we only selected a meaningful name for it.
The pull-request was merged after a few days.

\begin{figure}[!h]

  \includegraphics[width=0.95\linewidth]{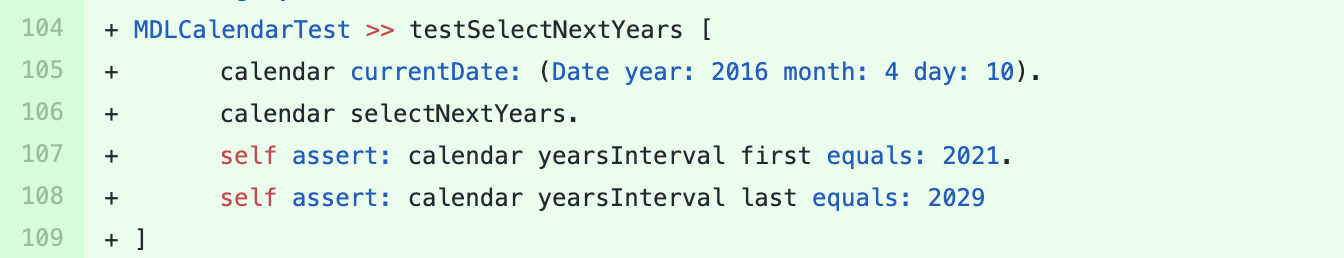}
  \includegraphics[width=0.95\linewidth]{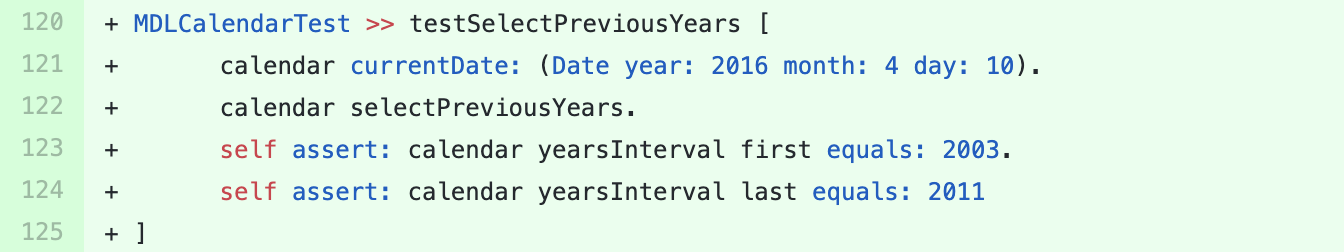}

  \caption{Test methods sent in a pull-request to the project MaterialDesignLite}
  \label{fig:pr-materialdesignlite}
\end{figure}

\paragraph{MaterialDesignLite.}
We sent a pull-request\footnote{\url{https://github.com/DuneSt/MaterialDesignLite/pull/308}} to this project containing the suggestion  for adding two new test methods  in the test class \inlinecode{MDLCalendarTest}.
The suggested test methods are shown in figure \ref{fig:pr-materialdesignlite}. 

Both of test methods are similar and are created by adding a new method call to the test input.
The tests are created for the \textit{Calendar} widget and verify correctness of \inlinecode{\#selectPreviousYears} and \inlinecode{\#selectNextYears} methods.
In these test methods, the oracle-reduction step removed most of the assertions and it only
 preserved the first assertion killing the mutant: \inlinecode{self assert: calendar yearsInterval fourth equals: 2006}.
We replaced the assertions with more human readable assertions (asserting \inlinecode{first} and \inlinecode{last} of the interval).
The pull-request was merged the day after.

\begin{figure}[!h]
  \includegraphics[width=0.95\linewidth]{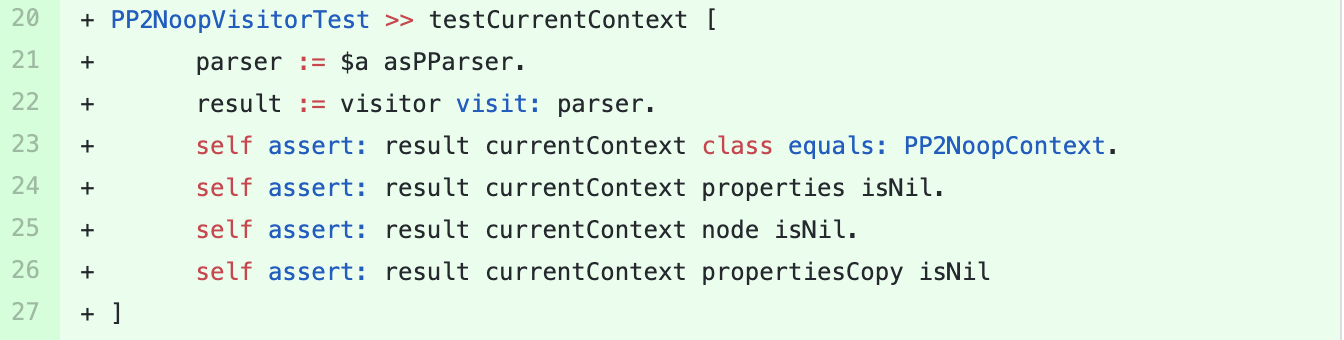}
  \caption{A test method sent in a pull-request to the project PetitParser2}
  \label{fig:pr-petitparser2}
\end{figure}

\paragraph{PetitParser2.}
We sent a pull-request\footnote{\url{https://github.com/kursjan/petitparser2}} to this project containing the suggestion  for adding a new test method  in the test class \inlinecode{PP2NoopVisitorTest}.
The suggested test method is shown in \figref{fig:pr-petitparser2}. 

The test method tests the value of \inlinecode{currentContext} in \inlinecode{result} object.
This test method resulted from assertion amplification combined with oracle-reduction.
The test had two assertions: 
\inlinecode{self deny: visitor isRoot} and \inlinecode{self assert: visitor currentContext class equals: PP2NoopContext}.
We added back some of removed assertions relating to \inlinecode{currentContext} and also removed the \inlinecode{self deny: visitor isRoot} to make the test more focused.
The pull-request is not merged up to the date of  writing (\today).

\begin{figure}[!h]
  \includegraphics[width=0.95\linewidth]{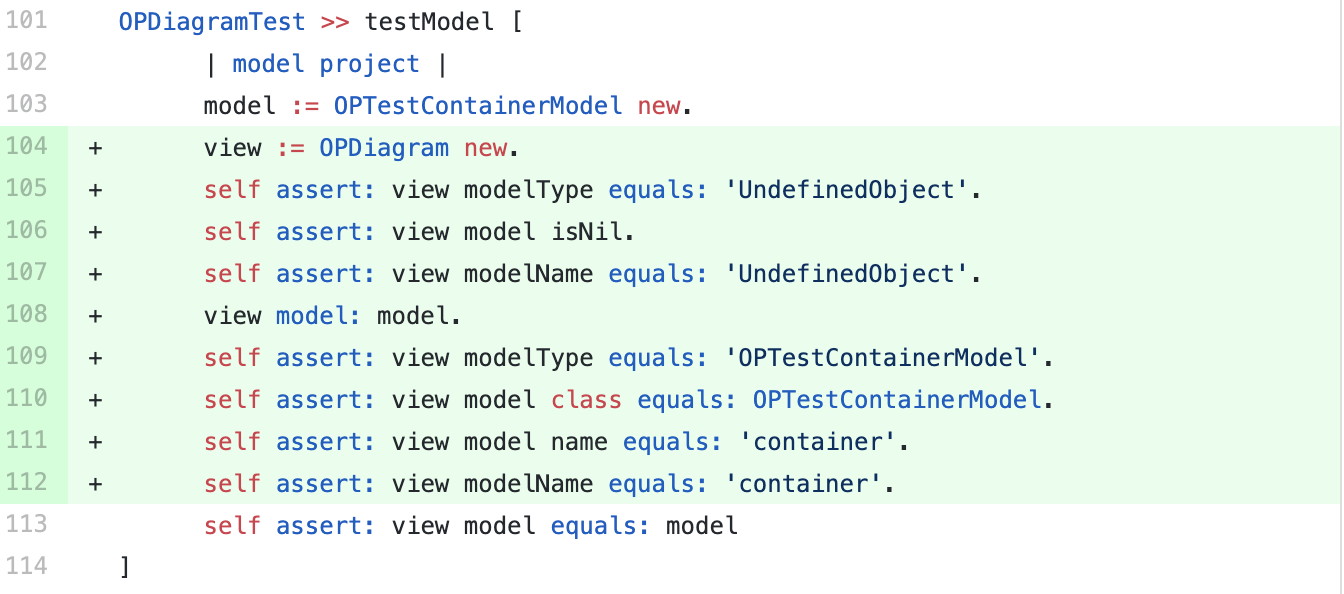}
  \caption{Changes on an existing test method - OpenPonk}
  \label{fig:pr-openponk}
\end{figure}

\paragraph{OpenPonk.}
We sent a pull-request\footnote{\url{https://github.com/OpenPonk/openponk}} to this project containing the suggestion  for adding a set of new lines in an existing test method  in the test class \inlinecode{OPDiagramTest}.
The suggested test method is shown in \figref{fig:pr-openponk}. 

The original test method is presented in \lstref{listing:openponk-original}.

\begin{minipage}{\textwidth}
\begin{lstlisting}[
         captionpos=b,   caption=Original test method - OpenPonk,
	label=listing:openponk-original]
OPDiagramTest >> testModel [
	| model project |
	model := OPTestContainerModel new.
	view := OPDiagram new model: model.
	self assert: view model equals: model
]	
\end{lstlisting}
\end{minipage}

\smallamp has broken the statement at line 4 in \lstref{listing:openponk-original} (the result is visible in lines 104 and 108 in \figref{fig:pr-openponk}) and then added a series of assertions.
Since this test is dedicated to test \inlinecode{model}, we kept all assertions reflecting the state of \inlinecode{model} and removed other assertions.
So, the assertions  in lines 104 to 107 verify the state of a freshly initialized \inlinecode{OPDiagram} object (where model is \inlinecode{nil}), and the assertions in lines 109 to 112 verify the public API through the accessor methods.
The pull-request is not merged up to the date of  writing this paper.

\begin{figure}[!h]
  \includegraphics[width=0.95\linewidth]{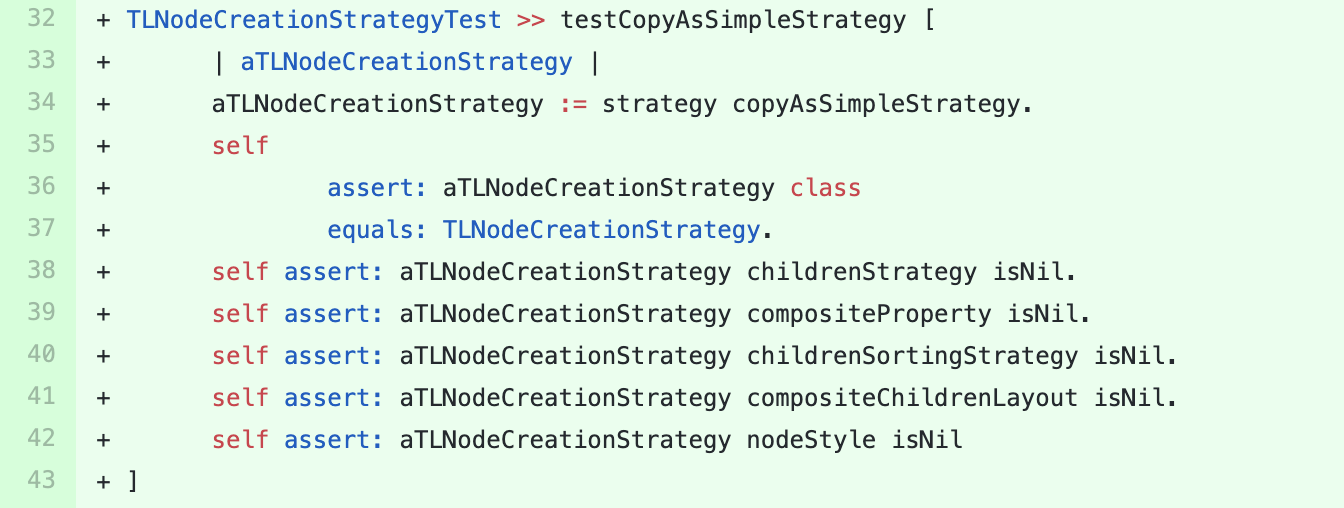}
  \caption{A test method suggested in a pull-request to the project Telescope}
  \label{fig:pr-telescope}
\end{figure}

\paragraph{Telescope.}
We sent a pull-request\footnote{\url{https://github.com/TelescopeSt/Telescope}} to this project containing the suggestion  for adding a new test method  in the test class \inlinecode{TLNodeCreationStrategyTest}.
The suggested test method is shown in \figref{fig:pr-telescope}. 

The test method verifies the state of the returned object from calling \inlinecode{copyAsSimpleStrategy}.
This method is never covered in the test class.
It also contain technical debt.
The call to \inlinecode{copyAsSimpleStrategy} is added by method-call-addition amplifier and the state of the returned value is asserted via assertion-amplification.
We kept all assertions related to the returned value, and removed all other superfluous lines to make the test more readable.
The pull-request is not merged up to the date of  writing this paper.

\hypobox{
\textbf{Answer to RQ1: }
We submitted \nSentPR pull requests through GitHub to propose amplified test methods to developers.
In \nMergedPR cases, our request was accepted by the developers and the test has been merged to the code base.
In the three remaining cases our pull request was ignored.
Moreover, we received qualitative feedback from developers acknowledging the relevance of amplified test methods.
}

\subsection{RQ2 --- \rqTwoKey}

We use the results in \tabref{tab:results} for answering the next research questions.
These tables present the result of test amplification on the all selected classes selected in our dataset.
In \tabref{tab:results}, the first 104 rows represent test amplification for test classes with high mutation coverage, while the remaining of the rows show the test classes with poor mutation coverage.
\smallamp algorithm (\algoref{alg:main}) has a stochastic nature, especially test input reduction (\secref{sec:input-reducing}) which heavily depends on randomness.
Therefore, we ran the algorithm three times on each test class to observe the effect of randomness on the results.
\addressedRvw{In addition, we ran the algorithm another time by disabling the profiling and the type sensitive operator for investigating the effectiveness of type profiler (denoted by \noProfilingText).}

The columns in this table indicate:

\begin{sidewaystable}
    \centering
    {\small
\caption{The result of test amplification by \smallamp on the 52 test classes. (Tests with high coverage)}
\label{tab:results}
\resizebox{\textwidth}{!}{
\begin{tabular}{$c^W^c^c^c^c^c^c^c^c^c^c^c^c^c}
Id &
  Class &
  \# Test  &
  \# loc  &  
  \% Mut. &
  \# New&
  \# Focused &
  \# Killed&
  \# Newly&
  \% Increase&
  \# Newly&
  \% Increase&
  \# Newly&
  \% Increase&
  Time \\
  &
  &
  methods&
  CUT&
  score &
  test &
  methods &
  mutants&
  killed&
  killed&
  mutant&
  killed only&
  killed type&
  killed type&
  (h:m:s)
  \\
  &
  &
  original &
  &
  &
  methods &
  &
  original &
  amplified&
  amplified&
  A-amp &
  A-amp &
  aided &
  aided
  \\
  \hline
1   & DataFrameJsonWriterTest              & 4  & 51  & 50    & 1  & 1  & 6   & 2  & 33.33  & 1 & 16.67 & 0  & 0.00                                          & 0:00:11 \\
2   &                                      &    &     &       & 1  & 1  & 6   & 2  & 33.33  & 1 & 16.67 & 0  & 0.00                                          & 0:00:11 \\
3   &                                      &    &     &       & 1  & 1  & 6   & 2  & 33.33  & 1 & 16.67 & 0  & 0.00                                          & 0:00:11 \\
4   &  \noProfiling                                 &    &     &       & 1  & 1  & 6   & 2  & 33.33  & 1 & 16.67 & 0  & 0.00                                          & 0:00:09 \\ \hline
5   & DataFrameCsvReaderTest               & 4  & 62  & 85    & 0  & 0  & 17  & 0  & 0.00   & 0 & 0     & 0  & 0.00                                          & 0:00:08 \\
6   &                                      &    &     &       & 0  & 0  & 17  & 0  & 0.00   & 0 & 0     & 0  & 0.00                                          & 0:00:07 \\
7   &                                      &    &     &       & 0  & 0  & 17  & 0  & 0.00   & 0 & 0     & 0  & 0.00                                          & 0:00:07 \\
8   &  \noProfiling                                       &    &     &       & 0  & 0  & 17  & 0  & 0.00   & 0 & 0     & 0  & 0.00                                          & 0:00:07 \\ \hline
9   & DataFrameJsonReaderTest              & 9  & 80  & 72.73 & 2  & 2  & 16  & 4  & 25.00  & 1 & 6.25  & 0  & 0.00                                          & 0:08:43 \\
10  &                                      &    &     &       & 2  & 2  & 16  & 4  & 25.00  & 1 & 6.25  & 0  & 0.00                                          & 0:08:57 \\
11  &                                      &    &     &       & 2  & 2  & 16  & 4  & 25.00  & 1 & 6.25  & 0  & 0.00                                          & 0:09:11 \\
12  &  \noProfiling                                       &    &     &       & 2  & 2  & 16  & 4  & 25.00  & 1 & 6.25  & 0  & 0.00                                          & 0:05:14 \\ \hline
13  & DataFrameTypeDetectorTest            & 15 & 279 & 94.12 & 0  & 0  & 64  & 0  & 0.00   & 0 & 0     & 0  & 0.00                                          & 0:02:02 \\
14  &                                      &    &     &       & 0  & 0  & 64  & 0  & 0.00   & 0 & 0     & 0  & 0.00                                          & 0:02:06 \\
15  &                                      &    &     &       & 0  & 0  & 64  & 0  & 0.00   & 0 & 0     & 0  & 0.00                                          & 0:02:06 \\
16  & \noProfiling                                        &    &     &       & 0  & 0  & 64  & 0  & 0.00   & 0 & 0     & 0  & 0.00                                          & 0:01:48 \\ \hline
17  & ZnMessageBenchmarkTest               & 2  & 187 & 80    & 0  & 0  & 24  & 0  & 0.00   & 0 & 0     & 0  & 0.00                                          & 0:01:23 \\
18  &                                      &    &     &       & 0  & 0  & 24  & 0  & 0.00   & 0 & 0     & 0  & 0.00                                          & 0:01:21 \\
19  &                                      &    &     &       & 0  & 0  & 24  & 0  & 0.00   & 0 & 0     & 0  & 0.00                                          & 0:01:24 \\
20  &  \noProfiling                                       &    &     &       & 0  & 0  & 24  & 0  & 0.00   & 0 & 0     & 0  & 0.00                                          & 0:01:05 \\ \hline
21  & ZnEasyTest                           & 10 & 65  & 66.67 & 0  & 0  & 10  & 0  & 0.00   & 0 & 0     & 0  & 0.00                                          & 0:46:25 \\
22  &                                      &    &     &       & 0  & 0  & 10  & 0  & 0.00   & 0 & 0     & 0  & 0.00                                          & 0:44:20 \\
23  &                                      &    &     &       & 0  & 0  & 10  & 0  & 0.00   & 0 & 0     & 0  & 0.00                                          & 0:46:35 \\
24  &  \noProfiling                                       &    &     &       & 0  & 0  & 10  & 0  & 0.00   & 0 & 0     & 0  & 0.00                                          & 0:28:16 \\ \hline
25  & GQLRequestGrammarTest                & 29 & 142 & 96.55 & 0  & 0  & 56  & 0  & 0.00   & 0 & 0     & 0  & 0.00                                          & 0:04:07 \\
26  &                                      &    &     &       & 0  & 0  & 56  & 0  & 0.00   & 0 & 0     & 0  & 0.00                                          & 0:04:09 \\
27  &                                      &    &     &       & 0  & 0  & 56  & 0  & 0.00   & 0 & 0     & 0  & 0.00                                          & 0:04:02 \\
28  &  \noProfiling                                       &    &     &       & 0  & 0  & 56  & 0  & 0.00   & 0 & 0     & 0  & 0.00                                          & 0:04:10 \\ \hline
29  & GQLSingleAnonimousQueryEvaluatorTest & 22 & 56  & 52.27 & 2  & 2  & 23  & 2  & 8.70   & 0 & 0     & 0  & 0.00                                          & 0:05:21 \\
30  &                                      &    &     &       & 2  & 2  & 23  & 2  & 8.70   & 0 & 0     & 0  & 0.00                                          & 0:05:26 \\
31  &                                      &    &     &       & 2  & 2  & 23  & 2  & 8.70   & 0 & 0     & 0  & 0.00                                          & 0:05:15 \\
32  &  \noProfiling                                       &    &     &       & 1  & 1  & 23  & 1  & 4.35   & 0 & 0     & 0  & 0.00                                          & 0:06:02 \\ \hline
33  & GQLSchemaGrammarTest                 & 57 & 40  & 93.2  & 7  & 7  & 96  & 7  & 7.29   & 0 & 0     & 6  & 6.25                                          & 2:16:15 \\
34  &                                      &    &     &       & 7  & 7  & 96  & 7  & 7.29   & 0 & 0     & 6  & 6.25                                          & 2:22:55 \\
35  &                                      &    &     &       & 7  & 7  & 96  & 7  & 7.29   & 0 & 0     & 6  & 6.25                                          & 2:34:02 \\
36  &  \noProfiling                                       &    &     &       & 0  & 0  & 96  & 0  & 0.00   & 0 & 0     & 0  & 0.00                                          & 0:11:01 \\ \hline
37  & WebGrammarTest                       & 14 & 28  & 90    & 2  & 2  & 18  & 2  & 11.11  & 0 & 0     & 2  & 11.11                                         & 0:46:51 \\
38  &                                      &    &     &       & 2  & 2  & 18  & 2  & 11.11  & 0 & 0     & 2  & 11.11                                         & 0:43:21 \\
39  &                                      &    &     &       & 2  & 2  & 18  & 2  & 11.11  & 0 & 0     & 2  & 11.11                                         & 0:42:40 \\
40  & \noProfiling                                        &    &     &       & 0  & 0  & 18  & 0  & 0.00   & 0 & 0     & 0  & 0.00                                          & 0:16:10 \\ \hline
41  & PP2BufferStreamTest                  & 16 & 188 & 83.87 & 0  & 0  & 78  & 0  & 0.00   & 0 & 0     & 0  & 0.00                                          & 0:00:34 \\
42  &                                      &    &     &       & 0  & 0  & 78  & 0  & 0.00   & 0 & 0     & 0  & 0.00                                          & 0:00:35 \\
43  &                                      &    &     &       & 0  & 0  & 78  & 0  & 0.00   & 0 & 0     & 0  & 0.00                                          & 0:00:36 \\
44  &  \noProfiling                                      &    &     &       & 0  & 0  & 78  & 0  & 0.00   & 0 & 0     & 0  & 0.00                                          & 0:00:31 \\
  \hline
\end{tabular}
}}
\end{sidewaystable}

\begin{sidewaystable}
    \centering
\ContinuedFloat
\captionsetup{list=off,format=cont}
\caption{The result of test amplification by \smallamp on the 52 test classes. (Tests with high coverage)}
\resizebox{\textwidth}{!}{
\begin{tabular}{$c^W^c^c^c^c^c^c^c^c^c^c^c^c^c}
Id &
  Class &
  \# Test  &
  \# loc  &    
  \% Mut. &
  \# New&
  \# Focused &
  \# Killed&
  \# Newly&
  \% Increase&
  \# Newly&
  \% Increase&
  \# Newly&
  \% Increase&
  Time \\
  &
  &
  methods&
  CUT&
  score &
  test &
  methods &
  mutants&
  killed&
  killed&
  mutant&
  killed only&
  killed type&
  killed type&
  (h:m:s)
  \\
  &
  &
  original &
  &
  &
  methods &
  &
  original &
  amplified&
  amplified&
  A-amp &
  A-amp &
  aided &
  aided
  \\
  \hline
45  & MDLPanelSwitcherButtonTest           & 6  & 100 & 53.85 & 0  & 0  & 7   & 0  & 0.00   & 0 & 0     & 0  & 0.00                                          & 0:00:46 \\
46  &                                      &    &     &       & 0  & 0  & 7   & 0  & 0.00   & 0 & 0     & 0  & 0.00                                          & 0:00:46 \\
47  &                                      &    &     &       & 0  & 0  & 7   & 0  & 0.00   & 0 & 0     & 0  & 0.00                                          & 0:00:47 \\
48  & \noProfiling                                     &    &     &       & 0  & 0  & 7   & 0  & 0.00   & 0 & 0     & 0  & 0.00                                          & 0:00:47 \\ \hline
49  & MDLPaginationComponentTest           & 11 & 199 & 71.19 & 0  & 0  & 42  & 0  & 0.00   & 0 & 0     & 0  & 0.00                                          & 0:00:27 \\
50  &                                      &    &     &       & 0  & 0  & 42  & 0  & 0.00   & 0 & 0     & 0  & 0.00                                          & 0:00:27 \\
51  &                                      &    &     &       & 0  & 0  & 42  & 0  & 0.00   & 0 & 0     & 0  & 0.00                                          & 0:00:26 \\
52  &  \noProfiling                                    &    &     &       & 0  & 0  & 42  & 0  & 0.00   & 0 & 0     & 0  & 0.00                                          & 0:00:27 \\ \hline
53  & RSUMLClassBuilderTest                & 2  & 23  & 50    & 1  & 1  & 1   & 1  & 100.00 & 1 & 100   & 0  & 0.00                                          & 0:05:00 \\
54  &                                      &    &     &       & 1  & 1  & 1   & 1  & 100.00 & 1 & 100   & 0  & 0.00                                          & 0:05:34 \\
55  &                                      &    &     &       & 1  & 1  & 1   & 1  & 100.00 & 1 & 100   & 0  & 0.00                                          & 0:05:23 \\
56  &  \noProfiling                                    &    &     &       & 1  & 1  & 1   & 1  & 100.00 & 1 & 100   & 0  & 0.00                                          & 0:03:30 \\ \hline
57  & RSLabelGeneratorTest                 & 2  & 353 & 82.38 & 0  & 0  & 173 & 0  & 0.00   & 0 & 0     & 0  & 0.00                                          & 0:11:19 \\
58  &                                      &    &     &       & 0  & 0  & 173 & 0  & 0.00   & 0 & 0     & 0  & 0.00                                          & 0:11:07 \\
59  &                                      &    &     &       & 0  & 0  & 173 & 0  & 0.00   & 0 & 0     & 0  & 0.00                                          & 0:12:21 \\
60  & \noProfiling                                     &    &     &       & 0  & 0  & 173 & 0  & 0.00   & 0 & 0     & 0  & 0.00                                          & 0:12:55 \\ \hline
61  & OPNullSerializerTest                 & 3  & 251 & 50    & 0  & 0  & 1   & 0  & 0.00   & 0 & 0     & 0  & 0.00                                          & 0:00:00 \\
62  &                                      &    &     &       & 0  & 0  & 1   & 0  & 0.00   & 0 & 0     & 0  & 0.00                                          & 0:00:00 \\
63  &                                      &    &     &       & 0  & 0  & 1   & 0  & 0.00   & 0 & 0     & 0  & 0.00                                          & 0:00:00 \\
64  &  \noProfiling                                    &    &     &       & 0  & 0  & 1   & 0  & 0.00   & 0 & 0     & 0  & 0.00                                          & 0:00:00 \\ \hline
65  & OPRTElementsConstraintTest           & 2  & 42  & 50    & 1  & 1  & 1   & 1  & 100.00 & 1 & 100   & 0  & 0.00                                          & 0:00:03 \\
66  &                                      &    &     &       & 1  & 1  & 1   & 1  & 100.00 & 1 & 100   & 0  & 0.00                                          & 0:00:03 \\
67  &                                      &    &     &       & 1  & 1  & 1   & 1  & 100.00 & 1 & 100   & 0  & 0.00                                          & 0:00:03 \\
68  & \noProfiling                                     &    &     &       & 1  & 1  & 1   & 1  & 100.00 & 1 & 100   & 0  & 0.00                                          & 0:00:03 \\ \hline
69  & PMBernoulliGeneratorTest             & 4  & 113 & 76.47 & 2  & 2  & 13  & 3  & 23.08  & 0 & 0     & 0  & 0.00                                          & 0:00:06 \\
70  &                                      &    &     &       & 3  & 3  & 13  & 2  & 15.38  & 0 & 0     & 0  & 0.00                                          & 0:00:07 \\
71  &                                      &    &     &       & 2  & 2  & 13  & 2  & 15.38  & 0 & 0     & 0  & 0.00                                          & 0:00:06 \\
72  & \noProfiling                                     &    &     &       & 3  & 3  & 13  & 3  & 23.08  & 0 & 0     & 0  & 0.00                                          & 0:00:05 \\ \hline
73  & PMBinomialGeneratorTest              & 3  & 164 & 83.33 & 1  & 1  & 10  & 1  & 10.00  & 1 & 10    & 0  & 0.00                                          & 0:01:01 \\
74  &                                      &    &     &       & 1  & 1  & 10  & 1  & 10.00  & 1 & 10    & 0  & 0.00                                          & 0:01:14 \\
75  &                                      &    &     &       & 1  & 1  & 10  & 1  & 10.00  & 1 & 10    & 0  & 0.00                                          & 0:01:18 \\
76  & \noProfiling                                     &    &     &       & 1  & 1  & 10  & 1  & 10.00  & 1 & 10    & 0  & 0.00                                          & 0:01:31 \\ \hline
77  & TLHideActionTest                     & 3  & 468 & 55.56 & 2  & 2  & 5   & 0  & 0.00   & 0 & 0     & 0  & 0.00                                          & 0:00:30 \\
78  &                                      &    &     &       & 3  & 3  & 5   & 3  & 60.00  & 0 & 0     & 1  & 20.00                                         & 0:00:33 \\
79  &                                      &    &     &       & 1  & 1  & 5   & 2  & 40.00  & 0 & 0     & 0  & 0.00                                          & 0:00:26 \\
80  & \noProfiling                                     &    &     &       & 1  & 1  & 5   & 2  & 40.00  & 0 & 0     & 0  & 0.00                                          & 0:00:14 \\ \hline
81  & TLExpandCollapseNodesActionTest      & 3  & 8   & 66.67 & 2  & 2  & 14  & 2  & 14.29  & 1 & 7.14  & 0  & 0.00                                          & 0:01:28 \\
82  &                                      &    &     &       & 2  & 2  & 14  & 2  & 14.29  & 0 & 0     & 0  & 0.00                                          & 0:00:58 \\
83  &                                      &    &     &       & 2  & 2  & 14  & 2  & 14.29  & 0 & 0     & 0  & 0.00                                          & 0:01:16 \\
84  & \noProfiling                                     &    &     &       & 1  & 1  & 14  & 1  & 7.14   & 0 & 0     & 0  & 0.00                                          & 0:00:31 \\ \hline
85  & DSDetectChannelCommandTest           & 5  & 51  & 50    & 0  & 0  & 3   & 0  & 0.00   & 0 & 0     & 0  & 0.00                                          & 0:00:40 \\
86  &                                      &    &     &       & 0  & 0  & 3   & 0  & 0.00   & 0 & 0     & 0  & 0.00                                          & 0:00:41 \\
87  &                                      &    &     &       & 0  & 0  & 3   & 0  & 0.00   & 0 & 0     & 0  & 0.00                                          & 0:00:39 \\
88  & \noProfiling                                     &    &     &       & 0  & 0  & 3   & 0  & 0.00   & 0 & 0     & 0  & 0.00                                          & 0:00:30 \\

\hline
\end{tabular}
}
\end{sidewaystable}

\begin{sidewaystable}
    \centering
    {\small
\ContinuedFloat
\captionsetup{list=off,format=cont}
\caption{The result of test amplification by \smallamp on the 52 test classes. \\
(Rows up to 104 are tests with high coverage, after 104 are tests with low coverage)}
\resizebox{\textwidth}{!}{
\begin{tabular}{$c^W^c^c^c^c^c^c^c^c^c^c^c^c^c}
Id &
  Class &
  \# Test  &
  \# loc  &      
  \% Mut. &
  \# New&
  \# Focused &
  \# Killed&
  \# Newly&
  \% Increase&
  \# Newly&
  \% Increase&
  \# Newly&
  \% Increase&
  Time \\
  &
  &
  methods&
  CUT &
  score &
  test &
  methods &
  mutants&
  killed&
  killed&
  mutant&
  killed only&
  killed type&
  killed type&
  (h:m:s)
  \\
  &
  &
  original &
  &
  &
  methods &
  &
  original &
  amplified&
  amplified&
  A-amp &
  A-amp &
  aided &
  aided
  \\
  \hline
  89  & DSUserTest                           & 14 & 62  & 56.52 & 1  & 1  & 13  & 4  & 30.77  & 1 & 7.69  & 0  & 0.00                                          & 0:13:53 \\
90  &                                      &    &     &       & 1  & 1  & 13  & 4  & 30.77  & 1 & 7.69  & 0  & 0.00                                          & 0:17:13 \\
91  &                                      &    &     &       & 1  & 1  & 13  & 4  & 30.77  & 1 & 7.69  & 0  & 0.00                                          & 0:14:45 \\
92  & \noProfiling                                     &    &     &       & 1  & 1  & 13  & 4  & 30.77  & 1 & 7.69  & 0  & 0.00                                          & 0:10:15 \\ \hline
93  & PhLAboutCommandTest                  & 1  & 111 & 75    & 0  & 0  & 3   & 0  & 0.00   & 0 & 0     & 0  & 0.00                                          & 0:00:01 \\
94  &                                      &    &     &       & 0  & 0  & 3   & 0  & 0.00   & 0 & 0     & 0  & 0.00                                          & 0:00:01 \\
95  &                                      &    &     &       & 0  & 0  & 3   & 0  & 0.00   & 0 & 0     & 0  & 0.00                                          & 0:00:01 \\
96  &  \noProfiling                                    &    &     &       & 0  & 0  & 3   & 0  & 0.00   & 0 & 0     & 0  & 0.00                                          & 0:00:01 \\ \hline
97  & PhLCopyImageCommandTest              & 1  & 25  & 66.67 & 0  & 0  & 4   & 0  & 0.00   & 0 & 0     & 0  & 0.00                                          & 0:00:02 \\
98  &                                      &    &     &       & 0  & 0  & 4   & 0  & 0.00   & 0 & 0     & 0  & 0.00                                          & 0:00:02 \\
99  &                                      &    &     &       & 0  & 0  & 4   & 0  & 0.00   & 0 & 0     & 0  & 0.00                                          & 0:00:01 \\
100 &  \noProfiling                                    &    &     &       & 0  & 0  & 4   & 0  & 0.00   & 0 & 0     & 0  & 0.00                                          & 0:00:02 \\ \hline
101 & WACookieTest                         & 16 & 34  & 50    & 1  & 1  & 33  & 4  & 12.12  & 1 & 3.03  & 0  & 0.00                                          & 0:02:33 \\
102 &                                      &    &     &       & 2  & 2  & 33  & 4  & 12.12  & 1 & 3.03  & 1  & 3.03                                          & 0:02:39 \\
103 &                                      &    &     &       & 2  & 2  & 33  & 4  & 12.12  & 1 & 3.03  & 1  & 3.03                                          & 0:02:30 \\
104 &  \noProfiling                                    &    &     &       & 1  & 1  & 33  & 4  & 12.12  & 1 & 3.03  & 0  & 0.00                                          & 0:01:45 \\ \hline \hline
105 & BlShortcutTest                       & 1  & 113 & 20    & 1  & 1  & 2   & 4  & 200.00 & 1 & 50    & 0  & 0.00                                          & 0:00:04 \\
106 &                                      &    &     &       & 1  & 1  & 2   & 4  & 200.00 & 1 & 50    & 0  & 0.00                                          & 0:00:04 \\
107 &                                      &    &     &       & 1  & 1  & 2   & 4  & 200.00 & 1 & 50    & 0  & 0.00                                          & 0:00:04 \\
108 &  \noProfiling                                    &    &     &       & 1  & 1  & 2   & 4  & 200.00 & 1 & 50    & 0  & 0.00                                          & 0:00:02 \\ \hline
109 & BlCompulsoryCombinationTest          & 4  & 271 & 11.76 & 1  & 1  & 4   & 2  & 50.00  & 0 & 0     & 1  & 25.00                                         & 0:00:34 \\
110 &                                      &    &     &       & 1  & 1  & 4   & 2  & 50.00  & 0 & 0     & 1  & 25.00                                         & 0:00:33 \\
111 &                                      &    &     &       & 1  & 1  & 4   & 2  & 50.00  & 0 & 0     & 1  & 25.00                                         & 0:00:36 \\
112 &  \noProfiling                                    &    &     &       & 0  & 0  & 4   & 0  & 0.00   & 0 & 0     & 0  & 0.00                                          & 0:00:23 \\ \hline
113 & BlLayoutExactResizerTest             & 8  & 36  & 40.54 & 3  & 3  & 15  & 3  & 20.00  & 1 & 6.67  & 2  & 13.33                                         & 0:01:08 \\
114 &                                      &    &     &       & 3  & 3  & 15  & 3  & 20.00  & 1 & 6.67  & 2  & 13.33                                         & 0:01:07 \\
115 &                                      &    &     &       & 3  & 3  & 15  & 3  & 20.00  & 1 & 6.67  & 2  & 13.33                                         & 0:01:06 \\
116 &  \noProfiling                                    &    &     &       & 2  & 2  & 15  & 2  & 13.33  & 2 & 13.33 & 0  & 0.00                                          & 0:00:54 \\ \hline
117 & BlInsetsTest                         & 10 & 30  & 44.14 & 14 & 12 & 64  & 49 & 76.56  & 2 & 3.12  & 3  & 4.69                                          & 0:03:10 \\
118 &                                      &    &     &       & 18 & 14 & 64  & 46 & 71.88  & 2 & 3.12  & 11 & 17.19                                         & 0:03:08 \\
119 &                                      &    &     &       & 18 & 18 & 64  & 50 & 78.13  & 2 & 3.12  & 4  & 6.25                                          & 0:03:13 \\
120 &  \noProfiling                                    &    &     &       & 19 & 19 & 64  & 50 & 78.13  & 2 & 3.12  & 0  & 0.00                                          & 0:02:01 \\ \hline
121 & ZnBivalentWriteStreamTest            & 2  & 39  & 44.44 & 0  & 0  & 8   & 0  & 0.00   & 0 & 0     & 0  & 0.00                                          & 0:00:04 \\
122 &                                      &    &     &       & 0  & 0  & 8   & 0  & 0.00   & 0 & 0     & 0  & 0.00                                          & 0:00:04 \\
123 &                                      &    &     &       & 0  & 0  & 8   & 0  & 0.00   & 0 & 0     & 0  & 0.00                                          & 0:00:04 \\
124 &  \noProfiling                                    &    &     &       & 0  & 0  & 8   & 0  & 0.00   & 0 & 0     & 0  & 0.00                                          & 0:00:04 \\ \hline
125 & ZnStatusLineTest                     & 7  & 220 & 28.57 & 1  & 1  & 16  & 1  & 6.25   & 0 & 0     & 1  & 6.25                                          & 0:00:41 \\
126 &                                      &    &     &       & 1  & 1  & 16  & 1  & 6.25   & 0 & 0     & 1  & 6.25                                          & 0:00:43 \\
127 &                                      &    &     &       & 1  & 1  & 16  & 1  & 6.25   & 0 & 0     & 1  & 6.25                                          & 0:00:44 \\
128 &  \noProfiling                                    &    &     &       & 0  & 0  & 16  & 0  & 0.00   & 0 & 0     & 0  & 0.00                                          & 0:00:39 \\ \hline

\end{tabular}
}}
\end{sidewaystable}

\begin{sidewaystable}
    \centering
\ContinuedFloat
\captionsetup{list=off,format=cont}
\caption{The result of test amplification by \smallamp on the 52 test classes. (Tests with low coverage)}
\resizebox{\textwidth}{!}{
\begin{tabular}{$c^W^c^c^c^c^c^c^c^c^c^c^c^c^c}
Id &
  Class &
  \# Test  &
  \# loc  &      
  \% Mut. &
  \# New&
  \# Focused &
  \# Killed&
  \# Newly&
  \% Increase&
  \# Newly&
  \% Increase&
  \# Newly&
  \% Increase&
  Time \\
  &
  &
  methods&
  CUT&
  score &
  test &
  methods &
  mutants&
  killed&
  killed&
  mutant&
  killed only&
  killed type&
  killed type&
  (h:m:s)
  \\
  &
  &
  original &
  &
  &
  methods &
  &
  original &
  amplified&
  amplified&
  A-amp &
  A-amp &
  aided &
  aided
  \\
  \hline
129 & GQLArgumentsTest                     & 16 & 83  & 47.06 & 0  & 0  & 8   & 0  & 0.00   & 0 & 0     & 0  & 0.00                                          & 0:31:10 \\
130 &                                      &    &     &       & 0  & 0  & 8   & 0  & 0.00   & 0 & 0     & 0  & 0.00                                          & 0:30:56 \\
131 &                                      &    &     &       & 0  & 0  & 8   & 0  & 0.00   & 0 & 0     & 0  & 0.00                                          & 0:32:28 \\
132 &  \noProfiling                                    &    &     &       & 0  & 0  & 8   & 0  & 0.00   & 0 & 0     & 0  & 0.00                                          & 0:27:17 \\   \hline  
133 & PP2ParsingGuardTest                  & 3  & 29  & 47.06 & 1  & 1  & 8   & 2  & 25.00  & 1 & 12.5  & 0  & 0.00                                          & 0:00:08 \\
134 &                                      &    &     &       & 1  & 1  & 8   & 2  & 25.00  & 1 & 12.5  & 0  & 0.00                                          & 0:00:07 \\
135 &                                      &    &     &       & 1  & 1  & 8   & 2  & 25.00  & 1 & 12.5  & 0  & 0.00                                          & 0:00:07 \\
136 &  \noProfiling                                    &    &     &       & 1  & 1  & 8   & 2  & 25.00  & 1 & 12.5  & 0  & 0.00                                          & 0:00:07 \\ \hline
137 & PP2BenchmarkTest                     & 3  & 24  & 15.91 & 1  & 1  & 7   & 6  & 85.71  & 1 & 14.29 & 0  & 0.00                                          & 0:09:15 \\
138 &                                      &    &     &       & 1  & 1  & 7   & 6  & 85.71  & 1 & 14.29 & 0  & 0.00                                          & 0:08:48 \\
139 &                                      &    &     &       & 1  & 1  & 7   & 6  & 85.71  & 1 & 14.29 & 0  & 0.00                                          & 0:08:31 \\
140 &  \noProfiling                                    &    &     &       & 1  & 1  & 7   & 6  & 85.71  & 1 & 14.29 & 0  & 0.00                                          & 0:08:51 \\ \hline
141 & MDLDialogTest                        & 4  & 103 & 14.29 & 0  & 0  & 1   & 0  & 0.00   & 0 & 0     & 0  & 0.00                                          & 0:00:00 \\
142 &                                      &    &     &       & 0  & 0  & 1   & 0  & 0.00   & 0 & 0     & 0  & 0.00                                          & 0:00:00 \\
143 &                                      &    &     &       & 0  & 0  & 1   & 0  & 0.00   & 0 & 0     & 0  & 0.00                                          & 0:00:00 \\
144 &  \noProfiling                                    &    &     &       & 0  & 0  & 1   & 0  & 0.00   & 0 & 0     & 0  & 0.00                                          & 0:00:00 \\ \hline
145 & MDLNestedListTest                    & 11 & 558 & 41.04 & 1  & 1  & 55  & 8  & 14.55  & 1 & 1.82  & 0  & 0.00                                          & 0:00:14 \\
146 &                                      &    &     &       & 1  & 1  & 55  & 8  & 14.55  & 1 & 1.82  & 0  & 0.00                                          & 0:00:15 \\
147 &                                      &    &     &       & 1  & 1  & 55  & 8  & 14.55  & 1 & 1.82  & 0  & 0.00                                          & 0:00:15 \\
148 &  \noProfiling                                    &    &     &       & 1  & 1  & 55  & 8  & 14.55  & 1 & 1.82  & 0  & 0.00                                          & 0:00:14 \\ \hline
149 & RSAthensRendererTest                 & 1  & 19  & 9.88  & 1  & 1  & 48  & 1  & 2.08   & 1 & 2.08  & 0  & 0.00                                          & 0:01:08 \\
150 &                                      &    &     &       & 1  & 1  & 48  & 1  & 2.08   & 1 & 2.08  & 0  & 0.00                                          & 0:01:10 \\
151 &                                      &    &     &       & 1  & 1  & 48  & 1  & 2.08   & 1 & 2.08  & 0  & 0.00                                          & 0:01:15 \\
152 &  \noProfiling                                    &    &     &       & 1  & 1  & 48  & 1  & 2.08   & 1 & 2.08  & 0  & 0.00                                          & 0:01:04 \\ \hline
153 & RSDraggableCanvasTest                & 6  & 84  & 35.71 & 0  & 0  & 10  & 0  & 0.00   & 0 & 0     & 0  & 0.00                                          & 0:00:34 \\
154 &                                      &    &     &       & 0  & 0  & 10  & 0  & 0.00   & 0 & 0     & 0  & 0.00                                          & 0:00:35 \\
155 &                                      &    &     &       & 0  & 0  & 10  & 0  & 0.00   & 0 & 0     & 0  & 0.00                                          & 0:00:36 \\
156 &  \noProfiling                                    &    &     &       & 0  & 0  & 10  & 0  & 0.00   & 0 & 0     & 0  & 0.00                                          & 0:00:32 \\ \hline
157 & OPNavigatorAdaptersTest              & 3  & 161 & 24    & 0  & 0  & 6   & 0  & 0.00   & 0 & 0     & 0  & 0.00                                          & 0:00:26 \\
158 &                                      &    &     &       & 0  & 0  & 6   & 0  & 0.00   & 0 & 0     & 0  & 0.00                                          & 0:00:24 \\
159 &                                      &    &     &       & 0  & 0  & 6   & 0  & 0.00   & 0 & 0     & 0  & 0.00                                          & 0:00:23 \\
160 &  \noProfiling                                    &    &     &       & 0  & 0  & 6   & 0  & 0.00   & 0 & 0     & 0  & 0.00                                          & 0:00:26 \\ \hline
161 & OPProjectTest                        & 2  & 40  & 28    & 0  & 0  & 7   & 0  & 0.00   & 0 & 0     & 0  & 0.00                                          & 0:00:03 \\
162 &                                      &    &     &       & 0  & 0  & 7   & 0  & 0.00   & 0 & 0     & 0  & 0.00                                          & 0:00:03 \\
163 &                                      &    &     &       & 0  & 0  & 7   & 0  & 0.00   & 0 & 0     & 0  & 0.00                                          & 0:00:03 \\
164 &  \noProfiling                                    &    &     &       & 0  & 0  & 7   & 0  & 0.00   & 0 & 0     & 0  & 0.00                                          & 0:00:04 \\ \hline
165 & PMExponentialDistributionTest        & 1  & 25  & 17.07 & 3  & 3  & 7   & 3  & 42.86  & 0 & 0     & 2  & 28.57                                         & 0:00:23 \\
166 &                                      &    &     &       & 2  & 2  & 7   & 2  & 28.57  & 0 & 0     & 1  & 14.29                                         & 0:00:15 \\
167 &                                      &    &     &       & 3  & 3  & 7   & 3  & 42.86  & 0 & 0     & 2  & 28.57                                         & 0:00:11 \\
168 &  \noProfiling                                    &    &     &       & 1  & 1  & 7   & 1  & 14.29  & 0 & 0     & 0  & 0.00                                          & 0:00:05 \\ \hline
169 & PMFixpointTest                       & 6  & 18  & 39.47 & 5  & 5  & 30  & 12 & 40.00  & 3 & 10    & 0  & 0.00                                          & 0:03:01 \\
170 &                                      &    &     &       & 4  & 4  & 30  & 9  & 30.00  & 3 & 10    & 0  & 0.00                                          & 0:03:03 \\
171 &                                      &    &     &       & 5  & 5  & 30  & 12 & 40.00  & 3 & 10    & 0  & 0.00                                          & 0:04:35 \\
172 &  \noProfiling                                    &    &     &       & 7  & 7  & 30  & 8  & 26.67  & 3 & 10    & 0  & 0.00                                          & 0:02:11 \\ \hline

\end{tabular}
}
\end{sidewaystable}

\begin{sidewaystable}
    \centering
\ContinuedFloat
\captionsetup{list=off,format=cont}
\caption{The result of test amplification by \smallamp on the 52 test classes. (Tests with low coverage)}
\resizebox{\textwidth}{!}{
\begin{tabular}{$c^W^c^c^c^c^c^c^c^c^c^c^c^c^c}
Id &
  Class &
  \# Test  &
  \# loc  &      
  \% Mut. &
  \# New&
  \# Focused &
  \# Killed&
  \# Newly&
  \% Increase&
  \# Newly&
  \% Increase&
  \# Newly&
  \% Increase&
  Time \\
  &
  &
  methods&
  CUT&
  score &
  test &
  methods &
  mutants&
  killed&
  killed&
  mutant&
  killed only&
  killed type&
  killed type&
  (h:m:s)
  \\
  &
  &
  original &
  &
  &
  methods &
  &
  original &
  amplified&
  amplified&
  A-amp &
  A-amp &
  aided &
  aided
  \\
  \hline
173 & TLDistributionMapTest                & 1  & 168 & 37.04 & 3  & 3  & 10  & 4  & 40.00  & 1 & 10    & 2  & 20.00                                         & 0:00:26 \\
174 &                                      &    &     &       & 3  & 3  & 10  & 4  & 40.00  & 1 & 10    & 2  & 20.00                                         & 0:00:26 \\
175 &                                      &    &     &       & 2  & 2  & 10  & 3  & 30.00  & 1 & 10    & 1  & 10.00                                         & 0:00:28 \\
176 &  \noProfiling                                    &    &     &       & 1  & 1  & 10  & 2  & 20.00  & 1 & 10    & 0  & 0.00                                          & 0:00:14 \\ 

\hline  
177 & TLLegendTest                         & 2  & 96  & 0     & 0  & 0  & 0   & 0  & -      & 0 & -     & 0  & - & 0:00:02 \\
178 &                                      &    &     &       & 0  & 0  & 0   & 0  & -      & 0 & -     & 0  & - & 0:00:02 \\
179 &                                      &    &     &       & 0  & 0  & 0   & 0  & -      & 0 & -     & 0  & - & 0:00:02 \\
180 &  \noProfiling                                    &    &     &       & 0  & 0  & 0   & 0  & -      & 0 & -     & 0  & - & 0:00:02 \\ \hline
181 & DSSendUserTextMessageItemTest        & 5  & 62  & 33.33 & 2  & 2  & 4   & 3  & 75.00  & 1 & 25    & 0  & 0.00                                          & 0:00:34 \\
182 &                                      &    &     &       & 2  & 2  & 4   & 3  & 75.00  & 1 & 25    & 0  & 0.00                                          & 0:00:36 \\
183 &                                      &    &     &       & 2  & 2  & 4   & 3  & 75.00  & 1 & 25    & 0  & 0.00                                          & 0:00:34 \\
184 &   \noProfiling                                   &    &     &       & 2  & 2  & 4   & 3  & 75.00  & 2 & 50    & 0  & 0.00                                          & 0:00:21 \\ \hline
185 & DSEmbedTest                          & 11 & 33  & 19.67 & 7  & 7  & 12  & 20 & 166.67 & 1 & 8.33  & 6  & 50.00                                         & 0:01:58 \\
186 &                                      &    &     &       & 7  & 7  & 12  & 20 & 166.67 & 1 & 8.33  & 6  & 50.00                                         & 0:02:02 \\
187 &                                      &    &     &       & 8  & 8  & 12  & 20 & 166.67 & 1 & 8.33  & 6  & 50.00                                         & 0:02:16 \\
188 &  \noProfiling                                    &    &     &       & 2  & 2  & 12  & 12 & 100.00 & 2 & 16.67 & 0  & 0.00                                          & 0:01:15 \\ \hline
189 & PhLLocalTemplateTest                 & 11 & 81  & 12    & 1  & 1  & 3   & 2  & 66.67  & 1 & 33.33 & 0  & 0.00                                          & 0:00:33 \\
190 &                                      &    &     &       & 1  & 1  & 3   & 2  & 66.67  & 1 & 33.33 & 0  & 0.00                                          & 0:00:32 \\
191 &                                      &    &     &       & 1  & 1  & 3   & 2  & 66.67  & 1 & 33.33 & 0  & 0.00                                          & 0:00:33 \\
192 &   \noProfiling                                   &    &     &       & 1  & 1  & 3   & 2  & 66.67  & 1 & 33.33 & 0  & 0.00                                          & 0:00:39 \\ \hline
193 & PhLDirectoryBasedImageRepositoryTest & 14 & 156 & 39.8  & 1  & 1  & 39  & 3  & 7.69   & 1 & 2.56  & 0  & 0.00                                          & 0:01:21 \\
194 &                                      &    &     &       & 1  & 1  & 39  & 3  & 7.69   & 1 & 2.56  & 0  & 0.00                                          & 0:01:17 \\
195 &                                      &    &     &       & 1  & 1  & 39  & 3  & 7.69   & 1 & 2.56  & 0  & 0.00                                          & 0:01:19 \\
196 &   \noProfiling                                   &    &     &       & 1  & 1  & 39  & 3  & 7.69   & 1 & 2.56  & 0  & 0.00                                          & 0:01:30 \\ \hline
197 & WAKeyGeneratorTest                   & 1  & 51  & 33.33 & 0  & 0  & 1   & 0  & 0.00   & 0 & 0     & 0  & 0.00                                          & 0:00:12 \\
198 &                                      &    &     &       & 0  & 0  & 1   & 0  & 0.00   & 0 & 0     & 0  & 0.00                                          & 0:00:09 \\
199 &                                      &    &     &       & 0  & 0  & 1   & 0  & 0.00   & 0 & 0     & 0  & 0.00                                          & 0:00:09 \\
200 &    \noProfiling                                  &    &     &       & 0  & 0  & 1   & 0  & 0.00   & 0 & 0     & 0  & 0.00                                          & 0:00:10 \\ \hline
201 & WAXmlCanvasTest                      & 1  & 142 & 33.33 & 0  & 0  & 1   & 0  & 0.00   & 0 & 0     & 0  & 0.00                                          & 0:00:00 \\
202 &                                      &    &     &       & 0  & 0  & 1   & 0  & 0.00   & 0 & 0     & 0  & 0.00                                          & 0:00:00 \\
203 &                                      &    &     &       & 0  & 0  & 1   & 0  & 0.00   & 0 & 0     & 0  & 0.00                                          & 0:00:01 \\
204 &   \noProfiling                                   &    &     &       & 0  & 0  & 1   & 0  & 0.00   & 0 & 0     & 0  & 0.00                                          & 0:00:01 \\ \hline
205 & WAErrorHandlerTest                   & 11 & 182 & 38.46 & 0  & 0  & 5   & 0  & 0.00   & 0 & 0     & 0  & 0.00                                          & 0:01:15 \\
206 &                                      &    &     &       & 0  & 0  & 5   & 0  & 0.00   & 0 & 0     & 0  & 0.00                                          & 0:01:13 \\
207 &                                      &    &     &       & 0  & 0  & 5   & 0  & 0.00   & 0 & 0     & 0  & 0.00                                          & 0:01:17 \\
208 &  \noProfiling                                    &    &     &       & 0  & 0  & 5   & 0  & 0.00   & 0 & 0     & 0  & 0.00                                          & 0:01:19 \\ 
\hline
\end{tabular}
}
\end{sidewaystable}
 
\begin{compactitem}
\item \textit{Id}:~Used as a reference for the row in the table.
\item \textit{Class}:~The name of the test class to be amplified.
\item \textit{\# Test methods original}:~The number of test method in the test class before test amplification.
\item \textit{\# loc CUT}:~The number of lines in the class under test.
\item \textit{\% Mut. score}:~The mutation score (percentage) of the test class before test amplification.
\item \textit{\# New test methods}:~The absolute number of newly generated test methods after test amplification.
\item \textit{\# Focused methods}:~The absolute number of focused methods in the generated test methods.
\item \textit{\# Killed mutants original}:~The absolute number of killed mutants by the test class before test amplification.
\item \textit{\# Newly killed amplified}:~The absolute number of newly killed mutants by the test class after test amplification.
\item \textit{\% Increase killed amplified}:~The increase (in percentage) of killed mutants by the test class after test amplification.
\item \textit{\# Newly mutant A-amp}:~The absolute number of newly killed mutants only by Assertion amplification ($N_{iteration} = 0$ in Algorithm \ref{alg:main}).
\item \textit{\% Increase killed only A-amp}:~The relative increase (in percentage) of killed mutants only by assertion amplification.
\item \textit{\# Newly killed type aided}:~The absolute number of newly killed mutants by type sensitive input amplifiers.
\item \textit{\% Increase killed type aided}:~The relative increase (in percentage) of increase killed mutants by type sensitive input amplifiers.
\item \textit{Time}:~The duration of test amplification process in the hours-minutes-seconds (h:m:s) format.
\end{compactitem}

\textbf{RQ2: \rqTwo}

For answering this research question, we use values in the column  \textit{\# Focused methods}.
We use the same definition for focused methods as \dspot:

\begin{quote}
\textit{Focus is defined as where at least 50\% of the newly killed mutants are located in a single method.} \citep{Danglot2019EMSE}
\end{quote}

Generating focused tests is important because analysing a focused test is easier (most mutants reside in the same method under test) hence should take less review time from developers.
For calculating this value, we use generated annotations by \smallamp on the newly generated test methods which show the details of the killed mutants by the method.

We see that almost all amplified tests are focussed.
Only on two cases (\inlinecode{BlInsetsTest}, \#117 and \#118) we see that some generated methods are not focused.

\hypobox{
\textbf{Answer to RQ2: }
We see at least one focused test method in all amplified cases.
This illustrates that amplified tests seldom overwhelm developers, hence should save valuable developer time.
}

\subsection{RQ3 --- \rqThreeKey}\label{sec:rq-mutation-coverage}

\textbf{RQ3: \rqThree}

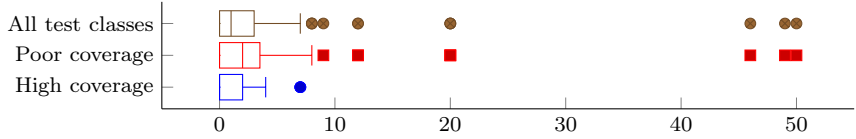
\begin{figure}
\caption{The distributions of the number of killed mutants}
  \label{fig:box-number}
  
 \begin{tikzpicture}
  \begin{axis}
    [
    height=3cm, 
    width=0.9\textwidth,
    axis y line*=left,
    axis x line*=bottom,
    ytick={1,2,3},
    yticklabels={High coverage, Poor coverage, All test classes},
    ]
    \addplot+[boxplot]
    	table [row sep=\\, y index=0]{
	data\\
	2	\\
2	\\
2	\\
0	\\
0	\\
0	\\
4	\\
4	\\
4	\\
0	\\
0	\\
0	\\
0	\\
0	\\
0	\\
0	\\
0	\\
0	\\
0	\\
0	\\
0	\\
2	\\
2	\\
2	\\
7	\\
7	\\
7	\\
2	\\
2	\\
2	\\
0	\\
0	\\
0	\\
0	\\
0	\\
0	\\
0	\\
0	\\
0	\\
1	\\
1	\\
1	\\
0	\\
0	\\
0	\\
0	\\
0	\\
0	\\
1	\\
1	\\
1	\\
3	\\
2	\\
2	\\
1	\\
1	\\
1	\\
0	\\
3	\\
2	\\
2	\\
2	\\
2	\\
0	\\
0	\\
0	\\
4	\\
4	\\
4	\\
0	\\
0	\\
0	\\
0	\\
0	\\
0	\\
4	\\
4	\\
4	\\
	};
    \addplot+[boxplot]
    	table [row sep=\\, y index=0]{
	data\\
	4	\\
4	\\
4	\\
2	\\
2	\\
2	\\
3	\\
3	\\
3	\\
49	\\
46	\\
50	\\
0	\\
0	\\
0	\\
1	\\
1	\\
1	\\
0	\\
0	\\
0	\\
2	\\
2	\\
2	\\
6	\\
6	\\
6	\\
0	\\
0	\\
0	\\
8	\\
8	\\
8	\\
1	\\
1	\\
1	\\
0	\\
0	\\
0	\\
0	\\
0	\\
0	\\
0	\\
0	\\
0	\\
3	\\
2	\\
3	\\
12	\\
9	\\
12	\\
4	\\
4	\\
3	\\
0	\\
0	\\
0	\\
3	\\
3	\\
3	\\
20	\\
20	\\
20	\\
2	\\
2	\\
2	\\
3	\\
3	\\
3	\\
0	\\
0	\\
0	\\
0	\\
0	\\
0	\\
0	\\
0	\\
0	\\
	};
    \addplot+[boxplot]
    	table [row sep=\\, y index=0]{
	data\\
	2	\\
2	\\
2	\\
0	\\
0	\\
0	\\
4	\\
4	\\
4	\\
0	\\
0	\\
0	\\
0	\\
0	\\
0	\\
0	\\
0	\\
0	\\
0	\\
0	\\
0	\\
2	\\
2	\\
2	\\
7	\\
7	\\
7	\\
2	\\
2	\\
2	\\
0	\\
0	\\
0	\\
0	\\
0	\\
0	\\
0	\\
0	\\
0	\\
1	\\
1	\\
1	\\
0	\\
0	\\
0	\\
0	\\
0	\\
0	\\
1	\\
1	\\
1	\\
3	\\
2	\\
2	\\
1	\\
1	\\
1	\\
0	\\
3	\\
2	\\
2	\\
2	\\
2	\\
0	\\
0	\\
0	\\
4	\\
4	\\
4	\\
0	\\
0	\\
0	\\
0	\\
0	\\
0	\\
4	\\
4	\\
4	\\
4	\\
4	\\
4	\\
2	\\
2	\\
2	\\
3	\\
3	\\
3	\\
49	\\
46	\\
50	\\
0	\\
0	\\
0	\\
1	\\
1	\\
1	\\
0	\\
0	\\
0	\\
2	\\
2	\\
2	\\
6	\\
6	\\
6	\\
0	\\
0	\\
0	\\
8	\\
8	\\
8	\\
1	\\
1	\\
1	\\
0	\\
0	\\
0	\\
0	\\
0	\\
0	\\
0	\\
0	\\
0	\\
3	\\
2	\\
3	\\
12	\\
9	\\
12	\\
4	\\
4	\\
3	\\
0	\\
0	\\
0	\\
3	\\
3	\\
3	\\
20	\\
20	\\
20	\\
2	\\
2	\\
2	\\
3	\\
3	\\
3	\\
0	\\
0	\\
0	\\
0	\\
0	\\
0	\\
0	\\
0	\\
0	\\
	};
  \end{axis}
  \end{tikzpicture}
\end{figure}

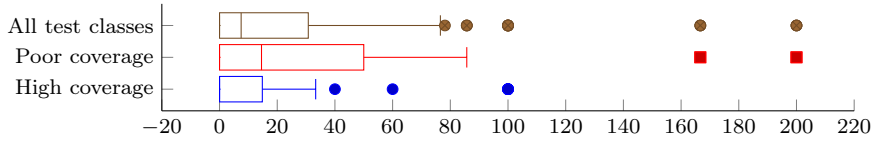
\begin{figure}
\caption{The distributions of the increase kills}
  \label{fig:box-inckill}
\begin{tikzpicture}
  \begin{axis}
    [
    height=3cm, 
    width=0.9\textwidth,
    axis y line*=left,
    axis x line*=bottom,
    ytick={1,2,3},
    yticklabels={High coverage, Poor coverage, All test classes},
    ]
    \addplot+[boxplot]
    	table [row sep=\\, y index=0]{
	data\\
	33.33	\\
33.33	\\
33.33	\\
0.00	\\
0.00	\\
0.00	\\
25.00	\\
25.00	\\
25.00	\\
0.00	\\
0.00	\\
0.00	\\
0.00	\\
0.00	\\
0.00	\\
0.00	\\
0.00	\\
0.00	\\
0.00	\\
0.00	\\
0.00	\\
8.70	\\
8.70	\\
8.70	\\
7.29	\\
7.29	\\
7.29	\\
11.11	\\
11.11	\\
11.11	\\
0.00	\\
0.00	\\
0.00	\\
0.00	\\
0.00	\\
0.00	\\
0.00	\\
0.00	\\
0.00	\\
100.00	\\
100.00	\\
100.00	\\
0.00	\\
0.00	\\
0.00	\\
0.00	\\
0.00	\\
0.00	\\
100.00	\\
100.00	\\
100.00	\\
23.08	\\
15.38	\\
15.38	\\
10.00	\\
10.00	\\
10.00	\\
0.00	\\
60.00	\\
40.00	\\
14.29	\\
14.29	\\
14.29	\\
0.00	\\
0.00	\\
0.00	\\
30.77	\\
30.77	\\
30.77	\\
0.00	\\
0.00	\\
0.00	\\
0.00	\\
0.00	\\
0.00	\\
12.12	\\
12.12	\\
12.12	\\
	};
    \addplot+[boxplot]
    	table [row sep=\\, y index=0]{
	data\\
	200.00	\\
200.00	\\
200.00	\\
50.00	\\
50.00	\\
50.00	\\
20.00	\\
20.00	\\
20.00	\\
76.56	\\
71.88	\\
78.13	\\
0.00	\\
0.00	\\
0.00	\\
6.25	\\
6.25	\\
6.25	\\
0.00	\\
0.00	\\
0.00	\\
25.00	\\
25.00	\\
25.00	\\
85.71	\\
85.71	\\
85.71	\\
0.00	\\
0.00	\\
0.00	\\
14.55	\\
14.55	\\
14.55	\\
2.08	\\
2.08	\\
2.08	\\
0.00	\\
0.00	\\
0.00	\\
0.00	\\
0.00	\\
0.00	\\
0.00	\\
0.00	\\
0.00	\\
42.86	\\
28.57	\\
42.86	\\
40.00	\\
30.00	\\
40.00	\\
40.00	\\
40.00	\\
30.00	\\
75.00	\\
75.00	\\
75.00	\\
166.67	\\
166.67	\\
166.67	\\
66.67	\\
66.67	\\
66.67	\\
7.69	\\
7.69	\\
7.69	\\
0.00	\\
0.00	\\
0.00	\\
0.00	\\
0.00	\\
0.00	\\
0.00	\\
0.00	\\
0.00	\\
	};
    \addplot+[boxplot]
    	table [row sep=\\, y index=0]{
	data\\
	33.33	\\
33.33	\\
33.33	\\
0.00	\\
0.00	\\
0.00	\\
25.00	\\
25.00	\\
25.00	\\
0.00	\\
0.00	\\
0.00	\\
0.00	\\
0.00	\\
0.00	\\
0.00	\\
0.00	\\
0.00	\\
0.00	\\
0.00	\\
0.00	\\
8.70	\\
8.70	\\
8.70	\\
7.29	\\
7.29	\\
7.29	\\
11.11	\\
11.11	\\
11.11	\\
0.00	\\
0.00	\\
0.00	\\
0.00	\\
0.00	\\
0.00	\\
0.00	\\
0.00	\\
0.00	\\
100.00	\\
100.00	\\
100.00	\\
0.00	\\
0.00	\\
0.00	\\
0.00	\\
0.00	\\
0.00	\\
100.00	\\
100.00	\\
100.00	\\
23.08	\\
15.38	\\
15.38	\\
10.00	\\
10.00	\\
10.00	\\
0.00	\\
60.00	\\
40.00	\\
14.29	\\
14.29	\\
14.29	\\
0.00	\\
0.00	\\
0.00	\\
30.77	\\
30.77	\\
30.77	\\
0.00	\\
0.00	\\
0.00	\\
0.00	\\
0.00	\\
0.00	\\
12.12	\\
12.12	\\
12.12	\\
200.00	\\
200.00	\\
200.00	\\
50.00	\\
50.00	\\
50.00	\\
20.00	\\
20.00	\\
20.00	\\
76.56	\\
71.88	\\
78.13	\\
0.00	\\
0.00	\\
0.00	\\
6.25	\\
6.25	\\
6.25	\\
0.00	\\
0.00	\\
0.00	\\
25.00	\\
25.00	\\
25.00	\\
85.71	\\
85.71	\\
85.71	\\
0.00	\\
0.00	\\
0.00	\\
14.55	\\
14.55	\\
14.55	\\
2.08	\\
2.08	\\
2.08	\\
0.00	\\
0.00	\\
0.00	\\
0.00	\\
0.00	\\
0.00	\\
0.00	\\
0.00	\\
0.00	\\
42.86	\\
28.57	\\
42.86	\\
40.00	\\
30.00	\\
40.00	\\
40.00	\\
40.00	\\
30.00	\\
75.00	\\
75.00	\\
75.00	\\
166.67	\\
166.67	\\
166.67	\\
66.67	\\
66.67	\\
66.67	\\
7.69	\\
7.69	\\
7.69	\\
0.00	\\
0.00	\\
0.00	\\
0.00	\\
0.00	\\
0.00	\\
0.00	\\
0.00	\\
0.00	\\
	};
  \end{axis}
  \end{tikzpicture}

\end{figure}

In 86/156 cases (55.12\%), \smallamp successfully amplified an existing test class.
The distribution of the number of killed mutants, and increase kill are presented in \figref{fig:box-number} and \figref{fig:box-inckill} for all test classes, highly covered as well as poorly covered ones.
The number of newly killed mutants in these classes (column \textit{\# Newly killed amplified}) varies from 1 up to 50 mutants (case \#119).
In the executions amplifying test classes having high coverage, \smallamp is able to amplify 38 out of 78 (48.7\%), and for the test classes having poor coverage this number is 48 out of 78 (61.5\%).
Therefore, we see more amplification in the classes with poor coverage.
The relative increase in mutation score (column \textit{\% Increase killed amplified}) varies from 2.08\% (cases 149-151) up to 200\% (cases 105-107).
It is also observable regarding to these metrics that amplification on classes with poor coverage is more successful.

Surprisingly, despite running \toolname{MuTalk} with its all mutation operators, the mutation testing framework did not manage to create any mutant for the class \inlinecode{TLLegendTest} (cases 177-179).
\toolname{MuTalk} mutation operators work statically and only a limited set of well-known transformations are provided in the tool.

\paragraph{The effect of randomness.}
In this section we report the effect of randomness on the results.
Based on the \algoref{alg:main}, the main randomness happens during the input-amplification (Line \ref{lst:line:inputamp}) and oracle reduction (Line \ref{lst:line:reduce}) steps.
Therefore, we expect to see the minimum difference in the results generated by assertion amplification (Line \ref{lst:line:assertamp1}).
The column 11 (\textit{\# Newly mutant A-amp}) shows the absolute number of killed mutants only by assertion amplification.
These values are identical in executions for all classes except the case 81 (\inlinecode{TLExpandCollapseNodesActionTest}).
The reason for this exception is that a specific mutant may be killed by input amplification in a test method, and if it is not killed, it will be killed by assertion amplification in the next test method.
Based on the information presented in \tabref{tab:results}, regardless of time, the same result are achieved from different executions in 43 out of 52 test classes. 

\addressedRvw{
For a deeper investigation, we randomly select 5 test classes from the cases that are affected by randomness, and 5 test classes from the cases without an observable change.
Then, we run \smallamp on these classes for 10 times (10 class $\times$ 10 times = 100 runs).
\tabref{tab:randomness} shows the results of this experiment.
In the column with title \textit{X10}, we report the number of newly killed mutants for each run in order, which is the most important metric for amplification.
In the column \textit{X3}, we also echo the values from column \textit{\# Newly killed amplified} in \tabref{tab:results} to make the comparison easier.
The next two columns compare the \textit{Median} and \textit{Average} values in these two columns.
The first 5 rows in this table are the cases affected by randomness, and the next 5 rows are cases without an observable effect.
}

\addressedRvw{
When we compare the values in columns \textit{X10} with \textit{X3}, we still do not see any visible effect from randomness in the rows 6 to 10.
In the first 5 rows, we see the median values and average values in both experiments are similar.
In three cases (rows 2, 3 and 4), the values of \textit{X3} did not achieve the maximum number of killed mutants seen in \textit{X10}. 
In one case (row 1), we see some runs lacking any improvements in \textit{X10} while all of its runs in \textit{X3} shows a successful amplification.
To sum up, we see that the randomness has an effect on the result, but the impact is minimal and does not invalidate the findings.
In addition, repeating the analysis 3 times is justifiable since running 10 times adds little extra information for a large increase in processing time.
}

\begin{table}[]
\begin{small}
\caption{The result of running \smallamp on 10 test class for 10 times}
\label{tab:randomness}
\begin{tabular}{p{0.1cm}Wp{1.7cm}p{0.9cm}p{1.4cm}p{1.4cm}}\hline
  &
  Test class &
  X10&
  X3&
  Median \tiny{(X10, X3)}&
  Average  \tiny{(X10, X3) }
\\
\hline
1 & PMBernoulliGeneratorTest & 3, 3, \underline{0}, \underline{0}, 2, 2, 2, 3, 2, 3 & 3, 2, 2 & 2.0, 2.0 & \textbf{2.0, 2.33}\\
2 & TLHideActionTest & 0, \underline{4}, 1, 2, 2, \underline{4}, 2, 1, 2, 2 & 0, 3, 2 & 2.0, 2.0 & \textbf{2.0, 1.67}\\
3 & TLExpandCollapseNodesActionTest & 2, 2, 2, 2, 2, 2, \underline{3}, 2, 2, 2 & 2, 2, 2 & 2.0, 2.0 & \textbf{2.1, 2.0}\\
4 & PMExponentialDistributionTest & 3, 2, 1, \underline{4}, 3, 2, 2, 3, 3, 2 & 3, 2, 3 & \textbf{2.5, 3.0} & \textbf{2.5, 2.67}\\
5 & TLDistributionMapTest & 3, 4, 3, 4, 4, 4, 4, 3, 3, 4 & 4, 4, 3 & 4.0, 4.0 & \textbf{3.6, 3.67}\\
\hline
6 & PMBinomialGeneratorTest & 1, 1, 1, 1, 1, 1, 1, 1, 1, 1 & 1, 1, 1 & 1.0, 1.0 & 1.0, 1.0\\
7 & DSSendUserTextMessageItemTest & 3, 3, 3, 3, 3, 3, 3, 3, 3, 3 & 3, 3, 3 & 3.0, 3.0 & 3.0, 3.0\\
8 & GQLSchemaGrammarTest & 7, 7, 7, 7, 7, 7, 7, 7, 7, 7 & 7, 7, 7 & 7.0, 7.0 & 7.0, 7.0\\
9 & BlCompulsoryCombinationTest & 2, 2, 2, 2, 2, 2, 2, 2, 2, 2 & 2, 2, 2 & 2.0, 2.0 & 2.0, 2.0\\
10 & ZnMessageBenchmarkTest & 0, 0, 0, 0, 0, 0, 0, 0, 0, 0 & 0, 0, 0 & 0.0, 0.0 & 0.0, 0.0\\

\hline
\end{tabular}
\end{small}
\end{table}

\hypobox{
\textbf{Answer to RQ3: }
\smallamp successfully amplified 86 test classes out of 156 cases (55.12\%).
Even for stronger test classes, \smallamp  improved the mutation coverage in 38 out of 78 cases.
In test classes with poor coverage, test amplification becomes even more effective: \smallamp increased the mutation coverage in 48 out of 78 cases and the absolute and relative increase in mutation score was higher.
}

\subsection{RQ4 --- \rqFourKey}
\label{sec:rqfout}

\textbf{RQ4: \rqFour}

As we reported in \secref{sec:rq-mutation-coverage}, in 86/156 cases (55.12\%), the improvements are achieved from input-amplification and assertion-amplification cooperation.
In this research question, we study the results generated only by assertion-amplification, and also generated by the type sensitive operators.

\paragraph{Contribution of the assertion-amplification step.}
In this section, we filter all amplified test methods that are generated only by assertion amplification.
In other words, we only account the improvements from all amplified test methods that are selected from the first assertion amplification (Line \ref{lst:line:assertamp1} in \algoref{alg:main}).
The column 11 (\textit{\# Newly mutant A-amp}) shows the absolute number of killed mutants only by assertion amplification; column 12 (\textit{\% Increase killed only A-amp}) also shows the relative increase.
The reported results in \tabref{tab:results} shows that in 61/156 cases (39.1\%) at least 1 mutant is killed only by the assertion amplification step.
Improvements in four classes (cases 53-55; 65-67; 73-75; 149-151) achieved only by adding new assertions.

\paragraph{Contribution of the type sensitive input amplifiers.}
Here, we filter all amplified test methods that in at least one of its transformations, a type sensitive input amplifier (in our case \textit{method-call-adder}) is used.
While type-sensitive operators benefit the information provided by dynamic profiler step (\secref{sec:profiling}), the contribution of these operators is important because it can show the effectiveness of dynamic profiling.

Column 13 (\textit{\# Newly killed type aided}) shows the absolute number of newly killed mutants by the type sensitive input amplifiers.
Column 14 (\textit{\% Increase killed type aided}) also shows the relative increase.
We see that in 30/156 cases (19.2\%) the type sensitive input amplifiers contribute to the result.
Especially for 2 test classes (\inlinecode{WebGrammarTest} rows 37-39, and \inlinecode{ZnStatusLineTest} rows 125-127), \smallamp was able to amplify the tests only by the type sensitive input operators.

\addressedRvw{To assess the impact of type profiling, we quantified the effect of the steps that rely on type profiling.
We therefore extended the analysis with an additional evaluation step where we disabled the type profiler in the algorithm, as well as the type sensitive input amplifier (method addition amplifier) and ran the tool on all test classes.
The results for this experiment are mentioned in the forth row for each test class in \tabref{tab:results} (denoted by \noProfilingText).
We focus on cases in which type-sensitive input amplifiers improved the coverage in at least two of three runs (10 test classes, cases starting with 33, 37, 101, 109, 113, 117, 125, 165, 173, 185).
In 8 cases we see that disabling the profiling and also the type sensitive operators decrease the improvements and only in two cases we see no difference (case 101) or a slight improvements (case 117).
 }

\hypobox{
\textbf{Answer to RQ4: }
Our experiments demonstrate that amplifying the tests only using assertion amplification is less efficient than in combination with input amplification.
Moreover, the extra information generated by dynamic type profiling helps input amplification in killing more mutants.
}

\subsection{RQ5 --- \rqFiveKey}

\textbf{RQ5: \rqFive}

With this research question, we compare our results from quantitative and qualitative studies with corresponding results from \dspot reported in \cite{Danglot2019EMSE}.

\begin{table*}[!h]
\centering
\begin{tabular}{|c|l|ccc|c|}
\hline
\textbf{Id} & \textbf{Metric} & & \textbf{\smallamp} & & \textbf{\dspot}\\
 &  & \textbf{\#1} & \textbf{\#2} & \textbf{\#3} & \\
\hline
1 & Projects & 13 & 13 & 13 & 10 \\
2 & Test classes & 52 & 52 & 52 & 40 \\
3 & Test methods & 403 & 403 & 403 & 220\\
4\rowhighlightOne & New generated test methods & 71 & 76 & 75 & 471 \\
5 & \# Amplified test classes & 28 & 29 & 29 & 26 \\
6 & \% Amplified test classes & 53\% & 55\% & 55\% & 65\% \\
7\rowhighlightOne & \# Total killed mutants by original & 1102 & 1102 & 1102 & 7980 \\
8\rowhighlightOne & \# Total newly killed mutants & 156 & 151 & 157 & 1617 \\
9 & \% Total increase killed & 14.15\%& 13.70\% & 14.24\% & 20.26\% \\

\hline
\end{tabular}
\caption{Comparing results in \smallamp and \dspot}
\label{table:compare}
\end{table*}
 
\tabref{table:compare} shows how the results from \smallamp and \dspot are comparable.
\smallamp is validated against 52 test classes. 
It has successfully amplified 28, 29 and 29 of them ($\approx$55\%), while \dspot has been validated against 40 test classes of which 26 cases were improved (65\%).
The most notable differences between the results from \smallamp and \dspot are the number of mutants in two ecosystems and consequently the number of newly generated test methods (denoted by \rowhighlightOne in the \tabref{table:compare}).
These differences can be attributed to the use of two different mutation testing frameworks in two different languages. 
\smallamp uses \toolname{MuTalk} which has notably fewer mutation operators than the \dspot counterpart \toolname{PITest}.
To reduce the effect of the mutation testing framework, we calculate the relative increase in killed mutants within the two ecosystem as follows:
\[ \%Total.Inc.killed = 100 \times \frac{\#Total.Mutants.killed_{new}}{\#Total.Mutants.killed_{original}} \]

This value is shown in the row number 12 in the \tabref{table:compare} for two experiments.
It is 14.03\%  in total for \smallamp, and 20.26\% for \dspot.

We have also submitted \nSentPR pull-requests by using \smallamp outputs and \nMergedPR of them were merged by developers (\percPR), while Danglot \etal ~ submitted 19 pull requests derived from \dspot output and 13 of them merged successfully (68\%).

Finally, it is worth mentioning that \smallamp is configured as $N_{maxInput} = 10$ which means the reduce step (\algoref{alg:main}, line \ref{lst:line:reduce}) select 10 test-input in each iteration.
This value for \dspot is not reported in their paper.
Increasing this hyperparameter may improve the result, but it also may increase the execution time significantly.

\hypobox{
\textbf{Answer to RQ5: }
The results from \smallamp and \dspot in two different ecosystems are comparable.
\smallamp and \dspot have been successful in amplifying respectively 55\% and 65\% of their input test classes.
We also see \percPR and 68\% merged pull requests in the tests derived from \smallamp and \dspot outputs.
}

\subsection{RQ6 --- \rqSixKey}\label{sec:rq-time}

\textbf{RQ6: \rqSix}

Time costs is an important factor when we study the practicality of test amplification tools.
In this research question, we report and compare the execution time of \smallamp and the relative cost of each step.
\figref{fig:box-time-seconds} and \figref{fig:box-time-percentage} illustrate a series of box-plots derived from the recorded execution time during the experiments in \tabref{tab:results}.
In these figures, \texttt{Init.} refers to all initializing steps, includes the dynamic profiling to collect type information (\secref{sec:profiling}).
Here we also calculate an initial mutation score for the original test suite.
\texttt{IAmp} refers to the input amplification step.
This step loops over all input amplification operators (\secref{sec:input-amp}) and afterwards reduces to N\textsubscript{maxInputs} of current inputs and discarding the rest (\secref{sec:input-reducing}). 
\texttt{AAmp} represents the assertion amplification step (\secref{sec:assert-amp}), while \texttt{Sel.} selects all test methods that increase the mutation score (\secref{sec:selection}).
\texttt{Read.} concerns the post-processing steps to increase the readability of the generated tests, in particular the oracle reduction (\secref{sec:post-processing}).
Finally, \texttt{Tot.} shows the entire execution time for a class.

\begin{figure}
\caption{The distribution of absolute execution time (in seconds)}
\label{fig:box-time-seconds}
  
 \begin{tikzpicture}
  \begin{axis}
    [
    height=6cm, 
    width=0.95\textwidth,
    axis y line*=left,
    axis x line*=bottom,
    ytick={1,2,3,4,5,6},
    xmode=log,
    yticklabels={Init. , IAmp, AAmp, Sel., Read., Tot.},
    ]
    \addplot+[boxplot]
    	table [row sep=\\, y index=0]{
	data\\
0.27	\\
0.26	\\
0.25	\\
0.47	\\
0.42	\\
0.37	\\
0.60	\\
0.59	\\
0.60	\\
1.09	\\
1.10	\\
1.19	\\
11.43	\\
11.49	\\
11.54	\\
47.05	\\
46.90	\\
47.08	\\
84.17	\\
84.21	\\
80.83	\\
73.20	\\
73.60	\\
73.93	\\
156.84	\\
158.84	\\
162.62	\\
14.97	\\
15.56	\\
15.08	\\
1.13	\\
1.21	\\
1.18	\\
22.10	\\
22.10	\\
23.36	\\
11.02	\\
11.25	\\
10.92	\\
0.56	\\
0.18	\\
0.21	\\
205.63	\\
205.84	\\
206.01	\\
0.08	\\
0.08	\\
0.08	\\
0.10	\\
0.09	\\
0.08	\\
0.21	\\
0.22	\\
0.21	\\
1.68	\\
1.64	\\
1.62	\\
0.19	\\
0.18	\\
0.16	\\
0.36	\\
0.35	\\
0.35	\\
0.19	\\
0.19	\\
0.19	\\
0.63	\\
1.11	\\
0.64	\\
0.14	\\
0.13	\\
0.11	\\
0.18	\\
0.19	\\
0.18	\\
1.05	\\
1.05	\\
1.04	\\
0.21	\\
0.21	\\
0.21	\\
0.57	\\
0.62	\\
0.54	\\
12.52	\\
12.03	\\
11.96	\\
1.73	\\
1.75	\\
1.78	\\
0.38	\\
0.33	\\
0.30	\\
11.24	\\
11.20	\\
11.57	\\
20.27	\\
19.56	\\
21.89	\\
0.27	\\
0.25	\\
0.25	\\
8.79	\\
7.68	\\
8.85	\\
0.15	\\
0.15	\\
0.15	\\
1.74	\\
2.25	\\
2.19	\\
11.66	\\
13.65	\\
12.91	\\
0.57	\\
0.46	\\
0.47	\\
12.47	\\
11.96	\\
11.65	\\
0.42	\\
0.43	\\
0.40	\\
0.98	\\
0.49	\\
0.55	\\
2.69	\\
4.04	\\
4.30	\\
0.43	\\
0.44	\\
0.39	\\
0.15	\\
0.17	\\
0.16	\\
0.26	\\
0.25	\\
0.25	\\
1.02	\\
1.03	\\
0.98	\\
1.90	\\
1.84	\\
1.88	\\
4.90	\\
4.34	\\
4.15	\\
0.31	\\
0.63	\\
0.42	\\
0.10	\\
0.10	\\
0.09	\\
11.72	\\
11.58	\\
11.44	\\
	}[above]
node[right] at
(boxplot box cs: \boxplotvalue{median},0.5)
{\pgfmathprintnumber[precision=2]{\boxplotvalue{median}}}
node[right] at
(boxplot box cs: \boxplotvalue{upper whisker},1)
{\pgfmathprintnumber[precision=1]{\boxplotvalue{upper whisker}}};
    \addplot+[boxplot]
    	table [row sep=\\, y index=0]{
	data\\
	3.23	\\
3.05	\\
3.07	\\
1.06	\\
1.15	\\
0.92	\\
10.74	\\
11.47	\\
12.29	\\
19.55	\\
20.70	\\
19.85	\\
15.38	\\
13.70	\\
16.79	\\
355.23	\\
358.33	\\
316.68	\\
30.31	\\
33.80	\\
30.64	\\
28.87	\\
32.35	\\
31.12	\\
147.45	\\
171.75	\\
197.25	\\
433.04	\\
426.92	\\
367.34	\\
5.66	\\
6.09	\\
6.07	\\
0.06	\\
0.04	\\
0.05	\\
0.11	\\
0.10	\\
0.09	\\
5.80	\\
6.46	\\
7.10	\\
37.52	\\
38.41	\\
45.60	\\
0.02	\\
0.02	\\
0.03	\\
0.01	\\
0.02	\\
0.02	\\
1.39	\\
1.67	\\
1.43	\\
9.87	\\
11.70	\\
11.22	\\
2.36	\\
2.73	\\
2.22	\\
4.80	\\
5.13	\\
6.87	\\
7.62	\\
7.14	\\
7.41	\\
21.53	\\
22.26	\\
79.29	\\
0.01	\\
0.01	\\
0.01	\\
0.01	\\
0.01	\\
0.01	\\
57.81	\\
65.53	\\
60.40	\\
1.23	\\
0.95	\\
1.29	\\
6.55	\\
6.09	\\
7.82	\\
12.23	\\
12.52	\\
11.67	\\
10.04	\\
9.57	\\
10.39	\\
1.17	\\
1.15	\\
1.12	\\
2.88	\\
3.14	\\
3.58	\\
256.22	\\
242.43	\\
249.43	\\
1.27	\\
1.18	\\
1.34	\\
47.45	\\
62.64	\\
68.61	\\
0.03	\\
0.03	\\
0.03	\\
0.09	\\
0.09	\\
0.09	\\
0.45	\\
0.32	\\
0.74	\\
11.00	\\
12.54	\\
11.80	\\
0.03	\\
0.02	\\
0.03	\\
0.02	\\
0.01	\\
0.02	\\
17.98	\\
11.84	\\
7.45	\\
27.26	\\
27.61	\\
38.81	\\
0.91	\\
0.71	\\
1.21	\\
0.39	\\
0.49	\\
0.50	\\
4.33	\\
5.18	\\
5.11	\\
18.86	\\
23.30	\\
21.24	\\
0.08	\\
0.07	\\
0.08	\\
0.09	\\
0.08	\\
0.09	\\
1.72	\\
1.36	\\
1.40	\\
0.09	\\
0.06	\\
0.34	\\
4.60	\\
5.01	\\
5.12	\\
	}[above]
node[right] at
(boxplot box cs: \boxplotvalue{median},0.5)
{\pgfmathprintnumber[precision=1]{\boxplotvalue{median}}}
node[right] at
(boxplot box cs: \boxplotvalue{upper whisker},1)
{\pgfmathprintnumber[precision=1]{\boxplotvalue{upper whisker}}};
    \addplot+[boxplot]
    	table [row sep=\\, y index=0]{
	data\\
	4.28	\\
4.16	\\
4.45	\\
3.02	\\
2.82	\\
3.07	\\
494.37	\\
508.25	\\
520.21	\\
78.45	\\
79.80	\\
81.04	\\
7.05	\\
6.30	\\
5.93	\\
1051.84	\\
1130.75	\\
1087.30	\\
30.03	\\
30.95	\\
28.16	\\
69.42	\\
68.85	\\
63.13	\\
7543.84	\\
7891.72	\\
8501.89	\\
1926.51	\\
1745.95	\\
1746.15	\\
9.75	\\
10.04	\\
11.74	\\
0.24	\\
0.21	\\
0.23	\\
2.39	\\
2.36	\\
2.39	\\
283.04	\\
317.67	\\
304.56	\\
14.03	\\
13.99	\\
13.69	\\
0.11	\\
0.10	\\
0.10	\\
2.77	\\
2.82	\\
2.79	\\
3.56	\\
3.67	\\
3.58	\\
41.46	\\
52.04	\\
56.99	\\
24.01	\\
25.71	\\
20.19	\\
76.15	\\
45.92	\\
60.40	\\
20.57	\\
21.84	\\
20.45	\\
432.10	\\
471.02	\\
416.81	\\
0.97	\\
0.94	\\
0.95	\\
1.71	\\
1.68	\\
1.59	\\
47.99	\\
49.50	\\
51.54	\\
2.21	\\
2.05	\\
2.09	\\
16.69	\\
16.50	\\
17.51	\\
17.92	\\
18.34	\\
17.82	\\
147.40	\\
145.13	\\
145.98	\\
1.10	\\
1.00	\\
1.08	\\
3.75	\\
4.20	\\
4.29	\\
449.86	\\
445.83	\\
468.19	\\
4.82	\\
4.29	\\
3.99	\\
425.02	\\
385.95	\\
361.22	\\
0.14	\\
0.12	\\
0.12	\\
1.95	\\
1.88	\\
1.98	\\
1.05	\\
0.93	\\
1.23	\\
12.12	\\
12.41	\\
13.77	\\
0.32	\\
0.29	\\
0.28	\\
2.64	\\
2.69	\\
2.63	\\
1.95	\\
1.15	\\
1.26	\\
57.65	\\
53.16	\\
135.65	\\
22.36	\\
22.11	\\
23.80	\\
0.76	\\
0.77	\\
0.88	\\
19.45	\\
19.97	\\
18.62	\\
57.44	\\
56.92	\\
71.60	\\
10.15	\\
9.48	\\
10.18	\\
17.75	\\
16.35	\\
17.23	\\
6.40	\\
4.76	\\
4.79	\\
0.20	\\
0.16	\\
0.58	\\
32.70	\\
30.40	\\
33.37	\\
	}[above]
node[right] at
(boxplot box cs: \boxplotvalue{median},0.5)
{\pgfmathprintnumber[precision=1]{\boxplotvalue{median}}}
node[right] at
(boxplot box cs: \boxplotvalue{upper whisker},1)
{\pgfmathprintnumber[precision=1]{\boxplotvalue{upper whisker}}};	
     \addplot+[boxplot]
    	table [row sep=\\, y index=0]{
	data\\
	3.75	\\
3.58	\\
3.59	\\
3.23	\\
3.21	\\
3.22	\\
17.06	\\
17.08	\\
17.51	\\
22.79	\\
23.46	\\
23.63	\\
40.31	\\
40.14	\\
40.30	\\
1294.51	\\
1087.95	\\
1307.16	\\
16.92	\\
17.16	\\
15.70	\\
77.81	\\
79.55	\\
73.97	\\
324.37	\\
349.26	\\
374.21	\\
431.64	\\
407.85	\\
426.82	\\
16.15	\\
16.76	\\
16.32	\\
0.57	\\
0.57	\\
0.60	\\
2.77	\\
2.66	\\
2.67	\\
10.76	\\
9.94	\\
11.43	\\
216.93	\\
203.96	\\
271.02	\\
0.13	\\
0.10	\\
0.10	\\
0.11	\\
0.12	\\
0.11	\\
1.51	\\
1.45	\\
1.41	\\
8.23	\\
8.26	\\
8.12	\\
3.63	\\
4.13	\\
3.59	\\
6.48	\\
7.04	\\
8.59	\\
11.67	\\
11.86	\\
11.48	\\
364.01	\\
341.68	\\
362.61	\\
0.00	\\
0.10	\\
0.11	\\
0.00	\\
0.00	\\
0.00	\\
45.64	\\
42.49	\\
36.82	\\
1.06	\\
0.99	\\
1.04	\\
9.87	\\
9.89	\\
10.25	\\
12.13	\\
12.77	\\
12.77	\\
28.24	\\
28.64	\\
32.12	\\
1.87	\\
1.81	\\
1.86	\\
12.58	\\
13.20	\\
13.60	\\
1127.95	\\
1129.58	\\
1189.48	\\
1.98	\\
1.94	\\
1.95	\\
70.04	\\
68.42	\\
69.09	\\
0.39	\\
0.37	\\
0.38	\\
8.61	\\
9.13	\\
9.03	\\
41.70	\\
38.70	\\
42.46	\\
9.93	\\
10.06	\\
10.15	\\
0.76	\\
0.72	\\
0.70	\\
0.00	\\
0.00	\\
0.26	\\
2.49	\\
1.81	\\
1.83	\\
88.24	\\
94.02	\\
91.39	\\
2.41	\\
2.80	\\
2.36	\\
1.01	\\
1.06	\\
1.09	\\
10.60	\\
10.68	\\
10.62	\\
39.66	\\
40.45	\\
41.01	\\
19.60	\\
19.34	\\
19.77	\\
53.24	\\
52.09	\\
52.51	\\
3.40	\\
2.75	\\
2.52	\\
0.17	\\
0.16	\\
0.73	\\
14.72	\\
14.68	\\
14.82	\\
	}[above]
node[right] at
(boxplot box cs: \boxplotvalue{median},0.5)
{\pgfmathprintnumber[precision=1]{\boxplotvalue{median}}}
node[right] at
(boxplot box cs: \boxplotvalue{upper whisker},1)
{\pgfmathprintnumber[precision=1]{\boxplotvalue{upper whisker}}};
    \addplot+[boxplot]
    	table [row sep=\\, y index=0]{
	data\\
	0.26	\\
0.26	\\
0.25	\\
0.33	\\
0.34	\\
0.34	\\
0.63	\\
0.62	\\
0.60	\\
1.00	\\
1.08	\\
1.07	\\
9.63	\\
9.38	\\
9.56	\\
36.58	\\
36.79	\\
37.12	\\
86.51	\\
83.28	\\
87.15	\\
72.12	\\
71.95	\\
73.35	\\
2.57	\\
4.30	\\
6.69	\\
4.92	\\
5.00	\\
5.03	\\
1.91	\\
1.12	\\
1.28	\\
23.26	\\
23.53	\\
23.28	\\
11.13	\\
10.83	\\
10.89	\\
0.22	\\
0.25	\\
0.27	\\
205.42	\\
205.47	\\
205.41	\\
0.06	\\
0.07	\\
0.07	\\
0.11	\\
0.10	\\
0.11	\\
0.22	\\
0.22	\\
0.21	\\
0.73	\\
0.73	\\
0.72	\\
0.29	\\
0.27	\\
0.20	\\
0.46	\\
0.47	\\
0.52	\\
0.16	\\
0.17	\\
0.17	\\
15.18	\\
197.83	\\
26.34	\\
0.10	\\
0.09	\\
0.09	\\
0.17	\\
0.17	\\
0.18	\\
0.97	\\
0.97	\\
1.08	\\
0.22	\\
0.21	\\
0.21	\\
0.53	\\
0.57	\\
0.59	\\
13.42	\\
12.21	\\
12.68	\\
2.62	\\
3.21	\\
2.81	\\
0.28	\\
0.31	\\
0.35	\\
11.39	\\
11.79	\\
11.49	\\
16.46	\\
18.61	\\
19.43	\\
0.26	\\
0.26	\\
0.26	\\
3.96	\\
3.72	\\
3.38	\\
0.13	\\
0.13	\\
0.14	\\
1.66	\\
1.67	\\
2.07	\\
13.67	\\
16.89	\\
17.79	\\
0.42	\\
0.42	\\
0.47	\\
12.66	\\
11.47	\\
11.29	\\
0.41	\\
0.39	\\
0.36	\\
0.49	\\
0.53	\\
0.52	\\
6.09	\\
5.09	\\
5.40	\\
0.50	\\
0.52	\\
0.44	\\
0.14	\\
0.13	\\
0.15	\\
0.29	\\
0.28	\\
0.29	\\
1.31	\\
1.27	\\
1.29	\\
1.69	\\
1.76	\\
1.78	\\
5.21	\\
5.08	\\
5.23	\\
0.19	\\
0.18	\\
0.18	\\
0.09	\\
0.08	\\
0.08	\\
11.26	\\
11.94	\\
12.39	\\
	}[above]
node[right] at
(boxplot box cs: \boxplotvalue{median},0.5)
{\pgfmathprintnumber[precision=2]{\boxplotvalue{median}}}
node[right] at
(boxplot box cs: \boxplotvalue{upper whisker},1)
{\pgfmathprintnumber[precision=1]{\boxplotvalue{upper whisker}}};
\addplot+[boxplot]
    	table [row sep=\\, y index=0]{
	data\\
	11.79	\\
11.31	\\
11.61	\\
8.10	\\
7.93	\\
7.91	\\
523.39	\\
538.00	\\
551.21	\\
122.90	\\
126.14	\\
126.77	\\
83.79	\\
81.00	\\
84.12	\\
2785.20	\\
2660.72	\\
2795.35	\\
247.93	\\
249.39	\\
242.48	\\
321.42	\\
326.29	\\
315.50	\\
8175.06	\\
8575.86	\\
9242.66	\\
2811.09	\\
2601.28	\\
2560.42	\\
34.60	\\
35.21	\\
36.58	\\
46.23	\\
46.45	\\
47.52	\\
27.42	\\
27.18	\\
26.96	\\
300.38	\\
334.49	\\
323.56	\\
679.54	\\
667.67	\\
741.73	\\
0.40	\\
0.36	\\
0.39	\\
3.10	\\
3.15	\\
3.11	\\
6.90	\\
7.23	\\
6.84	\\
61.97	\\
74.38	\\
78.68	\\
30.48	\\
33.02	\\
26.35	\\
88.24	\\
58.89	\\
76.72	\\
40.21	\\
41.19	\\
39.69	\\
833.45	\\
1033.90	\\
885.69	\\
1.23	\\
1.27	\\
1.28	\\
2.06	\\
2.05	\\
1.95	\\
153.46	\\
159.53	\\
150.89	\\
4.92	\\
4.41	\\
4.83	\\
34.20	\\
33.67	\\
36.72	\\
68.23	\\
67.86	\\
66.90	\\
190.02	\\
188.30	\\
193.08	\\
4.80	\\
4.60	\\
4.71	\\
41.84	\\
43.53	\\
44.53	\\
1870.76	\\
1856.00	\\
1948.42	\\
8.60	\\
7.92	\\
7.79	\\
555.26	\\
528.41	\\
511.16	\\
0.85	\\
0.79	\\
0.82	\\
14.03	\\
15.03	\\
15.35	\\
68.54	\\
70.49	\\
75.13	\\
34.05	\\
35.89	\\
36.67	\\
26.25	\\
24.46	\\
23.95	\\
3.48	\\
3.52	\\
3.68	\\
23.89	\\
15.82	\\
11.60	\\
181.93	\\
183.92	\\
275.55	\\
26.61	\\
26.58	\\
28.20	\\
2.44	\\
2.63	\\
2.77	\\
34.92	\\
36.37	\\
34.89	\\
118.28	\\
122.98	\\
136.12	\\
33.42	\\
32.49	\\
33.68	\\
81.19	\\
77.93	\\
79.21	\\
12.01	\\
9.69	\\
9.30	\\
0.65	\\
0.56	\\
1.82	\\
75.00	\\
73.61	\\
77.15	\\
	}[above]
node[right] at
(boxplot box cs: \boxplotvalue{median},0.5)
{\pgfmathprintnumber[precision=1]{\boxplotvalue{median}}}
node[right] at
(boxplot box cs: \boxplotvalue{upper whisker},1)
{\pgfmathprintnumber[precision=1]{\boxplotvalue{upper whisker}}};
  \end{axis}
  \end{tikzpicture}
\end{figure}

\begin{figure}
\caption{The relative distributions of the time-cost (percentage)}
  \label{fig:box-time-percentage}
  
 \begin{tikzpicture}
  \begin{axis}
    [
    height=5cm, 
    width=0.9\textwidth,
    axis y line*=left,
    axis x line*=bottom,
    ytick={1,2,3,4,5},
    yticklabels={Init. , IAmp, AAmp, Sel., Read.},
    ]
    \addplot+[boxplot]
    	table [row sep=\\, y index=0]{
	data\\
	2.27	\\
2.31	\\
2.18	\\
5.74	\\
5.25	\\
4.71	\\
0.12	\\
0.11	\\
0.11	\\
0.89	\\
0.87	\\
0.94	\\
13.64	\\
14.18	\\
13.71	\\
1.69	\\
1.76	\\
1.68	\\
33.95	\\
33.76	\\
33.34	\\
22.77	\\
22.56	\\
23.43	\\
1.92	\\
1.85	\\
1.76	\\
0.53	\\
0.60	\\
0.59	\\
3.28	\\
3.43	\\
3.22	\\
47.80	\\
47.57	\\
49.17	\\
40.18	\\
41.37	\\
40.48	\\
0.19	\\
0.05	\\
0.06	\\
30.26	\\
30.83	\\
27.77	\\
18.99	\\
21.15	\\
20.78	\\
3.26	\\
2.99	\\
2.41	\\
3.10	\\
3.06	\\
3.06	\\
2.70	\\
2.21	\\
2.06	\\
0.62	\\
0.55	\\
0.60	\\
0.40	\\
0.59	\\
0.45	\\
0.47	\\
0.47	\\
0.47	\\
0.08	\\
0.11	\\
0.07	\\
11.59	\\
9.85	\\
8.76	\\
8.77	\\
9.12	\\
9.42	\\
0.69	\\
0.66	\\
0.69	\\
4.23	\\
4.81	\\
4.37	\\
1.67	\\
1.84	\\
1.47	\\
18.35	\\
17.72	\\
17.88	\\
0.91	\\
0.93	\\
0.92	\\
7.81	\\
7.19	\\
6.31	\\
26.87	\\
25.72	\\
25.99	\\
1.08	\\
1.05	\\
1.12	\\
3.16	\\
3.12	\\
3.18	\\
1.58	\\
1.45	\\
1.73	\\
17.55	\\
18.64	\\
18.35	\\
12.38	\\
14.98	\\
14.24	\\
17.02	\\
19.37	\\
17.18	\\
1.69	\\
1.28	\\
1.27	\\
47.51	\\
48.89	\\
48.62	\\
11.98	\\
12.15	\\
10.97	\\
4.11	\\
3.12	\\
4.70	\\
1.48	\\
2.20	\\
1.56	\\
1.62	\\
1.65	\\
1.38	\\
5.99	\\
6.47	\\
5.81	\\
0.73	\\
0.69	\\
0.72	\\
0.86	\\
0.84	\\
0.72	\\
5.69	\\
5.67	\\
5.57	\\
6.03	\\
5.56	\\
5.24	\\
2.55	\\
6.48	\\
4.48	\\
15.49	\\
18.49	\\
4.77	\\
15.63	\\
15.73	\\
14.83	\\
	}[above]
node[right] at
(boxplot box cs: \boxplotvalue{median},0.5)
{\pgfmathprintnumber[precision=1]{\boxplotvalue{median}}\%}
node[right] at
(boxplot box cs: \boxplotvalue{upper whisker},1)
{\pgfmathprintnumber[precision=1]{\boxplotvalue{upper whisker}}\%};
    \addplot+[boxplot]
    	table [row sep=\\, y index=0]{
	data\\
	27.37	\\
26.93	\\
26.45	\\
13.04	\\
14.45	\\
11.60	\\
2.05	\\
2.13	\\
2.23	\\
15.91	\\
16.41	\\
15.66	\\
18.36	\\
16.91	\\
19.96	\\
12.75	\\
13.47	\\
11.33	\\
12.23	\\
13.55	\\
12.63	\\
8.98	\\
9.91	\\
9.86	\\
1.80	\\
2.00	\\
2.13	\\
15.40	\\
16.41	\\
14.35	\\
16.36	\\
17.30	\\
16.59	\\
0.13	\\
0.09	\\
0.11	\\
0.42	\\
0.36	\\
0.35	\\
1.93	\\
1.93	\\
2.19	\\
5.52	\\
5.75	\\
6.15	\\
4.56	\\
4.40	\\
7.53	\\
0.45	\\
0.48	\\
0.68	\\
20.21	\\
23.14	\\
20.94	\\
15.93	\\
15.72	\\
14.26	\\
7.75	\\
8.26	\\
8.43	\\
5.44	\\
8.70	\\
8.95	\\
18.96	\\
17.33	\\
18.67	\\
2.58	\\
2.15	\\
8.95	\\
1.06	\\
0.47	\\
0.86	\\
0.34	\\
0.39	\\
0.36	\\
37.67	\\
41.07	\\
40.03	\\
25.01	\\
21.51	\\
26.63	\\
19.14	\\
18.09	\\
21.31	\\
17.92	\\
18.45	\\
17.44	\\
5.28	\\
5.08	\\
5.38	\\
24.36	\\
24.92	\\
23.79	\\
6.89	\\
7.22	\\
8.03	\\
13.70	\\
13.06	\\
12.80	\\
14.79	\\
14.93	\\
17.22	\\
8.55	\\
11.85	\\
13.42	\\
3.65	\\
3.27	\\
3.65	\\
0.61	\\
0.63	\\
0.55	\\
0.65	\\
0.45	\\
0.99	\\
32.29	\\
34.95	\\
32.19	\\
0.12	\\
0.09	\\
0.11	\\
0.46	\\
0.40	\\
0.46	\\
75.24	\\
74.82	\\
64.23	\\
14.98	\\
15.01	\\
14.09	\\
3.41	\\
2.69	\\
4.28	\\
15.91	\\
18.66	\\
18.10	\\
12.41	\\
14.25	\\
14.64	\\
15.94	\\
18.95	\\
15.60	\\
0.25	\\
0.23	\\
0.22	\\
0.11	\\
0.10	\\
0.11	\\
14.28	\\
14.06	\\
15.03	\\
14.42	\\
10.05	\\
18.59	\\
6.13	\\
6.80	\\
6.64	\\
	}[above]
node[right] at
(boxplot box cs: \boxplotvalue{median},0.5)
{\pgfmathprintnumber[precision=1]{\boxplotvalue{median}}\%}
node[right] at
(boxplot box cs: \boxplotvalue{upper whisker},1)
{\pgfmathprintnumber[precision=1]{\boxplotvalue{upper whisker}}\%};
    \addplot+[boxplot]
    	table [row sep=\\, y index=0]{
	data\\
	36.31	\\
36.76	\\
38.30	\\
37.26	\\
35.54	\\
38.78	\\
94.46	\\
94.47	\\
94.38	\\
63.84	\\
63.27	\\
63.92	\\
8.41	\\
7.78	\\
7.05	\\
37.77	\\
42.50	\\
38.90	\\
12.11	\\
12.41	\\
11.61	\\
21.60	\\
21.10	\\
20.01	\\
92.28	\\
92.02	\\
91.99	\\
68.53	\\
67.12	\\
68.20	\\
28.16	\\
28.52	\\
32.09	\\
0.52	\\
0.45	\\
0.47	\\
8.71	\\
8.67	\\
8.88	\\
94.23	\\
94.97	\\
94.13	\\
2.07	\\
2.10	\\
1.85	\\
28.10	\\
27.75	\\
26.23	\\
89.26	\\
89.55	\\
89.74	\\
51.60	\\
50.77	\\
52.27	\\
66.91	\\
69.97	\\
72.44	\\
78.76	\\
77.88	\\
76.60	\\
86.30	\\
77.96	\\
78.73	\\
51.16	\\
53.01	\\
51.52	\\
51.84	\\
45.56	\\
47.06	\\
79.18	\\
74.23	\\
74.43	\\
82.75	\\
82.11	\\
81.21	\\
31.27	\\
31.03	\\
34.16	\\
44.90	\\
46.49	\\
43.19	\\
48.80	\\
49.00	\\
47.68	\\
26.27	\\
27.02	\\
26.64	\\
77.57	\\
77.07	\\
75.61	\\
23.00	\\
21.75	\\
22.88	\\
8.96	\\
9.66	\\
9.64	\\
24.05	\\
24.02	\\
24.03	\\
56.03	\\
54.23	\\
51.23	\\
76.54	\\
73.04	\\
70.67	\\
16.96	\\
14.86	\\
14.34	\\
13.87	\\
12.50	\\
12.89	\\
1.54	\\
1.32	\\
1.63	\\
35.60	\\
34.58	\\
37.56	\\
1.23	\\
1.20	\\
1.19	\\
75.83	\\
76.33	\\
71.65	\\
8.17	\\
7.25	\\
10.82	\\
31.69	\\
28.90	\\
49.23	\\
84.03	\\
83.18	\\
84.41	\\
31.16	\\
29.32	\\
31.59	\\
55.69	\\
54.92	\\
53.36	\\
48.56	\\
46.29	\\
52.60	\\
30.38	\\
29.17	\\
30.22	\\
21.86	\\
20.98	\\
21.75	\\
53.32	\\
49.16	\\
51.48	\\
30.21	\\
28.01	\\
31.96	\\
43.60	\\
41.31	\\
43.25	\\
	}[above]
node[right] at
(boxplot box cs: \boxplotvalue{median},0.5)
{\pgfmathprintnumber[precision=1]{\boxplotvalue{median}}\%}
node[right] at
(boxplot box cs: \boxplotvalue{upper whisker},1)
{\pgfmathprintnumber[precision=1]{\boxplotvalue{upper whisker}}\%};	
    \addplot+[boxplot]
    	table [row sep=\\, y index=0]{
	data\\
	31.84	\\
31.69	\\
30.94	\\
39.87	\\
40.45	\\
40.68	\\
3.26	\\
3.17	\\
3.18	\\
18.55	\\
18.60	\\
18.64	\\
48.10	\\
49.55	\\
47.91	\\
46.48	\\
40.89	\\
46.76	\\
6.82	\\
6.88	\\
6.48	\\
24.21	\\
24.38	\\
23.45	\\
3.97	\\
4.07	\\
4.05	\\
15.36	\\
15.68	\\
16.67	\\
46.68	\\
47.58	\\
44.60	\\
1.24	\\
1.23	\\
1.27	\\
10.10	\\
9.78	\\
9.92	\\
3.58	\\
2.97	\\
3.53	\\
31.92	\\
30.55	\\
36.54	\\
32.91	\\
27.75	\\
26.75	\\
3.39	\\
3.68	\\
3.60	\\
21.95	\\
19.99	\\
20.67	\\
13.28	\\
11.11	\\
10.33	\\
11.92	\\
12.50	\\
13.60	\\
7.34	\\
11.95	\\
11.20	\\
29.02	\\
28.79	\\
28.92	\\
43.67	\\
33.05	\\
40.94	\\
0.00	\\
8.12	\\
8.84	\\
0.00	\\
0.00	\\
0.00	\\
29.74	\\
26.64	\\
24.40	\\
21.50	\\
22.44	\\
21.45	\\
28.85	\\
29.36	\\
27.93	\\
17.78	\\
18.81	\\
19.08	\\
14.86	\\
15.21	\\
16.64	\\
39.03	\\
39.32	\\
39.60	\\
30.05	\\
30.33	\\
30.54	\\
60.29	\\
60.86	\\
61.05	\\
22.98	\\
24.50	\\
25.02	\\
12.61	\\
12.95	\\
13.52	\\
46.41	\\
46.98	\\
46.66	\\
61.33	\\
60.77	\\
58.82	\\
60.84	\\
54.90	\\
56.52	\\
29.18	\\
28.04	\\
27.69	\\
2.90	\\
2.94	\\
2.93	\\
0.00	\\
0.00	\\
7.02	\\
10.43	\\
11.44	\\
15.76	\\
48.50	\\
51.12	\\
33.17	\\
9.05	\\
10.54	\\
8.36	\\
41.29	\\
40.52	\\
39.20	\\
30.35	\\
29.37	\\
30.44	\\
33.53	\\
32.89	\\
30.13	\\
58.63	\\
59.52	\\
58.69	\\
65.57	\\
66.84	\\
66.30	\\
28.28	\\
28.42	\\
27.08	\\
26.53	\\
28.90	\\
40.08	\\
19.62	\\
19.95	\\
19.21	\\
	}[above]
node[right] at
(boxplot box cs: \boxplotvalue{median},0.5)
{\pgfmathprintnumber[precision=1]{\boxplotvalue{median}}\%}
node[right] at
(boxplot box cs: \boxplotvalue{upper whisker},1)
{\pgfmathprintnumber[precision=1]{\boxplotvalue{upper whisker}}\%};
\addplot+[boxplot]
    	table [row sep=\\, y index=0]{
	data\\
	2.21	\\
2.32	\\
2.13	\\
4.09	\\
4.31	\\
4.24	\\
0.12	\\
0.11	\\
0.11	\\
0.82	\\
0.86	\\
0.84	\\
11.49	\\
11.58	\\
11.36	\\
1.31	\\
1.38	\\
1.33	\\
34.89	\\
33.39	\\
35.94	\\
22.44	\\
22.05	\\
23.25	\\
0.03	\\
0.05	\\
0.07	\\
0.18	\\
0.19	\\
0.20	\\
5.51	\\
3.17	\\
3.49	\\
50.31	\\
50.66	\\
48.98	\\
40.59	\\
39.82	\\
40.38	\\
0.07	\\
0.07	\\
0.08	\\
30.23	\\
30.77	\\
27.69	\\
15.44	\\
18.96	\\
18.70	\\
3.65	\\
3.30	\\
3.57	\\
3.13	\\
3.04	\\
3.07	\\
1.18	\\
0.99	\\
0.92	\\
0.95	\\
0.81	\\
0.77	\\
0.52	\\
0.79	\\
0.67	\\
0.40	\\
0.41	\\
0.42	\\
1.82	\\
19.13	\\
2.97	\\
8.16	\\
7.33	\\
7.11	\\
8.14	\\
8.39	\\
9.01	\\
0.63	\\
0.61	\\
0.72	\\
4.37	\\
4.74	\\
4.37	\\
1.54	\\
1.70	\\
1.61	\\
19.68	\\
17.99	\\
18.95	\\
1.38	\\
1.71	\\
1.45	\\
5.79	\\
6.82	\\
7.41	\\
27.23	\\
27.07	\\
25.80	\\
0.88	\\
1.00	\\
1.00	\\
3.05	\\
3.22	\\
3.35	\\
0.71	\\
0.70	\\
0.66	\\
15.43	\\
16.25	\\
17.01	\\
11.82	\\
11.12	\\
13.50	\\
19.95	\\
23.96	\\
23.68	\\
1.25	\\
1.16	\\
1.29	\\
48.23	\\
46.88	\\
47.15	\\
11.72	\\
11.13	\\
9.90	\\
2.05	\\
3.37	\\
4.49	\\
3.35	\\
2.77	\\
1.96	\\
1.89	\\
1.95	\\
1.57	\\
5.66	\\
5.03	\\
5.30	\\
0.82	\\
0.76	\\
0.84	\\
1.10	\\
1.03	\\
0.95	\\
5.05	\\
5.41	\\
5.30	\\
6.42	\\
6.51	\\
6.60	\\
1.57	\\
1.88	\\
1.92	\\
13.34	\\
14.54	\\
4.61	\\
15.02	\\
16.22	\\
16.06	\\
	}[above]
node[right] at
(boxplot box cs: \boxplotvalue{median},0.5)
{\pgfmathprintnumber[precision=1]{\boxplotvalue{median}}\%}
node[right] at
(boxplot box cs: \boxplotvalue{upper whisker},1)
{\pgfmathprintnumber[precision=1]{\boxplotvalue{upper whisker}}\%};
  \end{axis}
  \end{tikzpicture}
\end{figure}

\figref{fig:box-time-seconds} shows how the execution time is distributed for total execution time and also for each step in seconds.
The horizontal axis presents the number of seconds in logarithmic scale.
The diagram includes also the values for the median and the upper whisker.

Regarding total execution time (\texttt{Tot.}), half of the executions finished in less than 36.7 seconds (the median value).
Furthermore, the diagram shows that the majority of amplification (upper whisker) finished in less than 334.4 seconds (5 minutes and 34 seconds).
However, we see 25 outliers that refer to the instances that finished in more than 335 seconds.
If we set a fixed time budget, for instance a 10 minutes budget for each class, the test amplification process for these classes will not terminate properly.
This show the importance of considering time budget management in test amplification tools.

The median value for other steps are: initializing 1 second, input amplification 4.8 seconds, assertion amplification 13.7 seconds, selection by mutation testing 10.1 seconds and post-processing steps 0.7 seconds.

\figref{fig:box-time-percentage} illustrates the relative proportion of time test amplification dedicates to each step.
So, the execution time for each step is divided to the total amplification time to compare the steps relatively.

The profiling step and the oracle reductions steps are the fastest steps. 
The median value for each of these steps are respectively 3.2\% and 3.4\%.
Therefore, we can say profiling and oracle reduction steps do not add much time overhead to the overall process.
Next, the input amplification step takes about 11.3\%.
A considerable portion of execution time is spent during assertion amplification (median 38.8\%, upper whisker 95\%).
The median execution time related to selection step, in which mutation testing is ran, is 24.5\%.
We suspect the execution time for mutation testing would be more if \toolname{MuTalk} would generate more mutants.

\hypobox{
\textbf{Answer to RQ6: }
The majority of classes in our dataset has been amplified in less than 5 minutes and 34 seconds.
However, in some cases the execution takes longer with a maximum of 2 hours and 34 minutes.
Therefore, a time budget management mechanism is needed in the test amplification tools.
In the execution time for each steps, we see that the profiling and oracle reductions steps (the extra steps compared to the original \dspot algorithm) do not add much overhead to the overall process.
}

\section{Threats to Validity} \label{sec:threats}

As in any empirical research, we identify factors that may jeopardise the validity of our results and the actions we took to reduce or alleviate the risk.
Consistent with the guidelines provided by ~\cite{Wohl00a}, we organise them into four categories.

\paragraph{Construct validity.} Do we measure what was intended?
For RQ1 (\rqOneKey), we manually selected test methods which we considered valuable additions to the project.
And we provided a motivation for the pull request based on a human interpretation of the extra mutants killed.
Thus, the percentage of accepted pull requests is a flattered result.
If we would have submitted all amplified test methods the results would be far lower.
We consider this risk as minimal, because \smallamp at the current stage should never be seen as a fully automated code synthesizer tool but rather as a recommender system supporting the human-in-the-loop.

For RQ2 to RQ4 we heavily rely on mutation coverage as a proxy for the corner cases the amplified tests are supposed to cover.
There is an ongoing debate in the mutation testing community of whether mutation operators can serve as proxies for actual faults.
Today, there is no alternative so we settled with mutation coverage.
But if ever another measure for test effectiveness comes along we need to revise the results.

\paragraph{Internal validity.} Are there unknown factors which might affect the outcome of the analyses?
For RQ1 (\rqOneKey), we don't have any knowledge about the policy the projects had concerning pull requests submitted by outsiders.
In the Pharo community, most developers know one another and are likely to trust contributions.
However, for our study it was the first author who submitted the pull-requests and at that point in time he was a newcomer in the community.
For the three pull requests which were ignored, we don't know whether this newcomer submission played a role.

\paragraph{External validity.} To what extent is it possible to generalise the findings?
We demonstrated that test amplification is feasible, even for dynamically typed language.
We have constructed a proof-by-construction for the Pharo/Smalltalk ecosystem.
However, we cannot make any claims regarding other dynamically typed languages (Python, Ruby, Javascript, \ldots).
We are quite confident that type profiling is the key to make test amplification successful in such a context.
However, coding conventions are equally important and this may jeopardise the kind of input amplification operators that work.

\paragraph{Reliability (a.k.a. Conclusion Validity).} Is the result dependent on the tools?
As mentioned earlier, we heavily rely on mutation coverage as calculated by \toolname{MuTalk}.
\toolname{MuTalk} lacks several mutation operators compared to the \toolname{PITest} tool used in the \dspot paper.
This implies that \smallamp will generate fewer test cases and that the newly killed mutants will also be generally lower.
We mitigated this threat by always reporting the absolute number combined with the relative increase (in percentage).

The other threat to conclusion validity in the impact of randomness.
\smallamp works based on applying random transformations on the tests.
Most importantly, the actual amplified tests surviving the Input Reduction step (see \secrefpage{sec:input-reducing}) may vary from one run to another.
We applied the tool on different classes in different projects from various domains, and achieved amplified tests in all circumstances.
In addition to the variety in the projects, we ran the tool three times on each test class.
Thus, the impact of randomness should be small at best.

\section{Related Work} \label{sec:relatedworks}

Test amplification systems can vary based on the engineering goal.
In addition to the amplification of the code coverage \citep{thummalapenta2011retrofitting, yoo2012test} or mutation score \citep{baudry2005genetic, patrick2017kd}, researchers have used test amplification for other goals like fault detection \citep{milani2014leveraging, pezze2013generating}, oracle improvement \citep{fraser2011generating, Xie_2006}, fault localization \citep{robetaler2012isolating, xuan2015crash}, and incompatible environments detection \citep{CAMP2019}.

A test amplification system may use search based techniques \citep{baudry2005genetic, yoo2012test, rojas2016seeding, xu2010factors} or symbolic and concolic execution techniques \citep{xu2009directed, yoshida2016fsx, thummalapenta2011retrofitting}.
The results of the amplification can be added to existing test suite \citep{baudry2005genetic, yoo2012test} or just modifying the current tests \citep{Xie_2006}.
They also may consider only new changes \citep{xu2010factors, xu2009directed} rather than working on the snapshot of the entire project.

Our work is pretty close to \dspot \citep{Baudry2015corr, Danglot2019EMSE} where \smallamp is an adaption of \dspot into a dynamic language.
This work, same as \dspot, also can be categorised under genetic improvement \citep{Petke2018Genetic} where it takes advantage of existing test suite as well as an automated search algorithm in order to find an improved version of test code.

\addressedRvw{
Brandt and Zaidman~\citep{brandt2021developer} use a lighter version of \dspot to increase the instruction coverage.
They also provide an IDE plugin to make the developers interplay with the test amplification tool possible.
Nijkamp et al.~\citep{nijkamp2021naming} and Oosterbroek et. al~\citep{oosterbroek2021removing} address the readability of the amplified tests by choosing proper names and removing redundant statements.
}

\paragraph{Dynamically-typed languages.} As we argued earlier, test case generation is well studied in statically typed languages \citep{randoop, Fraser2013Whole, dynamosa} but there are only a few academic works that target dynamically typed languages. We list the ones we were aware of below.

\cite{Lukasczyk2020Automated} introduce automated unit test generation for Python, and the tool \toolname{Pynguin} which works based on techniques used in statically typed languages more specifically \toolname{EvoSuite} \citep{Fraser2013Whole} and \toolname{Randoop} \citep{randoop}.
\toolname{Pynguin} circumvents the lack of type information by assuming that the system under test contains type information added by developers in terms of type annotations.

\cite{JSEFT} created a tool names \toolname{JSEFT}, which generate unit tests for javascript functions and events by a record and replay technique.
\toolname{JSEFT} relies on web crawling to collect traces from javascript executions.
They create test methods by replaying the executions and adding new oracles using a mutation-based process \citep{Fraser2012MutationDriven}.

\cite{Wibowo2015Lua} use a genetic algorithm to generate unit test code for the Lua scripting language.
The algorithm starts from a random initialized population and then evolve by crossover and mutation operators.
The tool only generates assertions for primitive data type values.
\smallamp on the other hand used recursive assertion generation to deal with non-primitive types.

\cite{Mairhofer11} introduce \toolname{RuTeG}, a test generation for Ruby based on genetic algorithms. 
For each method under test, \toolname{RuTeG} statically processes the parse tree and collects arguments and the list of methods invoked on each argument.
Then they use predefined and customized data generators to generate a part of code that is valid based on data collected from the parse tree.
\toolname{RuTeG} does not improve an existing test suite, but rather generates the whole test suite itself.

\paragraph{Mutation testing.} 
Using mutation testing as an actionable target for strengthening an existing test suite is used at large scale at Google \citep{petrovic2021practical, Petrovic2018SEIP} and Facebook \citep{beller2021use}.
These papers are close in spirit to \dspot and \smallamp as both create new test casess to increase mutation coverage.
However, these works use professional developers to generate new tests (manual test amplification) while \dspot and \smallamp illustrate that a recommender system is feasible.

\section{Future Work} \label{sec:futureworks}

In this section, we present some open problems and the ways how \smallamp and test amplification tools can be improved in the future.

\paragraph{Test amplification ecosystem.}\label{sec:test-amp-ci}
In the current implementation of \smallamp, we run the tool on the whole project.
This way of using the tool has some drawbacks:
since developers should run it manually, they need a knowledge about the concept, the process and also the tool interface;
it may take a long time to amplify all classes in the project;
the tool will reamplify some parts of the project on each execution, which will increase the execution time;
and finally, developers need to deal with the output manually, they need to understand it, polish the interesting tests, and merge them manually in the code base.
Imposing such extra work on developers is likely to make them loose their interest in using the tool often.

In the future, we will integrate the \smallamp with a build system, for instance GitHub Actions.
The build systems will run the tool automatically on the specified events such as on each push, or pull-request or periodically.
Additionally, \smallamp will amplify only the recent changes on each run.
It means that only the mutants in the changed parts will be generated which will reduce the cost of amplification significantly.
In this case, the amplification will be run in the project level instead of running class-by-class.
So, finding an exact relation between the production classes and the test classes will also lost its importance: all test methods covering a changed part can be included in the amplification.

Furthermore, we will build a web-based test editor dashboard to visualize the test amplification outputs, and also a GitHub-Bot to synchronize the Build system output, code base and the dashboard.
The developer can use this dashboard to assess the outputs, and also customize the amplified test.
The tests after the polishment will be reevaluated automatically, and if it is green, it can be pushed to the code base.
So, developers don't need to overwhelm themselves with tedious tasks and can benefit from test amplification in an ecosystem automated by bots.

\paragraph{Extended use-cases for dynamic profiling.}\label{sec:profiling-usecases}
Dynamic profiling is more than merely a type inference solution.
It can be generalized to collect various information about unit-under-test based on dynamically running existing test suite.
In some cases, statically typed languages also can benefit the profiling mechanism.
We enumerate two of these use-cases:
\begin{itemize}
\item
Pure methods detection:
An impure method is a method that looks like an accessor but calling it causes a change in the state of the object~\citep{sualcianu2005purity}.
In the scope of \smallamp, identifying pure/impure methods is important during oracle reduction.
A new profiler can be implemented using the method proxies to serialize the object state before and after method invocations.
If the state is changed, we can infer that the method is impure.

\item
Providing information for advanced input amplifiers: 
In this paper we proposed a basic algorithm for test-input reduction (\secref{sec:input-reducing}).
By addressing the test-input explosion problem, test amplification tools can benefit from wider range of input amplifiers.
Advanced input-amplifiers can exploit the profiling step to collect useful information dynamically and increase their knowledge about the program under test.
For example, a profiler can collect all literal values from the covered methods and use them in literal values amplification operator.
Another example can be object transplantation between test methods.
A profiler can collect patterns of how objects are created and manipulated and this information can be used in an input-amplifier.

\end{itemize}

\paragraph{Test method models, best practices and structured strings.}

Unit tests in object-oriented languages ideally are structured as a sequence of statements that instantiate an object, manipulates its state, and asserts expected values.
However, not every test fits this model in real projects. 
Developers use helper methods, customized assertions, structured strings, some optimizations like grouping the tests or parallelizing them.
For instance, a developer may write tests in XML files and load each file in a test method, so these tests heavily depend on parsing structured strings.

If a test does not fit with the ideal test model, the current algorithm of test amplification still can be used, but it may be less successful in producing strong results.
We leave identifying and adopting best practices of test methods as an open problem.
Additionally, mutating a structured string by understanding its syntax can also be interesting future work.

\paragraph{Using patterns to guide test amplification.}

As an important future work, we suggest using heuristics to guide the amplification algorithm.
Large scale manual test amplification histories like the work at Google \citep{petrovic2021practical, Petrovic2018SEIP} can be analyzed to answer questions like: 
How often do developers write new test methods? 
Are new test methods similar to an existing test?
If they update an existing test method, what is the relation of the updated test method and the mutant to be killed? 
What transformations are applied to the test?

Answering these questions leads us to find some patterns in how real developers kill mutants.
These patterns can help the tool to prioritize some test methods and input amplifiers for killing a particular mutant.
Recent advances in deep learning or other program synthesis techniques are promising and can be helpful in making test amplification tools more intelligent \citep{abdi2019adopting}.

\paragraph{Reducing the mutation testing burden.}

Test amplification generates tests to optimize mutation coverage, however calculating the mutation score is a time-consuming process.
During test amplification, this mutation score is calculated multiple times for each test method: in lines 6, and 14 of \algorefpage{alg:main}.
However, for a test to kill a mutant, it first must reach the statement, then infect the program state, propagate to the output where it must be revealed by an assertion~\citep{lu2020semi, vera2019suggestions}.

We can optimize the calculation of the mutation score by using a hierarchical coverage measurement. 
For example, we can first run a code coverage tool, then we can run an extreme transformation \citep{Descartes2018}, afterwards we only mutate the covered parts.

Another technique for reducing the mutation testing burden is to use mutation operators that are learned from known common bugs like \toolname{MutationMonkey} \citep{beller2021use}.

\paragraph{Using multiple type-inference mechanisms.} \label{sec:futue-multiple-type-infer}
The main drawback of using an existing test suite for dynamic profiling is if a method is not covered in the test, we can not collect its type information, hence can not add calls to such methods during input amplification.
Static type inference~\citep{DBLP:conf/dls/PluquetMW09} or live typing~\citep{Wilkinson2019Live} techniques can be helpful to empower \smallamp to generate method calls to such uncovered methods.

\paragraph{Involving readability metrics.}

Based on a previous study \citep{Pizza2020}, the most important aspects for developers in assessing the quality of a test suite are readability and maintainability.
Although the code coverage metrics are necessary for identifying the poor test suites, they are limited in distinguishing high-quality tests based on how practitioners perceive the test quality.
Since the goal of test amplification is to recommend new test cases ready to be merged into the code base, considering readability and maintainability metrics in the test amplification appears to be a critical next step.

\section{Conclusion} \label{sec:conclusion}

In this paper, we introduce \smallamp, an approach for test amplification in the dynamically typed language Pharo/Smalltalk.
The main algorithm of \smallamp is adapted from \dspot, a test amplification technique designed for Java programs.
In order to mitigate the lack of type information, we exploit profiling information, readily available by running the test suite.
We demonstrate that test amplification is feasible for dynamically typed languages, by replicating the experimental set-up of \dspot, including a qualitative and quantitative analysis of the improved test suite.

In our qualitative analysis, we submitted pull-requests of an amplified test by \smallamp to the GitHub repositories of the projects in our dataset. 
From \nSentPR pull-requests we submitted, \nMergedPR were merged (\percPR). 
The developers' comments on the pull-requests illustrate how valuable they perceive the new tests created by \smallamp.  
The results from our quantitative study show that \smallamp succeeds to amplify \nAmplified test classes out of 52, approximately 53\% of target classes, in 13 projects from our dataset.
The majority of the generated tests are focused, and all test amplification steps (including type profiling step) play a critical role.
The results from \smallamp and the results from \dspot are comparable.
We see \percPR~merged pull-requests and 53\% successfully amplified test classes in \smallamp, while for \dspot these values are 68\% and 65\%.
We also see that the value of total increase killed between two tools in two different ecosystems are similar (14\% in \smallamp and 20\% in \dspot).

In conclusion, the results of experiments show that by using a profiling step and collecting type information, we can successfully adopt a test amplification approach in a dynamically typed language.

\begin{acknowledgements}
This work is supported by (a) the Fonds de la Recherche Scientifique-FNRS and the Fonds Wetenschappelijk Onderzoek - Vlaanderen (FWO) under EOS Project 30446992 SECO-ASSIST (b) Flanders Make vzw, the strategic research centre for the manufacturing industry. Bergel is very grateful to Lam Research and Fondecyt Regular 1200067 for partially sponsoring this work.

This work originated during a sabbatical leave by Serge Demeyer in the RMOD lab.
St\'{e}phane Ducasse, Marcus Denker and Julien Delplanque helped a lot with the intricacies of Pharo/Smalltalk; Pavel Krivanek helped with \toolname{MuTalk}.
Finally, we thank the developers reviewing and discussing our pull requests, they really helped us in improving \smallamp.

\end{acknowledgements}

\section*{Declarations}

\subsection*{Funding and/or Conflicts of interests/Competing interests}
This work is supported financially by two public funding organisation in Belgium and Chile.
There is no direct or indirect industrial support for the research reported here.
We confirm that there are no conflicts of interest.

\subsection*{Research involving Human Participants and/or Animals}

Not applicable

\subsection*{Informed consent}

Not applicable

\bibliographystyle{spbasic}      \bibliography{Abdi2020refs}

\clearpage

\end{document}